\documentclass[prd]{revtex4}
\usepackage{graphicx}
\usepackage{lscape}
\usepackage{amssymb}
\usepackage{mathptmx}

\usepackage{graphicx}
\usepackage{amssymb,epsfig}

\begin{document}
\def\ds{\displaystyle}
\def\beq{\begin{equation}}
\def\eeq{\end{equation}}
\def\bea{\begin{eqnarray}}
\def\eea{\end{eqnarray}}
\def\beeq{\begin{eqnarray}}
\def\eeeq{\end{eqnarray}}
\def\ve{\vert}
\def\vel{\left|}
\def\ver{\right|}
\def\nnb{\nonumber}
\def\ga{\left(}
\def\dr{\right)}
\def\aga{\left\{}
\def\adr{\right\}}
\def\lla{\left<}
\def\rra{\right>}
\def\rar{\rightarrow}
\def\nnb{\nonumber}
\def\la{\langle}
\def\ra{\rangle}
\def\ba{\begin{array}}
\def\ea{\end{array}}
\def\tr{\mbox{Tr}}
\def\ssp{{\Sigma^{*+}}}
\def\sso{{\Sigma^{*0}}}
\def\ssm{{\Sigma^{*-}}}
\def\xis0{{\Xi^{*0}}}
\def\xism{{\Xi^{*-}}}
\def\qs{\la \bar s s \ra}
\def\qu{\la \bar u u \ra}
\def\qd{\la \bar d d \ra}
\def\qq{\la \bar q q \ra}
\def\gGgG{\la g^2 G^2 \ra}
\def\q{\gamma_5 \not\!q}
\def\x{\gamma_5 \not\!x}
\def\g5{\gamma_5}
\def\sb{S_Q^{cf}}
\def\sd{S_d^{be}}
\def\su{S_u^{ad}}
\def\ss{S_s^{??}}
\def\ll{\Lambda}
\def\lb{\Lambda_b}
\def\sbp{{S}_Q^{'cf}}
\def\sdp{{S}_d^{'be}}
\def\sup{{S}_u^{'ad}}
\def\ssp{{S}_s^{'??}}
\def\sig{\sigma_{\mu \nu} \gamma_5 p^\mu q^\nu}
\def\fo{f_0(\frac{s_0}{M^2})}
\def\ffi{f_1(\frac{s_0}{M^2})}
\def\fii{f_2(\frac{s_0}{M^2})}
\def\O{{\cal O}}
\def\sl{{\Sigma^0 \Lambda}}
\def\ar{&+& \!\!\!}
\def\es{\!\!\! &=& \!\!\!}
\def\ek{&-& \!\!\!}
\def\cp{&\times& \!\!\!}
\def\se{\!\!\! &\simeq& \!\!\!}
\def\hml{\hat{m}_{\ell}}
\def\rr{\hat{r}_{\Lambda}}
\def\ss{\hat{s}}
\def\tep{$B \rightarrow K_0^{*}(1430) \ell^+ \ell^-$}
\def\tepm{$B \rightarrow K_0^{*}(1430) \mu^+ \mu^-$}
\def\tept{$B \rightarrow K_0^{*}(1430) \tau^+ \tau^-$}
\def\simlt{\stackrel{<}{{}_\sim}}
\def\simgt{\stackrel{>}{{}_\sim}}
\newcommand{\shat}{\hat{s}}
\newcommand{\barB}{\overline{B}}
\newcommand{\barK}{\overline{K}}
\newcommand{\kone}{{K_1}}
\newcommand{\barkone}{{\overline{K}_1}}
\newcommand{\leftu}{\gamma^\mu L}
\newcommand{\leftd}{\gamma_\mu L}
\newcommand{\rightu}{\gamma^\mu R}
\newcommand{\rightd}{\gamma_\mu R}
\newcommand{\pA}{p_{\kone}}
\newcommand{\konel}{K_1(1270)}
\newcommand{\koneh}{K_1(1400)}
\newcommand{\barkonel}{\barK_1(1270)}
\newcommand{\barkoneh}{\barK_1(1400)}
\newcommand{\konea}{K_{1A}}
\newcommand{\koneb}{K_{1B}}
\newcommand{\barkonea}{\barK_{1A}}
\newcommand{\barkoneb}{\barK_{1B}}
\newcommand{\mkone}{m_{\kone}}
\newcommand{\konep}{K_1^+}
\newcommand{\konem}{K_1^-}
\newcommand{\konelm}{K_1^-(1270)}
\newcommand{\konehm}{K_1^-(1400)}
\newcommand{\konelp}{K_1^+(1270)}
\newcommand{\konehp}{K_1^+(1400)}
\newcommand{\konelz}{\overline{K}{}^0_1(1270)}
\newcommand{\konehz}{\overline{K}{}^0_1(1400)}
\newcommand{\degree}{^\circ}
\newcommand{\ket}[1]{{#1} \rangle}
\newcommand{\bra}[1]{\langle {#1}}
\newcommand{\ebar}{{\bar{e}}}
\newcommand{\sbar}{\bar{s}}
\newcommand{\cbar}{\bar{c}}
\newcommand{\bbar}{\bar{b}}
\newcommand{\qbar}{\bar{q}}\renewcommand{\l}{\ell}
\newcommand{\lbar}{\bar{\ell}}
\newcommand{\psibar}{\bar{\psi}}
\newcommand{\lpm}{\l^+\l^-}
\newcommand{\epm}{e^+e^-}
\newcommand{\mupm}{\mu^+\mu^-}
\newcommand{\taupm}{\tau^+\tau^-}
\newcommand{\hats}{\hat{s}}
\newcommand{\hatp}{\hat{p}}
\newcommand{\hatq}{\hat{q}}
\newcommand{\hatm}{\hat{m}}
\newcommand{\hatu}{\hat{u}}
\newcommand{\alphaem}{\alpha_{\rm em}}
\newcommand{\AFB}{A_{\rm FB}}
\newcommand{\barAFB}{\overline{A}_{\rm FB}}
\newcommand{\A}{{\cal A}}
\newcommand{\B}{{\cal B}}
\newcommand{\C}{{\cal C}}
\newcommand{\D}{{\cal D}}
\newcommand{\E}{{\cal E}}
\newcommand{\F}{{\cal F}}
\newcommand{\G}{{\cal G}}
\renewcommand{\H}{{\cal H}}
\newcommand{\T}{{\cal T}}
\renewcommand{\Re}{\mathop{\mbox{Re}}}
\renewcommand{\Im}{\mathop{\mbox{Im}}}
\newcommand{\eff}{{\rm eff}}
\newcommand{\GeV}{{\,\mbox{GeV}}}
\newcommand{\MeV}{{\,\mbox{MeV}}}
\newcommand{\Bm}{B^-}
\newcommand{\Bz}{\overline{B}{}^0}
\providecommand{\dfrac}[2]{\frac{\displaystyle
{#1}}{\displaystyle{#2}}}
\newcommand{\Br}{{\cal B}}

\title{
         {\Large
                 {\bf
 Polarized and Unpolarized Lepton Pair Forward-backward Asymmetries   in  $\overline{B}\rightarrow \overline{K}_{0}^{*}(1430) \ell^+\ell^-$ and  $\overline{B}\rightarrow \overline{K} \ell^+\ell^-$ Decays in
 Two Higgs Doublet Model
                  }
         }
      }

\author{F. Falahati$^{a}$\footnote{falahati@shirazu.ac.ir} and H. Abbasi}
\affiliation{$^{a}$Physics Department and Biruni Observatory, College of Sciences, Shiraz University, Shiraz 71454,
Iran}

\small{\begin{abstract}

In this paper we  shall focus on the effects of concrete models such as SM and Model III of 2HDM on the polarized and unpolarized  forward-backward  asymmetries of $\overline{B}\rightarrow \overline{K}_0^{*}(1430) \ell^+\ell^-$ and $\overline{B}\rightarrow \overline{K} \ell^+\ell^-$ decays.  The obtained results of these decay modes are compared  to each other. Also, we obtain the minimum required number of events for detecting each asymmetry and compare them with the number of  produced $B\bar{B}$ pairs  at the LHC or supposed to be produced at the Super-LHC. At the end, we conclude that the study of these asymmetries for  $\overline{B}\rightarrow \overline{K}_0^*(1430) \ell^+\ell^-$ and $\overline{B}\rightarrow \overline{K} \ell^+\ell^-$ processes
are very effective tools for establishing new
physics  in the future B-physics experiments.

\end{abstract}

\maketitle

Packs numbers: 12.60.-i, 13.30.-a, 14.20.Mr

\section{Introduction}
The flavor changing neutral current (FCNC) processes induced by $b\to s \ell^+
\ell^-(\ell=e,\mu,\tau)$ transitions
provide an important testing
ground to test the standard model (SM) at one loop
level, since they are forbidden in SM at tree level \cite{r1,r2}. Therefore  these decays are
very sensitive to the physics beyond the SM via the influence of new particles in the loops.

Although the branching ratios of FCNC decays are small in the SM,  interesting
results are yielded in developing experiments. The inclusive $b\to X_s \ell^+
\ell^-$ decay is observed
in BaBaR \cite{r3} and Belle collaborations. These collaborations also measured exclusive modes
$B\to K \ell^+\ell^-$ \cite{r4,r5,r6} and $B\to K^* \ell^+\ell^-$ \cite{r7}. These experimental results have
high agreement with theoretical predictions \cite{r8,r9,r10}.

There exists another  group of rare decays induced by $b\to s$ transition, such as $B\to K_2^*(1430) \ell^+\ell^-$ and $B\to K_0^*(1430) \ell^+\ell^-$
in which B meson decays into a tensor or  scalar meson, respectively. These decays are deeply investigated in SM in \cite{r11,r12} and the related transition form factors  are formulated within the framework of light front quark model \cite{r12,r13,r14} and  QCD sum rules method \cite{r15,r16}, respectively. Lately  these rare decays  have  been the matter of various
physical discussions in the frame work of some new physics models, such as  models including universal extra dimension
\cite{r17}, supersymmetry particles
\cite{r18} and the fourth-generation fermions
\cite{r19}. Generally, by studying the physical observables of these  decay modes  there would be a chance for  testing SM
or probing  possible NP models.  These physical quantities are for example the branching ratio, the
forward-backward asymmetry, the lepton polarization asymmetry,  the isospin asymmetry and etc.

The SM of electroweak
interactions has been strictly tested over the
past twenty years and shows an  excellent compatibility with all collider
data.  The dynamics of electroweak symmetry breaking, however, is not exactly known.
While the simplest possibility is the minimal Higgs
mechanism which suggests a single scalar SU(2) doublet,
many extensions of the SM predict a large  Higgs sector to contain
more scalars~\cite{r20,r21}.

Two conditions which tightly constrain the extensions of the SM Higgs sector are first the value of  rho parameter, $\rho \equiv M_W^2 / M_Z^2
\cos^2\theta_W \simeq 1$, where $M_W$ ($M_Z$) is the $W^{\pm}$ ($Z$)
boson mass and $\theta_W$ is the weak mixing angle and second the
absence of large flavor-changing neutral currents.  The first of these
conditions is spontaneously fulfilled by Higgs sectors that consists only
 SU(2) doublets (with the possibly  additional  singlets).  The
simplest such model that contains a charged Higgs boson is a
two-Higgs-doublet model (2HDM).  The second of these conditions is
spontaneously satisfied by models in which the masses of fermions  are produced trough couplings to exactly
one Higgs doublet; this is known as natural flavor
conservation and forbids the tree-level
flavor-changing neutral Higgs interactions phenomena.

Imposing natural flavor
conservation by considering an
{\it ad hoc} discrete symmetry ~\cite{r22}, there would be two different
ways to couple the SM fermions to two Higgs doublets. The Type-I and -II 2HDMs which have been studied extensively in the literatures, are such models~\cite{r21}. Without considering discrete symmetry  a more general form of  2HDM,
namely, model III has been obtained which allows  for the presence
of FCNC at tree level. Consistent with the low energy constraints,
the FCNC's involving the first two generations are highly
suppressed, and those involving the the third generation is not as
severely suppressed as the first two generations.
Also, in such a model there exists rich induced CP-violating sources
from a single CP phase of vacuum that is absent in the SM, model I
and model II. In order to consider the flavor-conserving limit of the Type III, we suppose that the two Yukawa
matrices for each fermion type to be diagonal in the same
fermion mixing basis~\cite{r23}. All three  structures of 2HDM  generally contain two scalar Higgs bosons $h^0$, $H^0$, one pseudoscalar Higgs boson $A^0$ and one charged Higgs boson $H^\pm$.

The aim of the present paper is to perform  a comprehensive study regarding the polarized and unpolarized forward-backward asymmetries of $\overline{B}\rightarrow \overline{K}_0^{*}(1430) \ell^+\ell^-$ decays in the SM and the Model III of 2HDM. Also, we consider the influences of such models on the same asymmetries of $\overline{B}\rightarrow \overline{K} \ell^+\ell^-$ decays.
In such a way, we study the sensitivity of results to the scalar  property or the pseudo-scalar property of produced mesons.

The  paper is organized as follows. In Section II, stating from the 2HDM form of four-Fermi interactions we derive
the expressions for the matrix elements of $B$ to a scalar   meson and $B$ to a pseudo-scalar meson,
here $\overline{B}\rightarrow \overline{K}_0^*(1430) \ell^+\ell^-$ and $\overline{B}\rightarrow \overline{K} \ell^+\ell^-$, respectively. Then the general expressions for the the polarized and unpolarized lepton pair forward-backward asymmetries have been extracted out. The
sensitivity of these polarizations and the corresponding averages to the model III 2HDM parameters   have been numerically
analyzed in Section III, In the final section a brief summery of our results is presented.

\section{Analytic Formulas }
\subsection{The Effective Hamiltonian for $\overline{B} \to \overline{K}  \ell^{+}\ell^{-}$  and $\overline{B} \to \overline{K}_0^*(1430)  \ell^{+}\ell^{-}$ transitions in SM and 2HDM}
The exclusive  decays $\overline{B} \to \overline{K}  \ell^{+}\ell^{-}$  and $\overline{B} \to \overline{K}_0^*(1430)  \ell^{+}\ell^{-}$ are described at quark level by $b\rar s \ell^{+}\ell^{-} $ transition.
The effective Hamiltonian, that is used to describe the $b\rar s \ell^{+}\ell^{-} $ transition in 2HDM models is:
\small{\bea {\cal
H}_{eff}(b\to s \ell^{+}\ell^{-}) = -\frac{4 G_F}{\sqrt 2} V_{tb} V^*_{ts}
\left\{\sum_{i=1}^{10} C_i( \mu ) O_i( \mu ) + \sum_{i=1}^{10}
C_{Q_i}( \mu ) Q_i( \mu ) \right\} ~, \eea}where the first part  is related to the effective Hamiltonian in the SM such that the respective  Wilson coefficients get extra terms due to the presence of charged Higgs bosons.  The second part which includes new operators is extracted from contributing the  massive neutral Higgs bosons to this decay.   All operators as well as the  related  Wilson coefficients   are given in ~\cite{r24, r25,r26}.
Now, using the
above  effective Hamiltonian, the one-loop matrix elements
of $b \rightarrow s \ell^+ \ell^-$   can be given as:
\bea {\cal M} &=& <s \ell^+  \ell^- |{\cal
H}_{\rm eff}|b> \nonumber\\
&=& -\frac{G_F \alpha}{2\sqrt 2 \pi} V_{tb} V^*_{ts}
\Bigg\{\tilde{C}_{9}^{\rm eff} \bar s \gamma_\mu (1- \gamma_5) b \,
\bar \ell \gamma^\mu \ell + \tilde{C}_{10} \bar s
\gamma_\mu (1- \gamma_5) b \, \bar \ell \gamma^\mu \gamma_5 \ell   \nnb \\
&-& 2C_7^{\rm eff}\frac{m_b}{q^2}\bar s i \sigma_{\mu \nu}q^\nu
(1+\gamma_5) b\,\bar \ell \gamma^\mu \ell-2C_7^{\rm
eff}\frac{m_s}{q^2}\bar s i \sigma_{\mu
\nu}q^\nu (1-\gamma_5) b \, \bar \ell \gamma^\mu \ell\nnb\\
 &+&C_{Q_1} \bar s (1 + \gamma_5) b \,\bar \ell \ell + C_{Q_2} \bar s
(1+\gamma_5) b \, \bar \ell \gamma_5 \ell \Bigg\}~. \label{me} \eea
The evolution of Wilson
coefficients $C_7^{\rm eff},~\tilde{C}_{9}^{\rm eff}$, $\tilde{C}_{10}$ from the higher scale $\mu=m_W$ to the lower scale $\mu=m_b$ is described by the renormalization group equation. These coefficients at the scale $\mu=m_b$ are calculated in \cite{r24,r25,r26} and $C_{Q_1}$ and $C_{Q_2}$ at the same scale to leading order are calculated in \cite{r26}. It should be noted that  the coefficient $\tilde{C}_9^{\rm eff}(\mu)$ can be decomposed into the following three parts:
\small{\bea
\tilde{C}_9^{\rm eff}(\mu)=\tilde{C}_9(\mu)+Y_{SD}(\hat{m}_c, \hat{s})+Y_{LD}(\hat{m}_c, \hat{s})~,
\eea}
where the parameters $\hat{m}_c$ and $\hat{s}$ are defined as $\hat{m}_c=m_c/m_b$, $\hat{s}=q^2/m_b^{2}$.
$Y_{SD}(\hat{m}_c, \hat{s})$ describes the short-distance
contributions from four-quark operators far away from the $c\bar{c}$ resonance regions, which can be calculated reliably in
the perturbative theory. The function $Y_{SD}(\hat{m}_c, \hat{s})$ is
given by:\small{\begin{eqnarray} Y_{SD}&=&
g(\hat{m}_c, \hat{s}) (3 C_1 + C_2 + 3 C_3 + C_4 + 3 C_5 + C_6)
\nonumber
\\&-& \frac{1}{2} g(1, \hat{s}) (4 C_3 + 4 C_4 + 3 C_5 + C_6) \nonumber
\\&-& \frac{1}{2} g(0, \hat{s}) (C_3 + 3 C_4) + \frac{2}{9} (3 C_3 +
C_4 + 3 C_5 + C_6), \end{eqnarray}} where the  explicit expressions
for the $g$ functions can be found  in ~\cite{r24}.
 The long-distance contributions $Y_{LD}(\hat{m}_c, \hat{s})$ from four-quark operators near the $c\bar{c}$ resonance
cannot be calculated from first principles of QCD and are usually parameterized in the form of a phenomenological
Breit-Wigner formula making use of the vacuum saturation approximation and quark-hadron duality. The function $Y_{LD}(\hat{m}_c, \hat{s})$ is 
given by \cite{ref7,ref8}:\small{\bea Y_{\rm LD} = {3\pi \over \alpha^2} C^{(0)} \sum_{V_i=\psi, \psi',
\cdots } k_i \, {\Gamma(V_i \rar \ell^+ \ell^-) m_{V_i}
\over m_{V_i}^2- q^2-im_{V_i}\Gamma_{V_i}} ~, \nnb \eea}
where
$\alpha$ is the fine structure constant and $C^{(0)}=(3 C_1 + C_2 + 3 C_3 + C_4 + 3 C_5 + C_6)$.
The phenomenological parameters $k_i$ for the
$\overline{B} \rar \overline{K}
\ell^+ \ell^-$ decay can be fixed from ${Br} (\overline{B}
\rar J/\psi \overline{K} \rar \overline{K} \ell^+ \ell^-) = {Br} (\overline{B}
\rar J/\psi \overline{K}) {Br} (J/\psi \rar \ell^+ \ell^-)$. For the lowest resonances $\psi$ and $\psi^{\prime}$ we will use $k_1=2.70$ and $k_2=3.51$,  respectively~\cite{r15}.
Also, for the $\overline{B} \rar \overline{K}_0^*(1430)
\ell^+ \ell^-$ decay such parameters can be determined by ${Br} (\overline{B}
\rar J/\psi \overline{K}_0^*(1430) \rar \overline{K}_0^*(1430) \ell^+ \ell^-) = {Br} (\overline{B}
\rar J/\psi \overline{K}_0^*(1430)) {Br} (J/\psi \rar \ell^+ \ell^-)$. However, since
the branching ratio of $\overline{B}
\rar J/\psi \overline{K}_0^*(1430)$ decay has not been measured yet, we assume that the values of
$k_i$ are in the order of one. Therefore, we use $k_1=1$ and $k_2=1$ in the following numerical calculations \cite{ref8}.

\subsection{Form factors for $\overline{B} \to \overline{K} \ell^{+}\ell^{-}$ transition}
The exclusive $\overline{B} \to \overline{K} \ell^{+}\ell^{-}$ decay is described in terms of
the matrix elements of the quark operators in Eq. (\ref{me}) over meson
states, which can be parameterized in terms of the form factors.
The needed  matrix elements for the calculation of  $\overline{B} \to \overline{K} \ell^{+}\ell^{-}$ decay are:
\small{\bea
&&\lla \overline{K}\vel \bar s \gamma_\mu (1 - \gamma_5)
b \ver \overline{B} \rra~,\nnb \\
&&\lla  \overline{K} \vel \bar s i\sigma_{\mu\nu} q^\nu
(1 \pm \gamma_5) b \ver\overline{B} \rra~, \nnb \\
&&\lla  \overline{K} \vel \bar s (1 + \gamma_5) b
\ver \overline{B} \rra~,
\label{me3}\eea}
which can be  obtained as follows:
\small{\bea \label{e8402} \lla
\overline{K}(p_{K}) \vel \bar{s}\gamma_\mu (1 \pm \gamma_5) b \ver
\overline {B}(p_B) \rra \es \big[f_+ (q^2) {(p_B + p_{K})}_\mu + f_-(q^2) q_\mu\big], \\\nonumber
\es f_+ (q^2) \big[{(p_B + p_{K})}_\mu-{(m_B^2-m_K^2)\over{q^2}}q_\mu \big]+ {(m_B^2-m_K^2)\over{q^2}} f_0(q^2) q_\mu,\\
\label{e8403} \lla  \overline{K}(p_{K}) \vel \bar{s}i
\sigma_{\mu\nu} q^\nu (1 \pm \gamma_5) b  \ver \overline{B}(p_B) \rra \es {- f_T (q^2)
\over m_B + m_{K}} \big[ {(p_B + p_{K})}_\mu q^2 - (m_B^2 -
m_{K}^2 ) q_\mu \big], \eea}
where $q = p_B-p_{K}$ is the momentum transfer. In deriving Eq. (\ref{e8402}) we have used the relationship
\small{\bea
\label{e8404} f_-(q^2)={(m_B^2-m_K^2)\over{q^2}} \big[f_0(q^2)-f_+(q^2)\big].
\eea}
Now,  multiplying  both sides of Eq. (\ref{e8402}) with $q^\mu$ and using the equation of
motion,   the expression in terms of form factors for $\lla \overline{K} \vel \bar s (1 \pm \gamma_5 ) b \ver \overline{B} \rra$
is calculated as:
\small{\bea
\label{e8405}
\lla \overline{K}(p_{K}) \vel \bar s (1 \pm \gamma_5) b \ver
\barB(p_B) \rra &=&\lla \overline{K}(p_{K}) \vel \bar s b \ver
\barB(p_B) \rra =
\frac{1}{m_{b}-m_{s}}[f_+ (q^2) (p_B + p_{K}).q +
f_-(q^2) q^{2}]~, \\\nonumber
&=& \frac{f_0 (q^2)}{m_{b}-m_{s}}(m_B^2-m_K^2) \\
\lla \overline{K}(p_{K}) \vel \bar s \gamma_5  b \ver
\barB(p_B) \rra&=&0~.
\eea}
For the form factors we have used the light cone QCD sum rules results
~\cite{r27} in which
the $q^2$ dependence of the semileptonic form factors,$f_0$ and $f_+$, is given by
\small{\bea
F(q^2)=\,{F(0)\over 1-a_F(q^2/m_{B}^2)+b_F(q^2/m_{B}^2)^2}~,
\eea}
where the values of parameters $F(0)$, $a_F$ and $b_F$ for the
$\overline{B} \to \overline{K} \ell^{+}\ell^{-}$ decay are listed in table \ref{tab:BK}. Also, the $q^2$ dependence of the penguin form factor, $f_T$, is obtained by
\small{\bea
\label{e8406} { f_T (q^2)
\over m_B + m_{K}}=\frac{1}{m_b}\big[(1+{{m_b^2-q^2}\over{q^2}})f_+-{{m_b^2-q^2}\over{q^2}}f_0\big].
\eea}
\begin{table}[t]
\begin{center}
\caption{Form factors for $\overline{B} \to \overline{K}$ transition obtained
in the LCQSR calculation are fitted to the
3-parameter form.} \label{tab:BK}
\begin{tabular}{clll}
\hline\hline
       $F$
    & $F(0)$
    & $a_F$
    & $b_F$
    \\
    \hline
$f_+^{\overline{B} \to \overline{K}}$ &$0.341 \pm 0.051$ & $1.410$ & $\phantom{-}0.406$ \\
$f_0^{\overline{B} \to \overline{K}}$ &$0.341 \pm 0.051$ & $0.410$ & $-0.361$\\
\end{tabular}
\end{center}
\end{table}

\subsection{Form factors for $\overline{B} \to \overline{K}_0^*(1430) \ell^{+}\ell^{-}$ transition}
 Like the exclusive $\overline{B} \to \overline{K} \ell^{+}\ell^{-}$ decay, the  $\overline{B} \to \overline{K}_0^*(1430) \ell^{+}\ell^{-}$ transition  is expressed by the matrix elements appeared in  Eq. (\ref{me3}) except $K$ is replaced by $K_0^*(1430)$. These  physical objects  could be parameterized
as:
\small {\bea \label{e90} \lla
\overline{K}_0^\ast(1430)(p_{K_0^\ast}) \vel \bar{s}\gamma_\mu (1 \pm \gamma_5) b \ver
\overline {B}(p_B) \rra \es \pm\big[f_+ (q^2) {(p_B + p_{K_0^\ast})}_\mu + f_-(q^2) q_\mu\big], \\
\label{e91} \lla  \overline{K}_0^\ast(1430)(p_{K_0^\ast}) \vel \bar{s}i
\sigma_{\mu\nu} q^\nu (1 \pm \gamma_5) b  \ver \overline{B}(p_B) \rra \es {\pm f_T (q^2)
\over m_B + m_{K_0^\ast}} \big[ {(p_B + p_{K_0^\ast})}_\mu q^2 - (m_B^2 -
m_{K_0^\ast}^2 ) q_\mu \big],\\
\label{e92}
\lla \overline{K}_0^\ast(1430)(p_{K_0^\ast}) \vel \bar s (1 \pm \gamma_5) b \ver
\barB(p_B) \rra &=&\pm\lla \overline{K}_0^\ast(1430)(p_{K_0^\ast}) \vel \bar s   \gamma_5 b \ver
\barB(p_B) \rra =
\mp\frac{1}{m_{b}+m_{s}}[f_+ (q^2) (p_B + p_{K_0^\ast}).q +
f_-(q^2) q^{2}]\\\nonumber\es\mp \frac{f_0 (q^2)}{m_{b}+m_{s}}(m_B^2-m_{K_0^*}^2) , \\
\label{e93}
\lla \overline{K}_0^\ast(1430)(p_{K_0^\ast}) \vel \bar s    b \ver
\barB(p_B) \rra&=&0~.
 \eea}
where $q =p_B -p_{K_0^\ast}$ and the function  $f_0(q^2)$ has been extracted from
the Eq. (\ref {e8404}).
For the form factors we have used the results of three-point QCD sum rules method
~\cite{r15} in which
the $q^2$ dependence of the all form factors is given by
\small{\bea
F(q^2)=\,{F(0)\over 1-a_F(q^2/m_{B}^2)+b_F(q^2/m_{B}^2)^2},
\eea}
where the values of parameters $F(0)$, $a_F$ and $b_F$ for the
$\overline{B} \to \overline{K}_0^*(1430) \ell^{+}\ell^{-}$ decay are exhibited in table \ref{tab:BK0star}.

\begin{table}[t]
\begin{center}
\caption{Form factors for $\overline{B} \to \overline{K}_0^*(1430)$ transition obtained
within three-point QCD sum rules are fitted to the
3-parameter form.} \label{tab:BK0star}
\begin{tabular}{clll}
\hline\hline
       $F$
    & $F(0)$
    & $a_F$
    & $b_F$
    \\
    \hline
$f_+^{\overline{B} \to \overline{K}_0^*}$ &$\phantom{-}0.31\pm 0.08$ & $0.81$ & $-0.21$ \\
$f_-^{\overline{B} \to \overline{K}_0^*}$ &$-0.31 \pm 0.07 $ & $0.80$ & $-0.36$\\
$f_T^{\overline{B} \to \overline{K}_0^*}$ &$-0.26 \pm 0.07 $ & $0.41$ & $-0.32$\\
\end{tabular}
\end{center}
\end{table}

\subsection{The differential decay rates and forward-backward asymmetries of $\overline{B} \to \overline{K}_0^{*}(1430) \ell^{+}\ell^{-}$}
Making use of  Eq.(\ref{me}) and the  definitions of form factors, the matrix element of the  $\overline{B} \to \overline{K}_0^{*}(1430) \ell^{+}\ell^{-}$
decay can be written as follows:
\begin{eqnarray}\label{ampl}
{\cal M} &=& \frac{G_F \alpha_{\rm em}}{4\sqrt{2}\pi} V_{ts}^*
V_{tb}^{}\, m_B \nnb \\ &&\Bigg\{
  [{\cal A}(p_B+p_{{K}_0^{*}}+{\cal B}q_{\mu})_{\mu}] \lbar \gamma^\mu \l
 + [{\cal C}(p_B+p_{{K}_0^{*}}+{\cal D}q_{\mu})_{\mu}]\lbar \gamma^\mu \gamma_5 \l
 +[{\cal Q}]\lbar \l+[{\cal N}]\lbar  \gamma_5 \l\Bigg\},
\end{eqnarray}
where the auxiliary functions $\A,
\cdots, {\cal Q}$ are listed  in the following:
\begin{eqnarray}
\A&=& -2\tilde{C}_{9}^{\rm eff}f_{+}(q^2)-4(m_b+m_s){C}_{7}^{\rm eff}{{f_T(q^2) }\over{m_B+m_{{K}_0^{*}}}},\label{fmA}\\
\B&=&  -2\tilde{C}_{9}^{\rm eff}f_{-}(q^2)+4(m_b+m_s){C}_{7}^{\rm eff}{{f_T(q^2) }\over{(m_B+m_{{K}_0^{*}}})q^2}(m_B^2-m_{{K}_0^{*}}^2),\\
\C&=&-2\tilde{C}_{10}f_{+}(q^2),\\
\D&=&-2\tilde{C}_{10}f_{-}(q^2),\\
{\cal Q}&=&-2C_{Q_{1}}f_{0}(q^2){(m_B^2-m_{{K}_0^{*}}^2)\over{m_b+m_s}},\\
{\cal N}&=&-2C_{Q_{2}}f_{0}(q^2){(m_B^2-m_{{K}_0^{*}}^2)\over{m_b+m_s}},\label{fmN}
\end{eqnarray}
with $q = p_B - p_{{K}_0^{*}} = p_{\ell^+} + p_{\ell^-} $.

The unpolarized differential decay rate  for the
$\barB\to \overline{K}_0^{*}(1430)\lpm$ decay in the rest frame of $B$ meson is given by:
\begin{eqnarray}
\frac{d \Gamma(\barB\to{{K}_0^{*}}\lpm)}{d \hats} = -\frac{G_F^2
\alphaem^2 m_B}{2^{14}\pi^5}
 \left|V_{tb}V_{ts}^*\right|^2 v\sqrt{\lambda}\Delta,
\end{eqnarray}
with
\bea\label{dgds1}\nnb \Delta &=&16 m_{\ell} m_{B}^2 (1-\hat{r}_{{{K}_0^{*}}})\mbox{\rm Re}[\C{\cal N^*}]+4\hats m_B^2 v^2|{\cal Q}|^2+16 \hats m_{\ell}^2 m_B^2  |\D|^2+32 m_{\ell}^2 m_B^2 (1-\hat{r}_{{{K}_0^{*}}})\mbox{\rm Re}[\C{\cal D^*}]\nnb\\
&+&16\hats m_{\ell} m_B^2\mbox{\rm Re}[\D{\cal N^*}] +2 \hats m_B^2 |{\cal N}|^2+\frac{4}{3} m_B^4 \lambda (3-v^2)|\A|^2\nnb\\
&+&\frac{4}{3}m_B^4 |\C|^2\{2\lambda-(1-v^2)(2\lambda-3(1-\hat{r}_{{{K}_0^{*}}})^2)\},
\end{eqnarray}
where $v=\sqrt{1-4{m}_\ell^2/q^2}$, $\hats=q^2/m_B^2$, $\hat{r}_{{K}_0^{*}}=m_{{K}_0^{*}}^2/m_B^2$ and $\lambda = 1 +
\hat{r}_{{K}_0^{*}}^2 + \hats^2 - 2\hats - 2\hat{r}_{{K}_0^{*}} (1+\hats)$.

The unpolarized and normalized differential forward-backward asymmetry of the
$\barB\to\overline{K}_0^{*}(1430\lpm$ decay in the center of mass frame of leptons  is defined by:

\bea
\label{e6312}
{\cal A}_{FB} = \frac{\ds \int_{0}^{1} \frac{d^2\Gamma}{d\hat{s} dz} -
\int_{-1}^{0} \frac{d^2\Gamma}{d\hat{s} dz}}
{\ds \int_{0}^{1} \frac{d^2\Gamma}{d\hat{s} dz} +
\int_{-1}^{0} \frac{d^2\Gamma}{d\hat{s} dz}}~,
\eea
where $z=\cos\theta$ and $\theta$ is the angle between three momenta of the $B$ meson and the negatively charged lepton ($\ell^-$) in the CM (center of mass) frame of leptons.

Using the above-mentioned definition, the result
can be written as follows:
\begin{eqnarray}
 {\cal A}_{FB}(\hats)&=&\frac{8 m_B^2 m_{\ell}v \sqrt{\lambda} }{\Delta} \mbox{\rm Re}[{\cal AQ^\ast}].
\end{eqnarray}

Having obtained the unpolarized and normalized differential forward-backward  asymmetry , let us now consider the
 normalized differential forward-backward  asymmetries associated with the polarized leptons.
 For this
purpose, we first define the following orthogonal unit vectors $s_i^{\pm\mu}$ in
the rest frame of $\ell^\pm$, where $i=L,N$ or $T$ are the abbreviations of
the longitudinal, normal and transversal spin projections, respectively:
\bea
\label{e6310}
s^{-\mu}_L \es \ga 0,\vec{e}_L^{\,-}\dr =
\ga 0,\frac{\vec{p}_{\ell^-}}{\vel\vec{p}_{\ell^-} \ver}\dr~, \nnb \\
s^{-\mu}_N \es \ga 0,\vec{e}_N^{\,-}\dr = \ga 0,\frac{\vec{p}_{{{K}_0^{*}}}\times
\vec{p}_{\ell^-}}{\vel \vec{p}_{{{K}_0^{*}}}\times \vec{p}_{\ell^-} \ver}\dr~, \nnb \\
s^{-\mu}_T \es \ga 0,\vec{e}_T^{\,-}\dr = \ga 0,\vec{e}_N^{\,-}
\times \vec{e}_L^{\,-} \dr~, \nnb \\
s^{+\mu}_L \es \ga 0,\vec{e}_L^{\,+}\dr =
\ga 0,\frac{\vec{p}_{\ell^+}}{\vel\vec{p}_{\ell^+} \ver}\dr~, \nnb \\
s^{+\mu}_N \es \ga 0,\vec{e}_N^{\,+}\dr = \ga 0,\frac{\vec{p}_{{{K}_0^{*}}}\times
\vec{p}_{\ell^+}}{\vel \vec{p}_{{{K}_0^{*}}}\times \vec{p}_{\ell^+} \ver}\dr~, \nnb \\
s^{+\mu}_T \es \ga 0,\vec{e}_T^{\,+}\dr = \ga 0,\vec{e}_N^{\,+}
\times \vec{e}_L^{\,+}\dr~,
\eea
where $\vec{p}_{\ell^{\mp}}$ and $\vec{p}_{K_0^{*}}$ are in the
CM frame of $\ell^- \,\ell^+$ system, respectively. Lorentz transformation is used to boost the  components of the lepton polarization
to the CM frame of the lepton pair as:
\bea
\label{e6311}
\ga s^{\mp\mu}_L \dr_{CM} \es \ga \frac{\vel\vec{p}_{\ell^{\mp}}\ver}{m_\ell}~,
\frac{E_{\ell} \vec{p}_{\ell^\mp}}{m_\ell \vel\vec{p}_{\ell^{\mp}} \ver}\dr~,\nnb\\
\ga s^{\mp\mu}_N \dr_{CM} \es \ga s^{\mp\mu}_N \dr_{RF}~,\nnb\\
\ga s^{\mp\mu}_T \dr_{CM} \es \ga s^{\mp\mu}_T \dr_{RF}~,
\eea
where  $RF$ refers to the rest frame of the corresponding lepton as well as $\vec{p}_{\ell^+} = - \vec{p}_{\ell^-}$ and $E_\ell$ and $m_\ell$ are the energy and mass
of leptons in the CM frame, respectively.

The polarized and normalized differential
forward--backward asymmetry can be defined as:
\bea
\label{e6313}
{\cal A}_{FB}^{ij}(\hat{s}) \es
\Bigg(\frac{d\Gamma(\hat{s})}{d\hat{s}} \Bigg)^{-1}
\Bigg\{ \int_0^1 dz - \int_{-1}^0 dz \Bigg\}
\Bigg\{
\Bigg[
\frac{d^2\Gamma(\hat{s},\vec{s}^{\,-} = \vec{i},\vec{s}^{\,+} = \vec{j})}
{d\hat{s} dz} -
\frac{d^2\Gamma(\hat{s},\vec{s}^{\,-} = \vec{i},\vec{s}^{\,+} = -\vec{j})}
{d\hat{s} dz}
\Bigg] \nnb \\
\ek
\Bigg[
\frac{d^2\Gamma(\hat{s},\vec{s}^{\,-} = -\vec{i},\vec{s}^{\,+} = \vec{j})}
{d\hat{s} dz} -
\frac{d^2\Gamma(\hat{s},\vec{s}^{\,-} = -\vec{i},\vec{s}^{\,+} = -\vec{j})}
{d\hat{s} dz}
\Bigg]
\Bigg\}\nnb \\ \nnb \\
\es
{\cal A}_{FB}(\vec{s}^{\,-}=\vec{i},\vec{s}^{\,+}=\vec{j})   -
{\cal A}_{FB}(\vec{s}^{\,-}=\vec{i},\vec{s}^{\,+}=-\vec{j})  -
{\cal A}_{FB}(\vec{s}^{\,-}=-\vec{i},\vec{s}^{\,+}=\vec{j})  \nnb \\
\ar
{\cal A}_{FB}(\vec{s}^{\,-}=-\vec{i},\vec{s}^{\,+}=-\vec{j})~,
\eea
where $\frac{d\Gamma(\hat{s})}{d\hat{s}}$ is calculated in the CM frame. Using these definitions for the double polarized $FB$ asymmetries, the following explicit forms for ${\cal A}_{FB}^{ij}$'s  are obtained:
\bea
\label{e6314}
{\cal A}_{FB}^{LL}&=&-{\cal A}_{FB}^{NN}=-{\cal A}_{FB}^{TT}={\cal A}_{FB}~,\\
\label{e6315}
{\cal A}_{FB}^{LN}&=&\frac{-16v {\lambda m_{\ell}m_B^3} }{3\sqrt{\hat{s}}\Delta}\mbox{\rm Im}[\A\C^*]~,\\
\label{e6316}
{\cal A}_{FB}^{NL} \es
{\cal A}_{FB}^{LN}~, \\
\label{e6317}
{\cal A}_{FB}^{LT} \es\frac{-16 {\lambda m_{\ell}m_B^3} }{3\sqrt{\hat{s}}\Delta}|\A|^2~,\\
\label{e6318}
{\cal A}_{FB}^{TL} \es
{\cal A}_{FB}^{LT}~, \\
\label{e6319}
{\cal A}_{FB}^{NT} \es\frac{ 8 m_B^2 m_{\ell}\sqrt{\lambda} }{\Delta}\mbox{\rm Im}\Big[-2{{m}_\ell\over {\hats}}(\A\C^*)(1-\hat{r}_{{{K}_0^{*}}})-(\A{\cal N}^*)+2{m}_\ell(\D\A^*) \Big]~,\\
\label{e6320}
{\cal A}_{FB}^{TN} \es -{\cal{ A}}_{FB}^{NT}~.
\eea

\subsection{The differential decay rates and forward-backward asymmetries of $\overline{B} \to \overline{K}\ell^{+}\ell^{-}$}
   Imposing $m_s=0$ in the whole afore-mentioned expressions for $\overline{B} \to \overline{K}_0^{*}(1430)\ell^{+}\ell^{-}$ and $\overline{B} \to \overline{K}\ell^{+}\ell^{-}$,  we could  obtain the similar expressions for $\overline{B} \to \overline{K}\ell^{+}\ell^{-}$ decay, such that all the above equations remain unchanged except   the definitions of the auxiliary functions (Eqs.(\ref{fmA}-\ref{fmN})). It is obvious from the matrix elements of the above-said decays, in order to obtain the auxiliary functions of the latter decay we should perform  the following substitutions:
\bea
f_+^{K_0^{*}}\to -f_+^{K},~~~~~~~~~~~~~f_-^{K_0^{*}}\to -f_-^{K},~~~~~~~~~~~~~f_T^{K_0^{*}}\to -f_T^{K}.\nnb\\
\eea

\section{Numerical Results and Discussion}
In this section we  shall focus on the concrete models such as SM and Model III of 2HDM. We study the effects of such models on the polarized and unpolarized  forward-backward  asymmetries and their averages for $\overline{B}\rightarrow \overline{K}_0^{*}(1430) \ell^+\ell^-$ and $\overline{B}\rightarrow \overline{K} \ell^+\ell^-$ decays. At the end, we compare the results of  different decay modes to  each other.  The  corresponding averages are defined by the following equation \cite{ref9}: \bea \la {\cal A}_{FB}^{ij} \ra = \frac{\ds
\int_{4 \hat{m}_\ell^2}^{(1-\sqrt{\hat{r}_{M}})^2} {\cal A}_{FB}^{ij}
\frac{d{\cal B}}{d \hat{s}} d \hat{s}} {\ds \int_{4
\hat{m}_\ell^2}^{(1-\sqrt{\hat{r}_{M}})^2} \frac{d{\cal B}}{d
\hat{s}} d \hat{s}}~, \eea
where the subscript M refers to $\overline{K}_{0}^*(1430)$ and $\overline{K}$ mesons.
The full kinematical interval of the dilepton invariant mass $q^2$ is $4
m_\ell^2 \le q^2 \le (m_B - m_M)^2$ for which the long
distance contributions (the charmonium resonances) can give substantial
effects by considering  the two
low lying resonances $J/\psi$ and $\psi^\prime$, in the interval of $8~GeV^2\le q^2 \le
14~GeV^2$. In order to decrease the hadronic uncertainties we
use the kinematical region of $q^2$ for muon as \cite{ref8}:
\bea
\begin{array}{cl}
\mbox{\rm I} & 4 m_\ell^2 \le q^2 \le (m_{J\psi} - 0.02~GeV)^2~,\\ \\
\mbox{\rm II} & (m_{J\psi} + 0.02~GeV)^2 \le q^2 \le
(m_{\psi^\prime} - 0.02~GeV)^2~, \\ \\
\mbox{\rm III} & (m_{\psi^\prime} + 0.02~GeV)^2 \le q^2 \le
(m_B-m_M)^2~, \nnb
\end{array} \nnb
\eea and for tau as: \bea
\begin{array}{cl}
\mbox{\rm I} & 4 m_\ell^2 \le q^2 \le (m_{\psi'} - 0.02~GeV)^2~,\\ \\
\mbox{\rm II} & (m_{\psi'} + 0.02~GeV)^2 \le q^2 \le
(m_B-m_M)^2. \nnb
\end{array} \nnb
\eea

 In   Model III of 2HDM apart from the masses of Higgs bosons,  two vertex parameters, $\lambda_{tt}$ and $\lambda_{bb}$,  are appeared in  the calculations of the related Feynnman diagrams.  Since  these coefficients can be complex, we can rewrite the following combination as:
 \beq
\lambda_{tt}\lambda_{bb} =
|\lambda_{tt}\lambda_{bb}|\,e^{i\theta}, \eeq in which the range of variations for
$|\lambda_{tt}|$, $|\lambda_{bb}|$ and the phase angle $\theta$ are
given by  the experimental limits of the  electric dipole
moments of  neutron(NEDM), $B^0 - \bar B^0$ mixing, $\rho
\,_0$, $R_b$ and $Br(b \rar s \gamma)$\cite{r21, r23, r28, r29}.
The experimental bounds on NEDM and $Br(b \rar s \gamma)$ as well as $M_{H^+}$ which are obtained at $\rm{LEP II}$  constrain  $\lambda_{tt}\lambda_{bb}$ to be closely equal to 1 and the phase angle $\theta$ to be between $60^{\circ}-90^{\circ}$. The next restriction which comes from the  experimental value of $x_d$ parameter, corresponding to the $B^0 - \bar B^0$ mixing, controls $|\lambda_{tt}|$ to be less than  $ 0.3$. Also, the experimental value of parameter $R_b$ which is defined as $R_b\equiv\frac {\Gamma(Z\rightarrow b\bar{b})}{\Gamma(Z\rightarrow\rm{hadrons})}$ affects on the magnitude of $|\lambda_{bb}|$ in such away this coefficient could be  around $50$. Using these restrictions and taking $\theta = \pi/2$, we consider the following three typical
parameter cases  throughout the
numerical analysis\cite{r23}:
\bea
\rm{ \, Case A}: \,\,\,|\lambda_{tt}|&=& 0.03; \,\,\,\,\,|\lambda_{bb}|=
100,\nnb\\ \rm{\, Case B}: \,\,\,|\lambda_{tt}|&=& 0.15; \,\,\,\,\,|\lambda_{bb}|=
50,\nnb\\ \rm{\, Case C}: \,\,\,|\lambda_{tt}|&=& 0.3; \,\,\,\,\,\,\,\,|\lambda_{bb}|=
30.
\eea

The other main input parameters are the form factors which are listed in tables \ref{tab:BK} and \ref{tab:BK0star}.  In addition, in  this study we have applied four sets of masses of Higgs bosons which are displayed in table\ref{tabmassset}\cite{r23}.
\begin{table}[ht]
\begin{center}
\caption{List of the values for the masses of the Higgs particles.} \label{tabmassset}
\begin{tabular}{cllll}
\hline\hline

    & $\rm{ m_{H^{\pm}}}$
    & $\rm{ m_{A^0}}$
    & $\rm{ m_{h^0}}$
    & $\rm{ m_{H^0}}$
    \\
    \hline
$\rm{mass}~\rm{ set-1}$ &$200 \rm{Gev} $ & $125 \rm{Gev}$ & $125 \rm{Gev}$ & $160 \rm{Gev}$ \\
$\rm{mass}~\rm{ set-2}$ &$160 \rm{Gev} $ & $125 \rm{Gev}$ & $125 \rm{Gev}$ & $160 \rm{Gev}$ \\
$\rm{mass}~\rm{ set-3}$ &$200 \rm{Gev} $ & $125 \rm{Gev}$ & $125 \rm{Gev}$ & $125 \rm{Gev}$ \\
$\rm{mass}~\rm{ set-4}$ &$160 \rm{Gev} $ & $125 \rm{Gev}$ & $125 \rm{Gev}$ & $125 \rm{Gev}$ \\
 \hline\hline
\end{tabular}
\end{center}
\end{table}

We have shown our analysis for the dependency of ${\cal A}_{FB}^{ij}$'s and their averages on
 the parameters of  Model III of 2HDM
 in a set of figures (\ref{AFBmKstar}-\ref{ANTtK}) and tables (\ref{masssetBK0smu12}-\ref{masssetBKtau34}), respectively. Moreover, in these tables the theoretical and experimental
uncertainties corresponding to the SM averages for  $\overline{B}\rightarrow \overline{K}_0^{*}(1430) \ell^+\ell^-$ and $\overline{B}\rightarrow \overline{K} \ell^+\ell^-$ decays  have been taken into account. It should also be mentioned finally that
the theoretical uncertainties are extracted from the hadronic uncertainties
related to the form factors and the experimental uncertainties
originate from the mass of quarks and hadrons and Wolfenstein
parameters. In the following analyses we
have just talked about the asymmetries whose predictions
are larger than 0.005 in 2HDM.

\begin{itemize}

\item \textbf { Analysis of ${\cal A}_{FB}$ asymmetries for $\overline{B}\to \overline{K}_0^{*} \mu^+ \mu^-$ and $\overline{B}\to \overline{K} \mu^+ \mu^-$ decays}: As it is obvious from   figure \ref{AFBmKstar} however the predictions of ${\cal A}_{FB}$ for $\overline{B}\to \overline{K}_0^{*} \mu^+ \mu^-$ in cases B and C for all mass sets coincide with that of SM which is zero throughout the domain $4 m_{\mu}^2<q^2< (m_B-m_{{K}_0^{*}})^2$, such coincidence is not generally seen in case A. In this case within the interval $m_{\psi^\prime}^2<q^2<(m_B-m_{{K}_0^{*}})^2$  a larger discrepancy between  the predictions of SM and 2HDM is observed compared with those predictions in the range $4 m_{\mu}^2<q^2< m_{\psi^\prime}^2$.  Also it is understood from these plots that  whenever the mass of $H^{\pm}$ increases or the mass of $H^{0}$ decreases this asymmetry shows  more sensitivity to the existence of new Higgs bosons in such a manner that the most deviation from the anticipation of SM   happens in the mass set 3 of the afore-mentioned case and range which is around $0.017$ occurring next to $q^2=(m_B-m_{{K}_0^{*}})^2$.  In contrast,  the magnitudes of averages related to tables \ref{masssetBK0smu12} and\ref{masssetBK0smu34} could not provide any signs for the presence of new Higgs bosons since those values are less than $0.005$ in both SM and 2HDM. It is also explicit from  figure \ref{AFBmK} and tables \ref{masssetBKmu12} and\ref{masssetBKmu34} that there are the same discussions regarding $\overline{B}\to \overline{K} \mu^+ \mu^-$ decay as those of $\overline{B}\to \overline{K}_0^{*} \mu^+ \mu^-$ decay except that the  dependency of ${\cal A}_{FB}$ on $q^2$ for $\overline{B}\to \overline{K}_0^{*} \mu^+ \mu^-$ decay indicates more sensitivity to the 2HDM parameters. For instance while the largest prediction of the former decay is about $0.017$, that of the latter decay is around $0.015$.
Based on this, experimental study of this observable for the $\mu$ channel of  $\overline{B}\to \overline{K}_0^{*}$ and $\overline{B}\to \overline{K}$ transitions can be suitable in looking for new Higgs bosons.

\item \textbf { Analysis of ${\cal A}_{FB}^{LN}$ asymmetries for $\overline{B}\to \overline{K}_0^{*} \mu^+ \mu^-$ and $\overline{B}\to \overline{K} \mu^+ \mu^-$ decays}:  It is seen from figure \ref{ALNmKstar} that for the  $\overline{B}\to \overline{K}_0^{*} \mu^+ \mu^-$ decay  the predictions of both mass sets 1 and 3 and both mass sets 2 and 4 are separately the same  and the deviation from the SM value in mass sets 2 and 4 is more than that in mass sets 1 and 3. Therefore, while this asymmetry is insensitive  to the variation of mass of $H^0$,   it is  susceptible to the change of mass of $H^{\pm}$, here the reduction of mass of such boson.  Also, the relevant plots show that this quantity is quite sensitive  to the variation of the parameters  $\lambda_{tt}$ and $\lambda_{bb}$. For example, by enhancing the magnitude of $|\lambda_{tt}\lambda_{bb}|$ the deviation from the SM value is increased. By combining the above analyses it is understood that the most deviation from the SM prediction occurs in the case C of mass sets 2 and 4.  Particularly at $q^2=4m_{\mu}^2$ in the afore-mentioned case and mass-sets,  a deviation around 30 times of the SM expectation is seen. In addition, it is found out through the corresponding tables  that the values of averages show the same dependencies as those of diagrams to the existence of new Higgs bosons so that the most distance between the SM prediction and that of 2HDM arises in the case C of mass sets 2 and 4 which is 16 times of the SM anticipation. It is also evident from  figure \ref{ALNmK} and tables \ref{masssetBKmu12} and\ref{masssetBKmu34} that there are the similar explanations concerning $\overline{B}\to \overline{K} \mu^+ \mu^-$ decay to those of $\overline{B}\to \overline{K}_0^{*} \mu^+ \mu^-$ decay except that two Higgs doublet scenario can flip the sign of ${\cal A}_{FB}^{LN}$ compared to the SM expectation in the latter decay in all cases and mass sets. The maximum deviations relative to the SM predictions which are observed in the respective diagrams and tables  take place in the case C of mass sets 2 and 4 which are closely $-55$ times  of the SM prediction  for the corresponding diagrams occurring at $q^2=4m_{\mu}^2$ and $-7$  times  of the SM prediction for the related averages. Therefore, it seems that the measurements of ${\cal A}_{FB}^{LN}$ and its average for each of decay modes and its sign for the latter decay mode could provide  appropriate ways to discover  new Higgs bosons.

\item \textbf { Analysis of ${\cal A}_{FB}^{NT}$ asymmetries for $\overline{B}\to \overline{K}_0^{*} \mu^+ \mu^-$ and $\overline{B}\to \overline{K} \mu^+ \mu^-$: decays}: It is found out from figure \ref{ANTmKstar} that whereas the predictions of 2HDM in the domain $4 m_{c}^2<q^2< (m_B-m_{{K}_0^{*}})^2$ for all mass sets and cases conform to that of SM, such conformity does not happen in the range $4 m_{\mu}^2<q^2< 4m_c^2$. In this range, by increasing $|\lambda_{tt}\lambda_{bb}|$ the difference between the SM and 2HDM predictions becomes greater. Also while this asymmetry is independent from the variation of mass of $H^0$,   it is  entirely sensitive to the reduction of mass of $H^{\pm}$ so that the predictions of mass set 1 resemble  those of mass set 3 and the predictions of mass set 2 resemble  those of mass set 4. The most deviation from the SM value arises in the case C of mass sets 2 and 4  which is 26 times of the SM anticipation, occurring at $q^2=4 m_{\mu}^2$. Moreover it is deduced from tables \ref{masssetBK0smu12} and \ref{masssetBK0smu34} that the interval between the average of SM and those of cases A and B is less than 0.005, so the SM and cases A and B predictions overlap with each other and those values could not be useful for finding new physics. However, the difference between the values of case C and that of SM is larger than those of the other cases such that equals with 0.005 and thus the average of this case for all mass sets could be suitable for discovering new Higgs bosons. In addition it is understood from figure \ref{ANTmK} that there are the same behaviors for $\overline{B}\to \overline{K} \mu^+ \mu^-$ as those for $\overline{B}\to \overline{K}_0^{*} \mu^+ \mu^-$ except that a change in the sign of ${\cal A}_{FB}^{NT}$ for the latter decay is seen such that the  most deviation from the SM anticipation is -26 times of the SM prediction. Also it is seen from tables \ref{masssetBKmu12} and \ref{masssetBKmu34} that the distance between the average of SM and those of all cases is less than 0.005 and hence the average of these cases could not be helpful for finding new Higgs bosons. Based on the above explanations, the measurement of this asymmetry for the afore-mentioned decays only in the range $4 m_{\mu}^2<q^2< 4m_c^2$ may  be promising  in looking for  new Higgs bosons.

\item \textbf { Analysis of ${\cal A}_{FB}$ asymmetries for $\overline{B}\to \overline{K}_0^{*} \tau^+ \tau^-$ and $\overline{B}\to \overline{K} \tau^+ \tau^-$ decays}: As it is obvious from   figure \ref{AFBtKstar} although the predictions of ${\cal A}_{FB}$ over the domain $4 m_{\tau}^2<q^2< (m_B-m_{{K}_0^{*}})^2$ for $\overline{B}\to \overline{K}_0^{*} \tau^+ \tau^-$ in cases B and C of all mass sets correspond to that of SM which is zero, such correspondence is not generally seen in case A. Also it is understood from  these plots that  during enhancing the mass of $H^{\pm}$  or reducing the mass of $H^{0}$  this asymmetry shows  more dependency to the existence of new Higgs bosons so that the most deviations from the anticipation of SM   arise in the mass set 3 of the afore-mentioned case. Asymmetries up to $\pm 0.12$ are possible as compared to SM prediction which occur around $q^2=m_{\psi^\prime}^2$. Moreover it is found out from  tables \ref{masssetBK0stau12} and \ref{masssetBK0stau34} that  the sensitivity of averages to the masses of Higgs bosons and cases is like the corresponding plots such that only the averages of case A can give  promising information about the existence of new Higgs bosons and the largest average for this asymmetry compared to SM prediction happens in the mass set 3  which is 0.083. As it clear from figure \ref{AFBtK} and tables \ref{masssetBKtau12} and \ref{masssetBKtau34}  there are similar expressions for $\overline{B}\to \overline{K} \tau^+ \tau^-$ to those for $\overline{B}\to \overline{K}_0^{*} \tau^+ \tau^-$ except that  the magnitudes of maximum  deviations of each of decay modes in the relevant diagrams and tables are different from  those of other decay. Asymmetries
up to -0.14 and +0.18 are possible as compared to SM prediction which occur at $q^2=m_{\psi^\prime}^2$ and $q^2=18$GeV, respectively. Therefore, study of this observable and its average in the experiments, for the $\tau$ channel of $\overline{B}\to \overline{K}_0^{*}$ and $\overline{B}\to \overline{K}$ transitions, can give inspiring facts about the existence of new Higgs bosons.

\item \textbf { Analysis of $ A_{FB}^{LN}$ asymmetries for $\overline{B}\to \overline{K}_0^* \tau^+ \tau^-$ and $\overline{B}\to \overline{K} \tau^+ \tau^-$ decays}: It is apparent from figure \ref{ALNtKstar} that for the  $\overline{B}\to \overline{K}_0^{*} \tau^+ \tau^-$ decay  the predictions of mass set 1 resemble those of mass set 3 and the predictions of mass set 2 resemble those of mass set 4  and the deviation from the SM value in mass sets 2 and 4 is larger than that in mass sets 1 and 3. Therefore, while the magnitude of this asymmetry does not change by varying  the mass of $H^0$,   it is  quite sensitive to the reduction of mass of $H^{\pm}$.  Also, it is explicit from the relevant plots  that this asymmetry is quite sensitive  to the changes of the parameters  $\lambda_{tt}$ and $\lambda_{bb}$. For example, during enhancing the magnitude of $|\lambda_{tt}\lambda_{bb}|$ the deviation from the SM value is increased. By adding up the above analyses it is understood that the most deviation from the SM prediction takes place in the case C of mass sets 2 and 4.  Specially at $q^2=m_{\psi^\prime}^2$ in the afore-mentioned case and mass-sets,  a deviation around 2.6 times of the SM expectation is observed. Besides, it is found out through the corresponding tables  that except for the averages of case A,  those of the other cases are not in the range of SM prediction and  their dependencies to the new Higgs boson parameters are like those of diagrams.  For example, the most distance between the SM prediction and that of 2HDM arises in the case C of mass sets 2 and 4 which is 2.5 times of the SM anticipation. It is also evident from  figure \ref{ALNtK} and tables \ref{masssetBKmu12} and\ref{masssetBKmu34} that there are the similar explanations concerning $\overline{B}\to \overline{K} \tau^+ \tau^-$ decay to those of $\overline{B}\to \overline{K}_0^{*} \tau^+ \tau^-$ decay except that  in the latter decay two Higgs doublet scenario can flip the sign of ${\cal A}_{FB}^{LN}$ compared to the SM expectation in cases B and C of all mass sets. The maximum deviations relative to the SM predictions which are observed in the respective diagrams and tables  take place in the case C of mass sets 2 and 4 which are closely $-1$ times  of the SM prediction  for the corresponding diagrams occurring at $q^2=m_{\psi^\prime}^2$ and $-1.2$  times  of the SM prediction for the related averages. Therefore, it seems that the measurement of ${\cal A}_{FB}^{LN}$ and its average for each of decay modes and its sign for the latter decay mode could provide a valuable tool in establishing  new Higgs bosons.

\item \textbf { Analysis of $ A_{FB}^{NT}$ asymmetries for $\overline{B}\to \overline{K}_0^* \tau^+ \tau^-$ and $\overline{B}\to \overline{K} \tau^+ \tau^-$ decays}: It is evident through figure \ref{ANTtKstar} and tables \ref{masssetBK0stau12} and \ref{masssetBK0stau34} that for the former decay the predictions of mass set 1 resemble  those of mass set 3 and likewise    the predictions of mass set 2 resemble  those of mass set 4. It is also revealed from the tables that the predictions  of each of mass sets have not lain on the SM range.
The most deviations compared to the SM predictions which are observed in the respective diagrams and tables  take place in the case C of mass sets 2 and 4 which are closely $2.6$ times  of the SM prediction  for the corresponding diagrams occurring at $q^2=m_{\psi^\prime}^2$ and $2.8$  times  of the SM prediction for the related averages.  It is also obvious from  figure \ref{ANTtK} and tables \ref{masssetBKtau12} and \ref{masssetBKtau34} that for the latter decay like the former decay the predictions of mass set 1 resemble  those of mass set 3 and likewise    the predictions of mass set 2 resemble  those of mass set 4. In addition, while the predictions of ${\cal A}_{FB}^{NT}$ over the domain $4 m_{\tau}^2<q^2< (m_B-m_{{K}_0^{*}})^2$  in SM and case A  are  positive,   those of cases B and C are completely negative. It is also explicit from the corresponding tables that the predictions  of each of mass sets have not lain on the SM range.
The most deviations compared to the SM predictions which are observed in the respective diagrams and tables  take place in the case C of mass sets 2 and 4 which are closely $-1.2$ times  of the SM prediction  for the corresponding diagrams occurring at $q^2=m_{\psi^\prime}^2$ and $-1.2$  times  of the SM prediction for the related averages.  So, the measurements  of $A^{NT}$ and its average for each of decay modes and its sign for the latter decay mode   can serve as good tests for discovering new Higgs bosons.
\end{itemize}
Finally, let us see briefly  whether
 the lepton polarization asymmetries are testable  or
 not. Experimentally, for measuring  an asymmetry $\lla {\cal A}_{ij}\rra$ of the decay with
 branching ratio $\cal{B}$ at $n\sigma$ level, the required number
 of events (i.e., the number of $B\bar{B}$) is given by the
 formula \bea N = \frac{n^2}{{\cal
B} s_1 s_2 \la {\cal A}_{ij} \ra^2}~,\nnb \eea  where $s_1$ and $s_2$ are
the efficiencies of the leptons. The values of the efficiencies of
the $\tau$--leptons differ from $50\%$ to $90\%$ for their various
decay modes\cite{r30} and the error in $\tau$--lepton polarization
is approximately  $(10 - 15)\%$ \cite{r31}. So, the error in
measurements of the $\tau$--lepton asymmetries is estimated to be
about $(20 - 30)\%$, and the error in obtaining the number of events
is about $50\%$.

Based on the above expression for  $N$, in order to detect the
polarized and unpolarized forward backward asymmetries in the $\mu$ and $\tau$ channels at
$3\sigma$ level, the lowest limit of required number of events are given
by(the efficiency of $\tau$--lepton is considered $0.5$):
\begin{itemize}
\item for $\overline{B} \rar \overline{K}_0^{*}(1430) \mu^+ \mu^-$ decay \bea N \sim \left\{
\begin{array}{llll}
10^{12 }  & (\mbox{\rm for} \lla {\cal A}_{FB} \rra, \lla {\cal A}_{FB}^{LL} \rra, \lla {\cal A}_{FB}^{TT} \rra, \lla {\cal A}_{FB}^{NN} \rra)~,\\
10^{11}  & (\mbox{\rm for} \lla {\cal A}_{FB}^{LN} \rra, \lla A_{FB}^{NL} \rra)~,\\
10^{9}  & (\mbox{\rm for} \lla {\cal A}_{FB}^{LT} \rra~,  \lla {\cal A}_{FB}^{TL} \rra)~,\\
10^{12}  & (\mbox{\rm for} \lla {\cal A}_{FB}^{NT} \rra,  \lla {\cal A}_{FB}^{TN} \rra)~,\\
\end{array} \right. \nnb \eea

\item for $\overline{B} \rar \overline{K} \mu^+ \mu^-$ decay \bea N \sim \left\{
\begin{array}{llll}
10^{12 }  & (\mbox{\rm for} \lla {\cal A}_{FB} \rra, \lla {\cal A}_{FB}^{LL} \rra, \lla {\cal A}_{FB}^{TT} \rra, \lla {\cal A}_{FB}^{NN} \rra)~,\\
10^{10}  & (\mbox{\rm for} \lla {\cal A}_{FB}^{LN} \rra, \lla {\cal A}_{FB}^{NL} \rra)~,\\
10^{9}  & (\mbox{\rm for} \lla {\cal A}_{FB}^{LT} \rra~,  \lla {\cal A}_{FB}^{TL} \rra)~,\\
10^{11}  & (\mbox{\rm for} \lla {\cal A}_{FB}^{NT} \rra,  \lla {\cal A}_{FB}^{TN} \rra)~,\\
\end{array} \right. \nnb \eea
\end{itemize}

\begin{itemize}
\item for $\overline{B} \rar \overline{K}_0^{*}(1430) \tau^+ \tau^-$ decay \bea N \sim \left\{
\begin{array}{llll}
10^{12 }  & (\mbox{\rm for} \lla {\cal A}_{FB} \rra, \lla {\cal A}_{FB}^{LL} \rra, \lla {\cal A}_{FB}^{TT} \rra, \lla {\cal A}_{FB}^{NN} \rra)~,\\
10^{14}  & (\mbox{\rm for} \lla {\cal A}_{FB}^{LN} \rra, \lla {\cal A}_{FB}^{NL} \rra)~,\\
10^{11}  & (\mbox{\rm for} \lla {\cal A}_{FB}^{LT} \rra~,  \lla {\cal A}_{FB}^{TL} \rra)~,\\
10^{12}  & (\mbox{\rm for} \lla {\cal A}_{FB}^{NT} \rra,  \lla {\cal A}_{FB}^{TN} \rra)~,\\
\end{array} \right. \nnb \eea

\item for $\overline{B} \rar \overline{K} \tau^+ \tau^-$ decay \bea N \sim \left\{
\begin{array}{llll}
10^{9 }  & (\mbox{\rm for} \lla {\cal A}_{FB} \rra, \lla {\cal A}_{FB}^{LL} \rra, \lla {\cal A}_{FB}^{TT} \rra, \lla {\cal A}_{FB}^{NN} \rra)~,\\
10^{11}  & (\mbox{\rm for} \lla {\cal A}_{FB}^{LN} \rra, \lla {\cal A}_{FB}^{NL} \rra)~,\\
10^{9}  & (\mbox{\rm for} \lla {\cal A}_{FB}^{LT} \rra~,  \lla {\cal A}_{FB}^{TL} \rra)~,\\
10^{10}  & (\mbox{\rm for} \lla {\cal A}_{FB}^{NT} \rra,  \lla {\cal A}_{FB}^{TN} \rra)~.\\
\end{array} \right. \nnb \eea
\end{itemize}

\section{Summary}
In short, in this paper by takeing into account the theoretical
and experimental uncertainties in the SM we have presented a comprehensive
analysis regarding the polarized and unpolarized forward backward asymmetries for
$\overline{B}\rightarrow \overline{K}_{0}^* \ell^+\ell^-$ and $\overline{B}\rightarrow \overline{K} \ell^+\ell^-$ decays
 using  Model III of 2HDM. At the same time we have compared the results of  both decay modes to  each other. Also, the minimum required
number of events for measuring each asymmetry  has been obtained and compared with
the number produced at the LHC experiments,
 containing ATLAS, CMS and LHCb,  ($\sim 10^{12}$ per year)  or expected to be produced at  the Super-LHC experiments ( supposed to be $\sim 10^{13}$ per year). In conclusion, the following results
have been obtained:
\par
i) For $\mu$ channel, only in ${\cal A}^{LN}$ and ${\cal A}^{NT}$ some sensitivities to the pseudo-scalar  property or scalar property of produced mesons have been observed. For example, while the sign of ${\cal A}^{LN}$ for $\overline{B}\rightarrow \overline{K}_{0}^{*}$ and $\overline{B}\rightarrow \overline{K}$ transitions in SM is positive that sign can change with the existence of Higgs bosons only in $\overline{B}\rightarrow \overline{K}$ transition. Also,  the sign of ${\cal A}^{NT}$ for each  decay mode is the opposite of that of the other decay mode. Since the effects of 2HDM generally on the $q^2$ dependency of ${\cal A}_{FB}$, ${\cal A}^{LN}$ and ${\cal A}^{NT}$ and the average of ${\cal A}^{LN}$  could be large and the  minimum  required number of $B
\bar{B}$ pairs for the measurement of those asymmetries at the LHC are smaller than $10^{12}$, so experimental studies of all mentioned asymmetries for each of decay modes can be suitable
for searching  Model III of  2HDM.
\par
ii) For $\tau$ channel,  in ${\cal A}_{FB}$, ${\cal A}^{LN}$ and ${\cal A}^{NT}$ some sensitivities to the pseudo-scalar  feature or scalar feature of products have been observed. For instance, while the 2HDM signs of ${\cal A}^{LN}$ and ${\cal A}^{NT}$  in $\overline{B}\rightarrow \overline{K}$ transition change compared to SM predictions which are positive for ${\cal A}^{LN}$ and  negative for ${\cal A}^{NT}$ such signs remain unchanged compared to SM predictions in $\overline{B}\rightarrow \overline{K}_{0}^{*}$ decay. Also for the ${\cal A}_{FB}$ of different decay modes the values of $q^2$ at which the most deviations from the SM predictions happen are not the same. Moreover although the effects of 2HDM generally on the $q^2$ dependency of ${\cal A}_{FB}$, ${\cal A}^{LN}$ and ${\cal A}^{NT}$ and their averages could be large  the  minimum  required number of events for detecting  such asymmetries at the LHC or SLHC impose some limitations for the measurements of those asymmetries. According to the above discussion for exploring  Model III of  2HDM  only experimental study of ${\cal A}_{FB}$ and ${\cal A}^{NT}$  are useful.
\par
Finally, it is worthwhile  mentioning that although  the muon polarization
is measured for stationary muons,  such experiments are very hard  to perform  in the near future. The tau
polarization can be studied by investigating the decay products of tau.
The measurement of tau polarization in this respect is
 easier than the polarization of muon.

\section{Acknowledgment}
The authors would like to thank V. Bashiry for his useful
discussions. Support of Research  Council of Shiraz University is
gratefully acknowledged.

\newpage

\begin{table}[t]
\begin{center}
\caption{The averaged  unpolarized and polarized forward-backward  asymmetries for
$\overline{B}\rightarrow \overline{K}_0^*(1430) \,\mu^{+}\mu^{-}$ in SM and 2HDM for   the mass sets 1 and 2 of Higgs bosons and the three cases A ($\theta=\pi/2$, $|\lambda_{tt}|=0.03$ and $|\lambda_{bb}|=100$), B ($\theta=\pi/2$, $|\lambda_{tt}|=0.15$ and $|\lambda_{bb}|=50$) and C ($\theta=\pi/2$, $|\lambda_{tt}|=0.3$ and $|\lambda_{bb}|=30$). The
errors shown for each asymmetry are due to the theoretical and
experimental uncertainties. The first ones are related to the
theoretical uncertainties and the second ones are due to experimental
uncertainties. The theoretical uncertainties come from the hadronic uncertainties
related to the form factors and the experimental uncertainties
originate from the mass of quarks and hadrons and Wolfenstein
parameters.}\label{masssetBK0smu12}
\begin{tabular}{clllllll}
\hline\hline
& \rm{SM} & \rm {Case A } & \rm{Case B }& \rm{Case C }& \rm {Case A }& \rm {Case B }& \rm {Case C }\\
&  & \rm { (Set 1)} & \rm{ (Set1)}& \rm{ (Set1)}& \rm { (Set 2)}& \rm { (Set 2)}& \rm { (Set 2)}\\
\hline

$ \rm{ \la {\cal A}_{FB}\ra}$ & $\phantom{-} 0.000^{+0.000+0.000}_{-0.000-0.000}$ & $+0.001 $ & $+0.000$& $+0.000$& $+0.001$& $+0.000$ & $+0.000$\\
$\rm{ \la {\cal A}_{FB}^{LN}\ra}$ & $+0.001^{+0.000+0.000}_{-0.000-0.000}$& $+0.006$ & $+0.012$ & $+0.014$& $+0.006$& $+0.014$& $+0.016$ \\
$\rm{ \la {\cal A}_{FB}^{LT}\ra}$ &$-0.072^{+0.002+0.004}_{-0.002-0.003} $ & $-0.072$ & $-0.073$ & $-0.072$& $-0.072$ & $-0.073$ & $-0.073$ \\
$\rm{ \la {\cal A}_{FB}^{NT}\ra}$ &$-0.000^{+0.000+0.000}_{-0.000-0.000} $  & $-0.002$ & $-0.004$ & $-0.005$  & $-0.002$ & $-0.004$ & $-0.005$ \\
 \hline\hline
\end{tabular}
\end{center}
\end{table}

\begin{table}[t]
\begin{center}
\caption{The same as TABLE \ref{masssetBK0smu12} but for the mass sets 3 and 4 of Higgs bosons.}\label{masssetBK0smu34}
\begin{tabular}{clllllll}
\hline\hline
& \rm{SM} & \rm {Case A } & \rm{Case B }& \rm{Case C }& \rm {Case A }& \rm {Case B }& \rm {Case C }\\
&  & \rm { (Set 3)} & \rm{ (Set3)}& \rm{ (Set3)}& \rm { (Set 4)}& \rm { (Set 4)}& \rm { (Set 4)}\\
\hline

$ \rm{ \la {\cal A}_{FB}\ra}$ & $\phantom{-} 0.000^{+0.000+0.000}_{-0.000-0.000}$ & $+0.003 $ & $+0.000$& $+0.000$& $+0.002$& $+0.000$ & $+0.000$\\
$\rm{ \la {\cal A}_{FB}^{LN}\ra}$ & $+0.001^{+0.000+0.000}_{-0.000-0.000}$& $+0.006$ & $+0.012$ & $+0.014$& $+0.006$& $+0.014$& $+0.016$ \\
$\rm{ \la {\cal A}_{FB}^{LT}\ra}$ &$-0.072^{+0.002+0.004}_{-0.002-0.003} $ & $-0.071$ & $-0.073$ & $-0.072$& $-0.072$ & $-0.073$ & $-0.073$ \\
$\rm{ \la {\cal A}_{FB}^{NT}\ra}$ &$-0.000^{+0.000+0.000}_{-0.000-0.000} $  & $-0.002$ & $-0.004$ & $-0.005$  & $-0.002$ & $-0.004$ & $-0.005$ \\
 \hline\hline
\end{tabular}
\end{center}
\end{table}

\begin{table}[t]
\begin{center}
\caption{The same as TABLE \ref{masssetBK0smu12}  but for the $\overline{B}\to \overline{K} \mu^+\mu^-$.}\label{masssetBKmu12}
\begin{tabular}{clllllll}
\hline\hline
& \rm{SM} & \rm {Case A } & \rm{Case B }& \rm{Case C }& \rm {Case A }& \rm {Case B }& \rm {Case C }\\
&  & \rm { (Set 1)} & \rm{ (Set1)}& \rm{ (Set1)}& \rm { (Set 2)}& \rm { (Set 2)}& \rm { (Set 2)}\\
\hline

$ \rm{ \la {\cal A}_{FB}\ra}$ & $\phantom{-} 0.000^{+0.000+0.000}_{-0.000-0.000}$ & $ +0.001 $ & $+0.000$& $+0.000$& $+0.001$& $+0.000$ & $+0.000$\\
$\rm{ \la {\cal A}_{FB}^{LN}\ra}$ & $+0.002^{+0.000+0.000}_{-0.000-0.000}$& $-0.003$ & $-0.009$ & $-0.011$& $-0.003$& $-0.011$& $-0.014$ \\
$\rm{ \la {\cal A}_{FB}^{LT}\ra}$ &$-0.051^{+0.018+0.000}_{-0.022-0.000} $ & $-0.051$ & $-0.052$ & $-0.052$& $-0.051$ & $-0.052$ & $-0.052$ \\
$\rm{ \la {\cal A}_{FB}^{NT}\ra}$ &$-0.000^{+0.000+0.000}_{-0.000-0.000} $  & $+0.000$ & $+0.003$ & $+0.003$  & $+0.001$ & $+0.003$ & $+0.004$ \\
 \hline\hline
\end{tabular}
\end{center}
\end{table}

\begin{table}[t]
\begin{center}
\caption{The same as TABLE \ref{masssetBKmu12} except for the mass sets 3 and 4 of Higgs bosons.}\label{masssetBKmu34}
\begin{tabular}{clllllll}
\hline\hline
& \rm{SM} & \rm {Case A } & \rm{Case B }& \rm{Case C }& \rm {Case A }& \rm {Case B }& \rm {Case C }\\
&  & \rm { (Set 3)} & \rm{ (Set3)}& \rm{ (Set3)}& \rm { (Set 4)}& \rm { (Set 4)}& \rm { (Set 4)}\\
\hline

$ \rm{ \la {\cal A}_{FB}\ra}$ & $\phantom{-} 0.000^{+0.000+0.000}_{-0.000-0.000}$ & $+0.002 $ & $+0.000$& $+0.000$& $+0.002$& $+0.000$ & $+0.000$\\
$\rm{ \la {\cal A}_{FB}^{LN}\ra}$ & $+0.002^{+0.000+0.000}_{-0.000-0.000}$& $-0.003$ & $-0.009$ & $-0.011$& $-0.003$& $-0.011$& $-0.014$ \\
$\rm{ \la {\cal A}_{FB}^{LT}\ra}$ &$-0.051^{+0.018+0.000}_{-0.022-0.000} $ & $-0.051$ & $-0.052$ & $-0.052$& $-0.051$ & $-0.052$ & $-0.052$ \\
$\rm{ \la {\cal A}_{FB}^{NT}\ra}$ &$-0.000^{+0.000+0.000}_{-0.000-0.000} $  & $+0.000$ & $+0.003$ & $+0.003$  & $+0.001$ & $+0.003$ & $+0.004$ \\
  \hline\hline
\end{tabular}
\end{center}
\end{table}
\
\begin{table}[t]
\begin{center}
\caption{The same as TABLE \ref{masssetBK0smu12} except for $\overline{B}\to \overline{K}_0^*(1430) \tau^+\tau^-$.}\label{masssetBK0stau12}
\begin{tabular}{clllllll}
\hline\hline
& \rm{SM} & \rm {Case A } & \rm{Case B }& \rm{Case C }& \rm {Case A }& \rm {Case B }& \rm {Case C }\\
&  & \rm { (Set 1)} & \rm{ (Set1)}& \rm{ (Set1)}& \rm { (Set 2)}& \rm { (Set 2)}& \rm { (Set 2)}\\
\hline

$ \rm{ \la {\cal A}_{FB}\ra}$ & $\phantom{-} 0.000^{+0.000+0.000}_{-0.000-0.000}$ & $+0.045 $ & $+0.002$& $+0.000$& $+0.026$& $+0.001$ & $+0.000$\\
$\rm{ \la {\cal A}_{FB}^{LN}\ra}$ & $+0.004^{+0.002+0.000}_{-0.004-0.000}$& $+0.005$ & $+0.008$ & $+0.009$& $+0.005$& $+0.009$& $+0.010$ \\
$\rm{ \la {\cal A}_{FB}^{LT}\ra}$ &$-0.176^{+0.080+0.008}_{-0.210-0.009} $ & $-0.160$ & $-0.180$ & $-0.178$& $-0.162$ & $-0.181$ & $-0.179$ \\
$\rm{ \la {\cal A}_{FB}^{NT}\ra}$ &$-0.043^{+0.008+0.003}_{-0.007-0.002} $  & $-0.063$ & $-0.099$ & $-0.109$  & $-0.067$ & $-0.108$ & $-0.120$ \\
 \hline\hline
\end{tabular}
\end{center}
\end{table}

\begin{table}[t]
\begin{center}
\caption{The same as TABLE \ref{masssetBK0stau12}  but for the mass sets 3 and 4 of Higgs bosons.}\label{masssetBK0stau34}
\begin{tabular}{clllllll}
\hline\hline
& \rm{SM} & \rm {Case A } & \rm{Case B }& \rm{Case C }& \rm {Case A }& \rm {Case B }& \rm {Case C }\\
&  & \rm { (Set 3)} & \rm{ (Set3)}& \rm{ (Set3)}& \rm { (Set 4)}& \rm { (Set 4)}& \rm { (Set 4)}\\
\hline

$ \rm{ \la {\cal A}_{FB}\ra}$ & $\phantom{-} 0.000^{+0.000+0.000}_{-0.000-0.000}$ & $+0.083$ & $+0.004$& $+0.000$& $+0.056$& $+0.003$ & $+0.000$\\
$\rm{ \la {\cal A}_{FB}^{LN}\ra}$ & $+0.004^{+0.002+0.000}_{-0.004-0.000}$& $+0.005$ & $+0.008$ & $+0.009$& $+0.005$& $+0.009$& $+0.010$ \\
$\rm{ \la {\cal A}_{FB}^{LT}\ra}$ &$-0.176^{+0.080+0.008}_{-0.210-0.009} $ & $-0.153$ & $-0.180$ & $-0.178$& $-0.158$ & $-0.181$ & $-0.179$ \\
$\rm{ \la {\cal A}_{FB}^{NT}\ra}$ &$-0.043^{+0.008+0.003}_{-0.007-0.002} $  & $-0.060$ & $-0.099$ & $-0.109$  & $-0.066$ & $-0.108$ & $-0.120$ \\
 \hline\hline
\end{tabular}
\end{center}
\end{table}

\begin{table}[t]
\begin{center}
\caption{The same as TABLE \ref{masssetBKmu12} except for $\overline{B}\to \overline{K} \tau^+\tau^-$.}\label{masssetBKtau12}
\begin{tabular}{clllllll}
\hline\hline
& \rm{SM} & \rm {Case A } & \rm{Case B }& \rm{Case C }& \rm {Case A }& \rm {Case B }& \rm {Case C }\\
&  & \rm { (Set 1)} & \rm{ (Set1)}& \rm{ (Set1)}& \rm { (Set 2)}& \rm { (Set 2)}& \rm { (Set 2)}\\
\hline

$ \rm{ \la {\cal A}_{FB}\ra}$ & $\phantom{-} 0.000^{+0.000+0.000}_{-0.000-0.000}$ & $+0.100 $ & $+0.005$& $+0.001$& $+0.059$& $+0.003$ & $+0.000$\\
$\rm{ \la {\cal A}_{FB}^{LN}\ra}$ & $+0.014^{+0.004+0.000}_{-0.004-0.000}$& $+0.004$ & $-0.008$ & $-0.013$& $+0.003$& $-0.012$& $-0.017$ \\
$\rm{ \la {\cal A}_{FB}^{LT}\ra}$ &$-0.181^{+0.046+0.003}_{-0.053-0.003} $ & $-0.147$ & $-0.178$ & $-0.175$& $-0.156$ & $-0.179$ & $-0.176$ \\
$\rm{ \la {\cal A}_{FB}^{NT}\ra}$ &$-0.054^{+0.003+0.000}_{-0.004-0.000} $  & $-0.017$ & $+0.033$ & $+0.049$  & $-0.013$ & $+0.048$ & $+0.067$ \\
 \hline\hline
\end{tabular}
\end{center}
\end{table}

\begin{table}[t]
\begin{center}
\caption{The same as TABLE \ref{masssetBKtau12} but for the mass sets 3 and 4 of Higgs bosons.}\label{masssetBKtau34}
\begin{tabular}{clllllll}
\hline\hline
& \rm{SM} & \rm {Case A } & \rm{Case B }& \rm{Case C }& \rm {Case A }& \rm {Case B }& \rm {Case C }\\
&  & \rm { (Set 3)} & \rm{ (Set3)}& \rm{ (Set3)}& \rm { (Set 4)}& \rm { (Set 4)}& \rm { (Set 4)}\\
\hline

$ \rm{ \la {\cal A}_{FB}\ra}$ & $\phantom{-} 0.000^{+0.000+0.000}_{-0.000-0.000}$ & $+0.155 $ & $+0.009$& $+0.002$& $+0.119$& $+0.006$ & $+0.002$\\
$\rm{ \la {\cal A}_{FB}^{LN}\ra}$ & $+0.014^{+0.004+0.000}_{-0.004-0.000}$& $+0.003$ & $-0.008$ & $-0.013$& $+0.003$& $-0.012$& $-0.017$ \\
$\rm{ \la {\cal A}_{FB}^{LT}\ra}$ &$-0.181^{+0.046+0.003}_{-0.053-0.003} $ & $-0.118$ & $-0.178$ & $-0.175$& $-0.139$ & $-0.178$ & $-0.176$ \\
$\rm{ \la {\cal A}_{FB}^{NT}\ra}$ &$-0.054^{+0.003+0.000}_{-0.004-0.000} $  & $-0.014$ & $+0.033$ & $+0.049$  & $-0.011$ & $+0.048$ & $+0.067$ \\
 \hline\hline
\end{tabular}
\end{center}
\end{table}

\begin{figure}[ht]
  \centering
  \setlength{\fboxrule}{2pt}
        \centering
                     \includegraphics[height=2in]{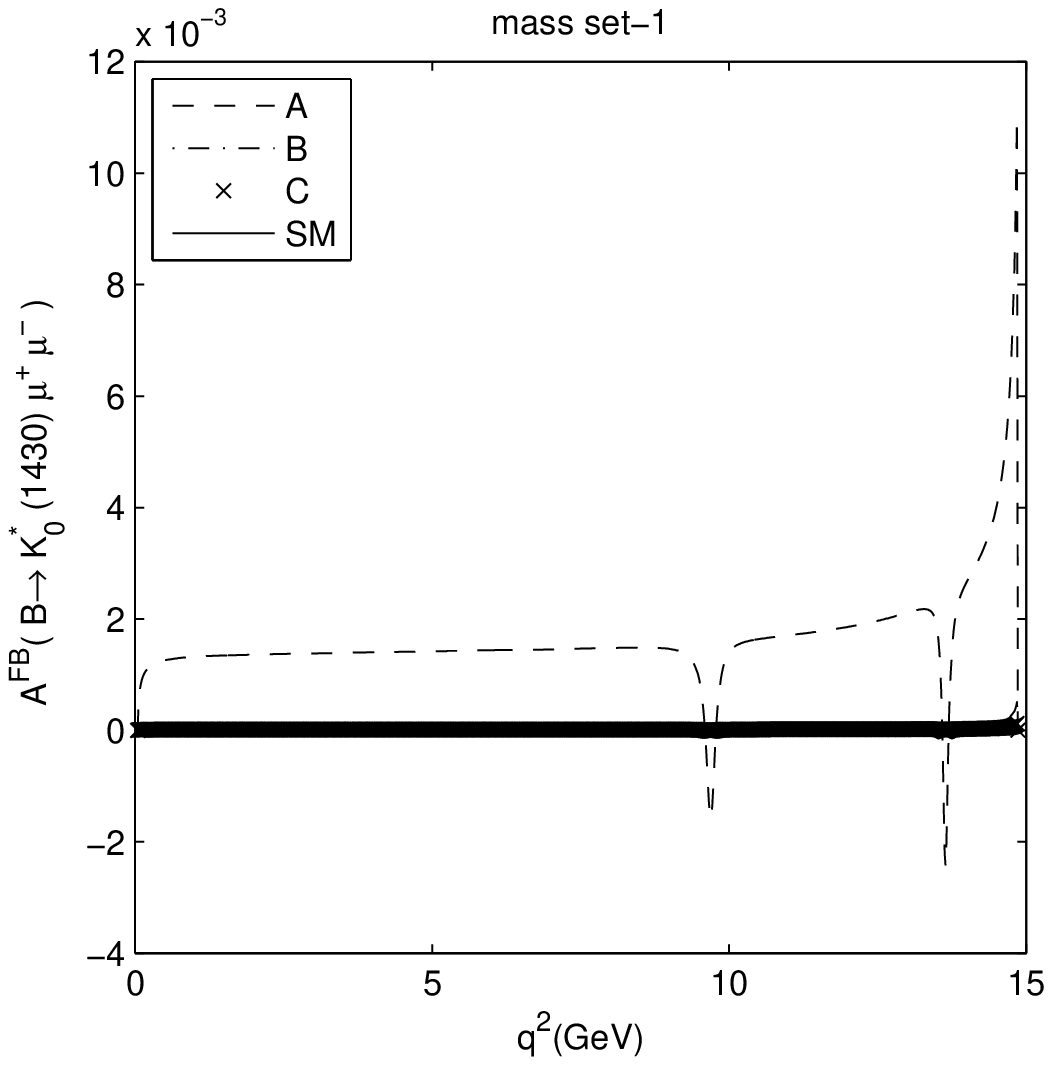}~
             \includegraphics[height=2in]{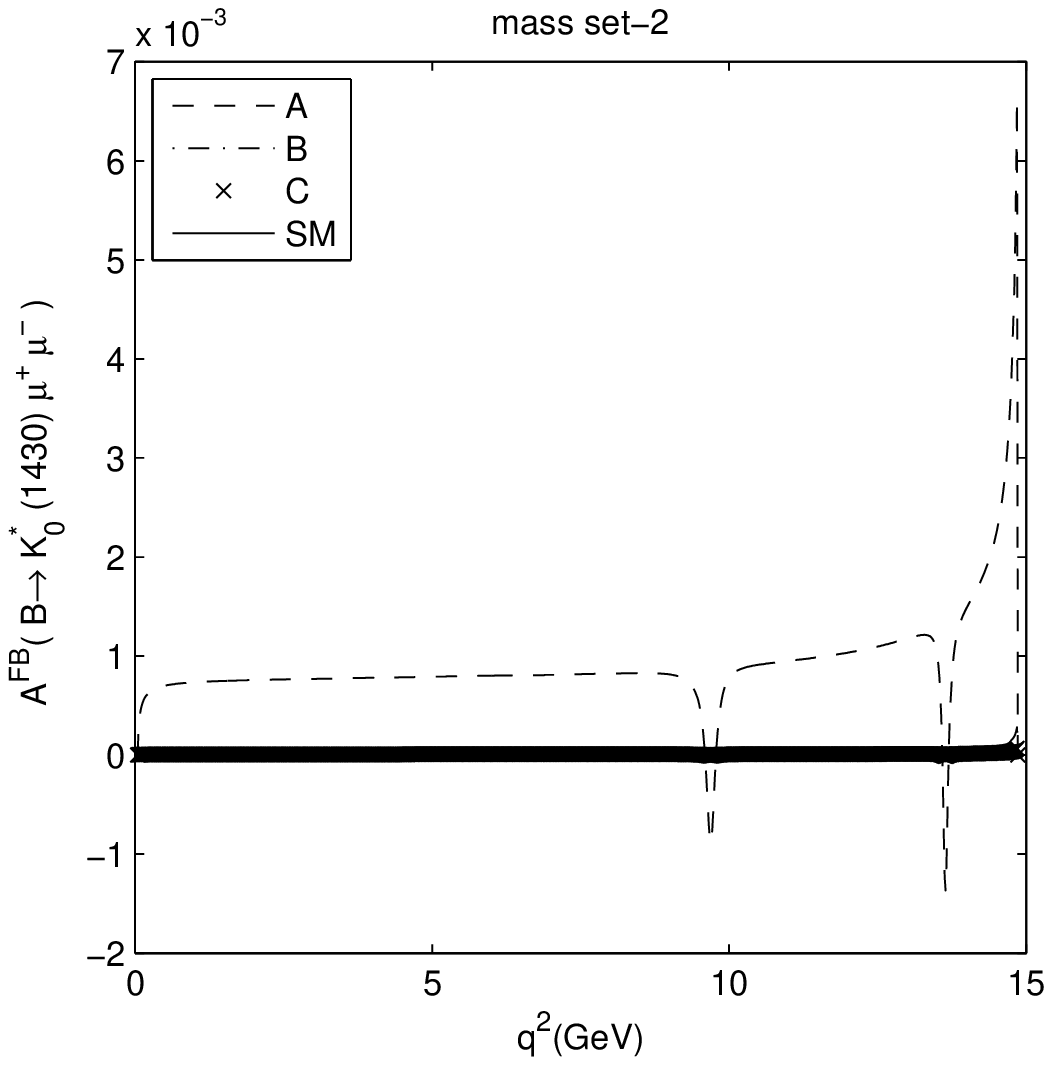}
             \includegraphics[height=2in]{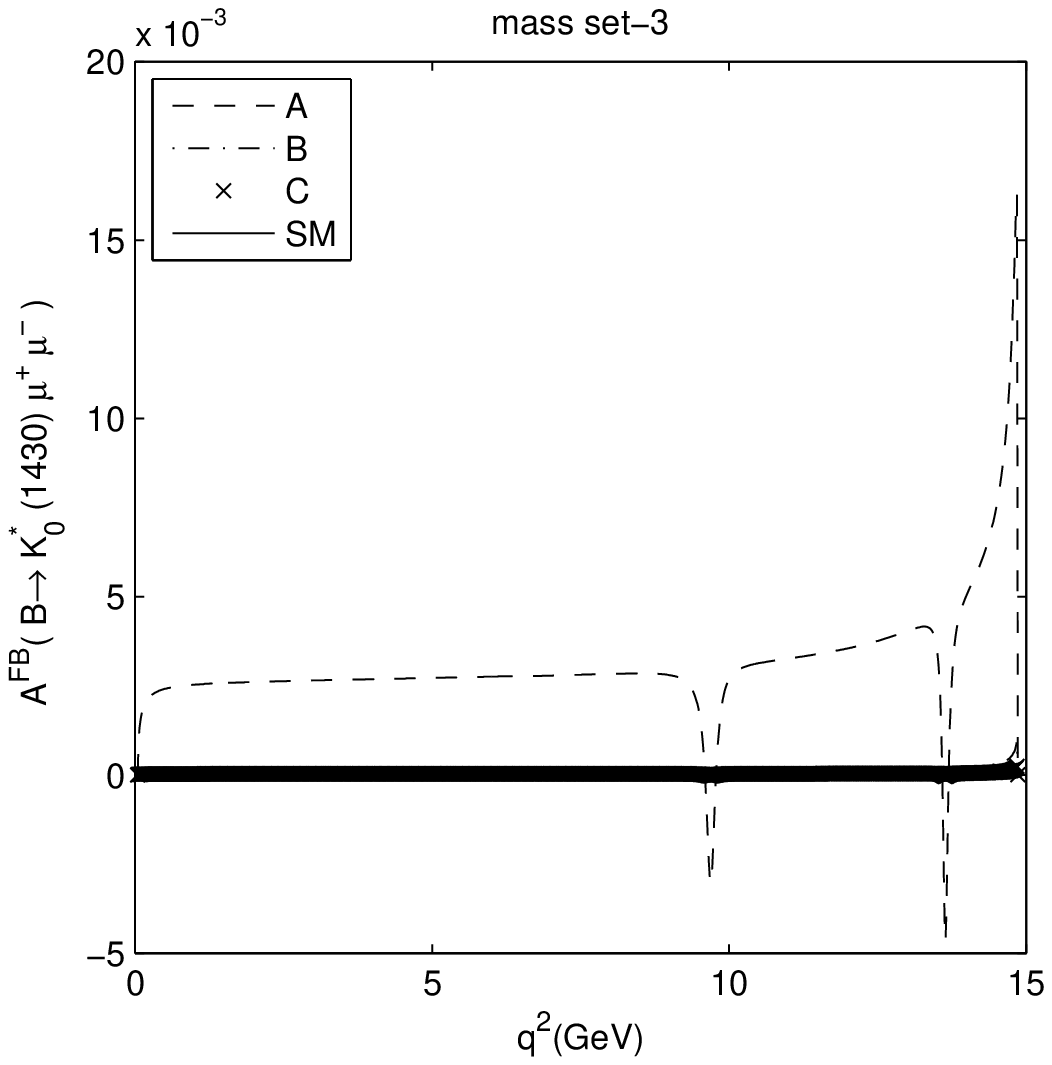}
             \includegraphics[height=2in]{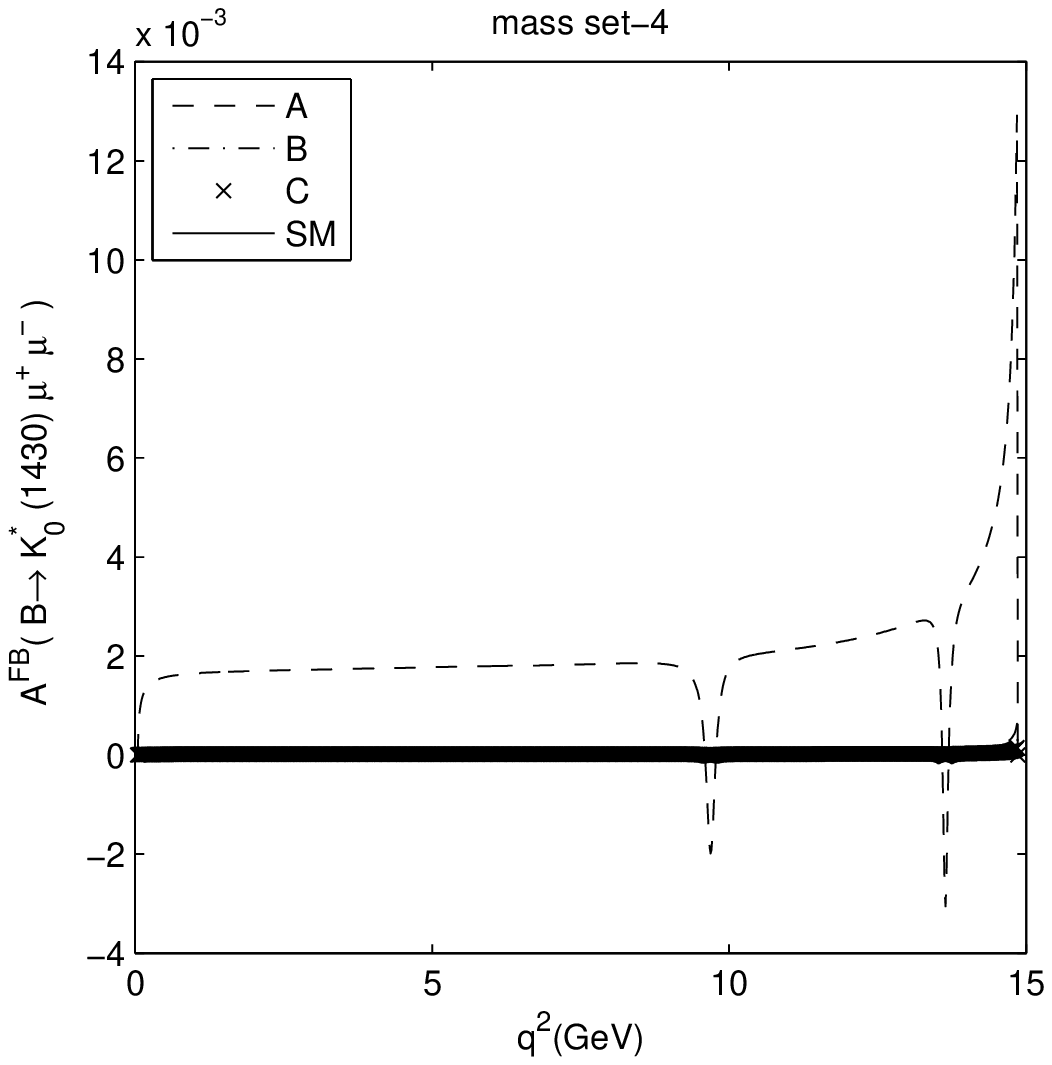}
               \caption{The dependence of the $ {\cal A}_{FB}$ polarization  on $q^2$  and the three typical cases of 2HDM, i.e.
               cases A, B and C and SM  for  the $\mu$  channel of  $\overline{B}\to\overline{K}_0^{*}$ transition for the  mass sets 1, 2, 3  and 4. } \label{AFBmKstar}
    \end{figure}
    \begin{figure}[ht]
  \centering
  \setlength{\fboxrule}{2pt}
        \centering
                     \includegraphics[height=2in]{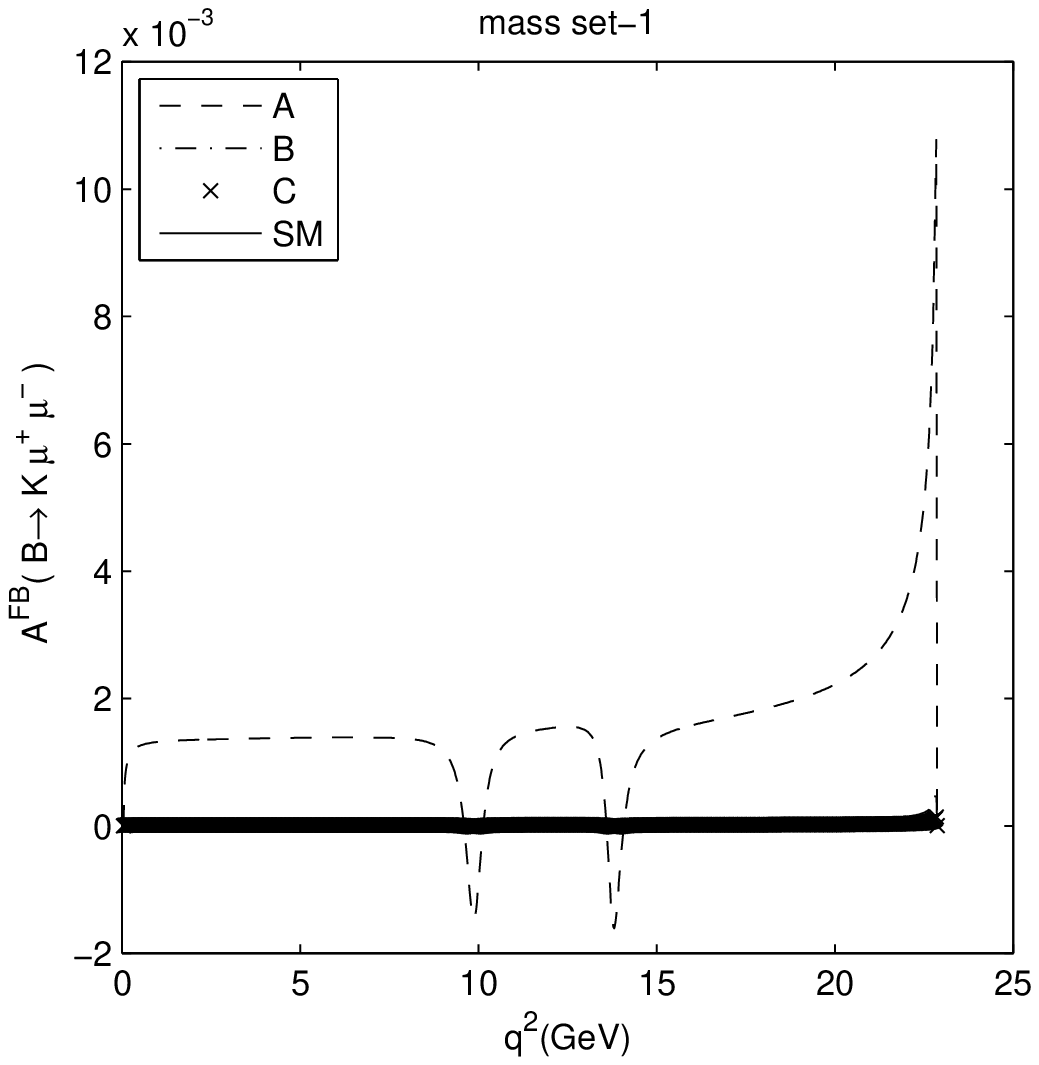}~
             \includegraphics[height=2in]{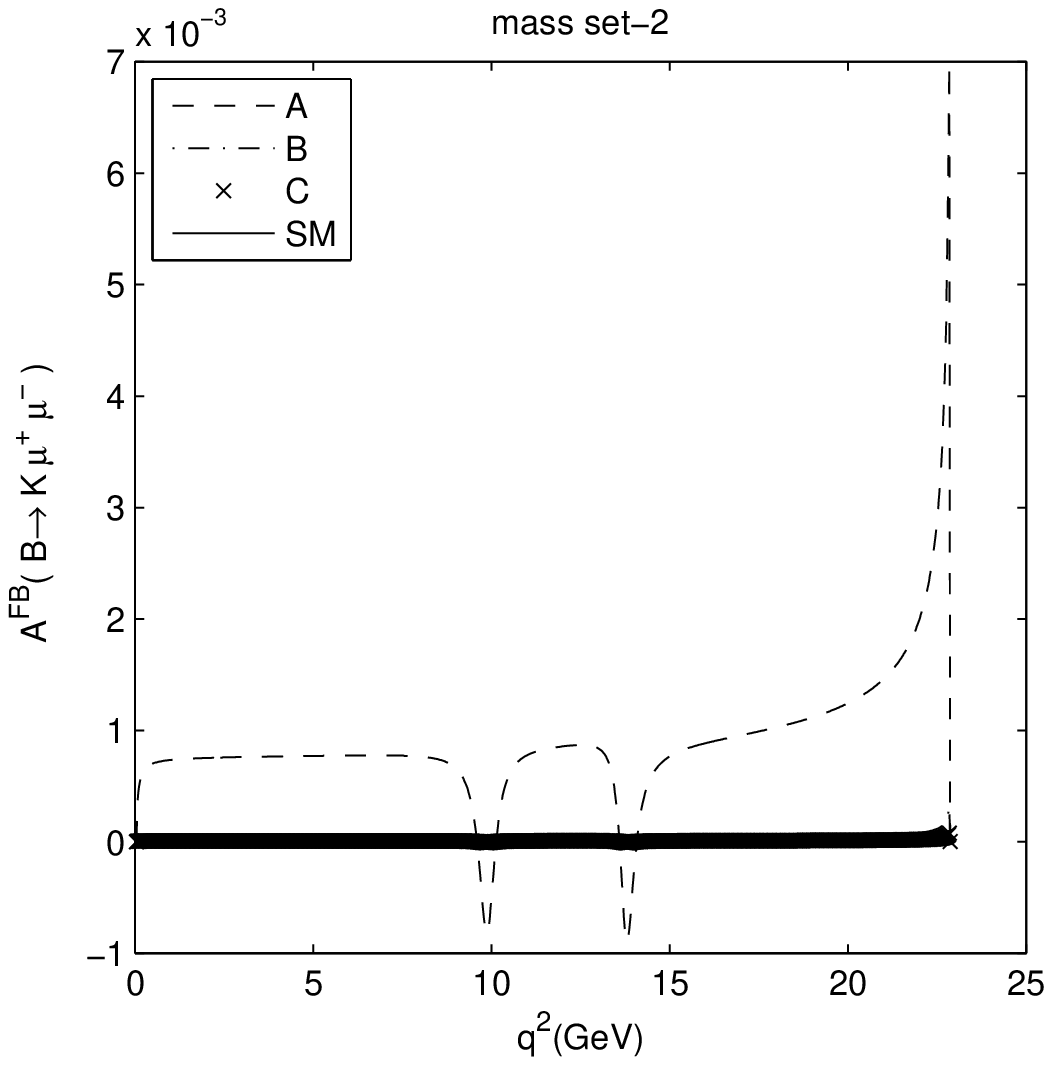}
             \includegraphics[height=2in]{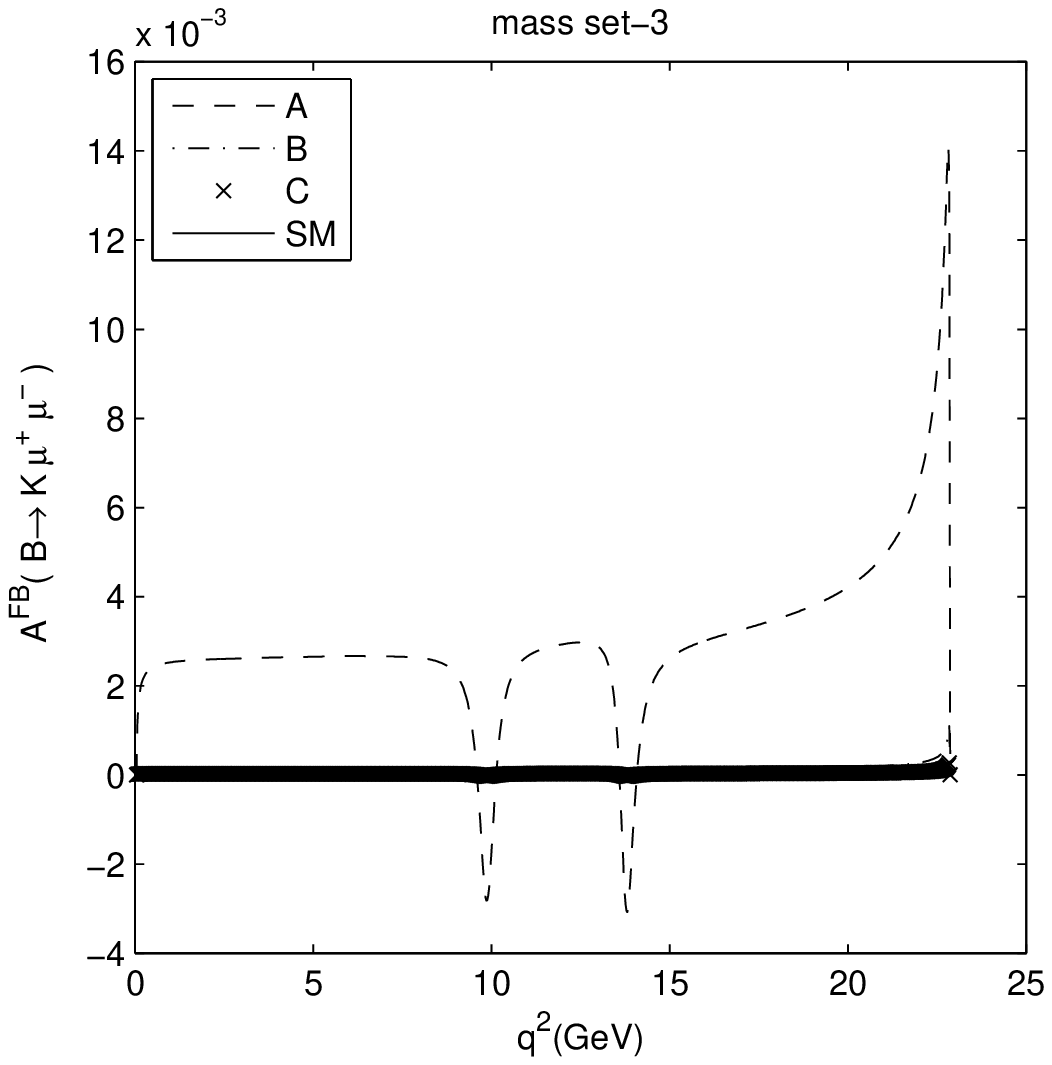}
             \includegraphics[height=2in]{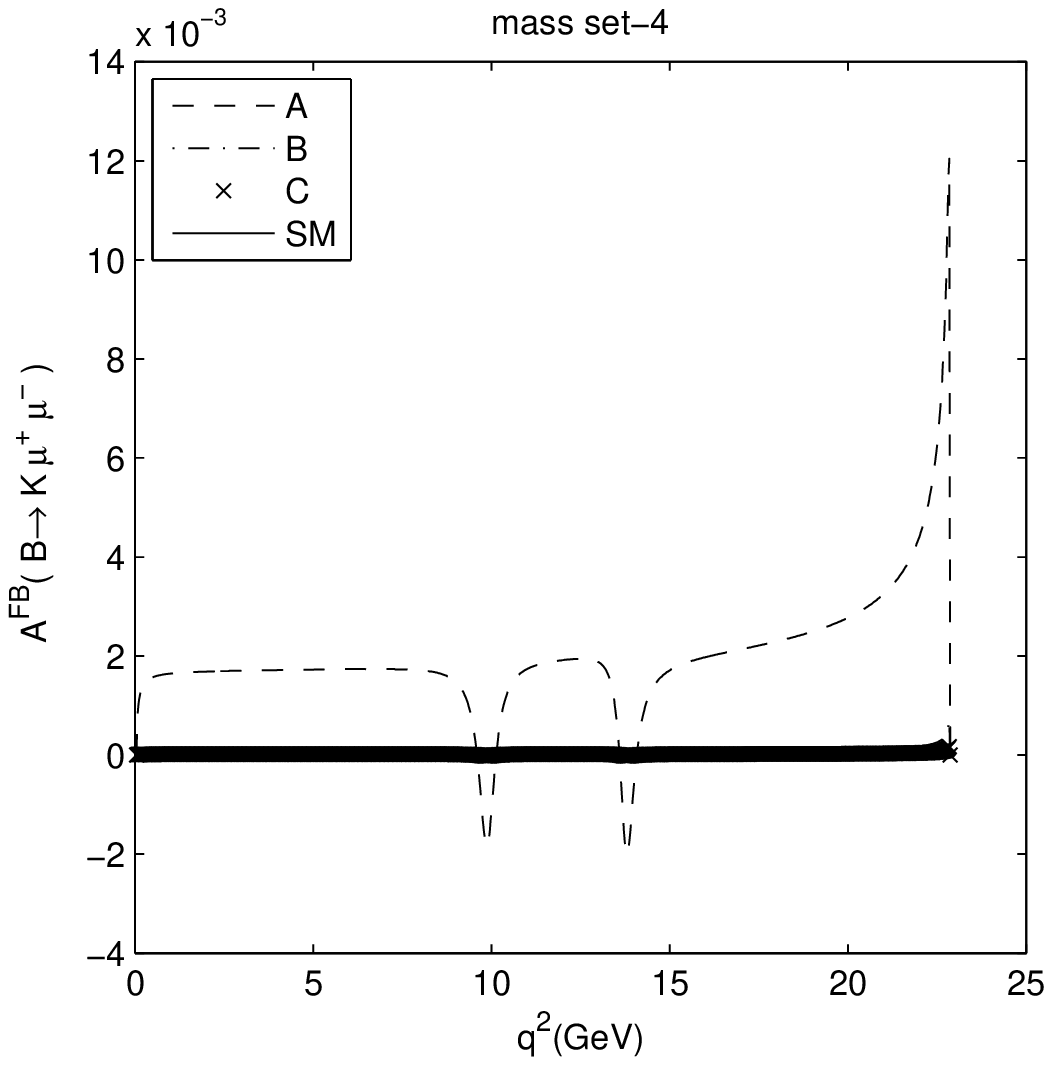}
               \caption{The dependence of the $ {\cal A}_{FB}$ polarization  on $q^2$  and the three typical cases of 2HDM, i.e.
               cases A, B and C and SM  for  the $\mu$  channel of $\overline{B}\to \overline{K}$ transition for the  mass sets 1, 2, 3  and 4. } \label{AFBmK}
    \end{figure}
\begin{figure}[ht]
    \setlength{\fboxrule}{2pt}
        \centering
                       \includegraphics[height=2in]{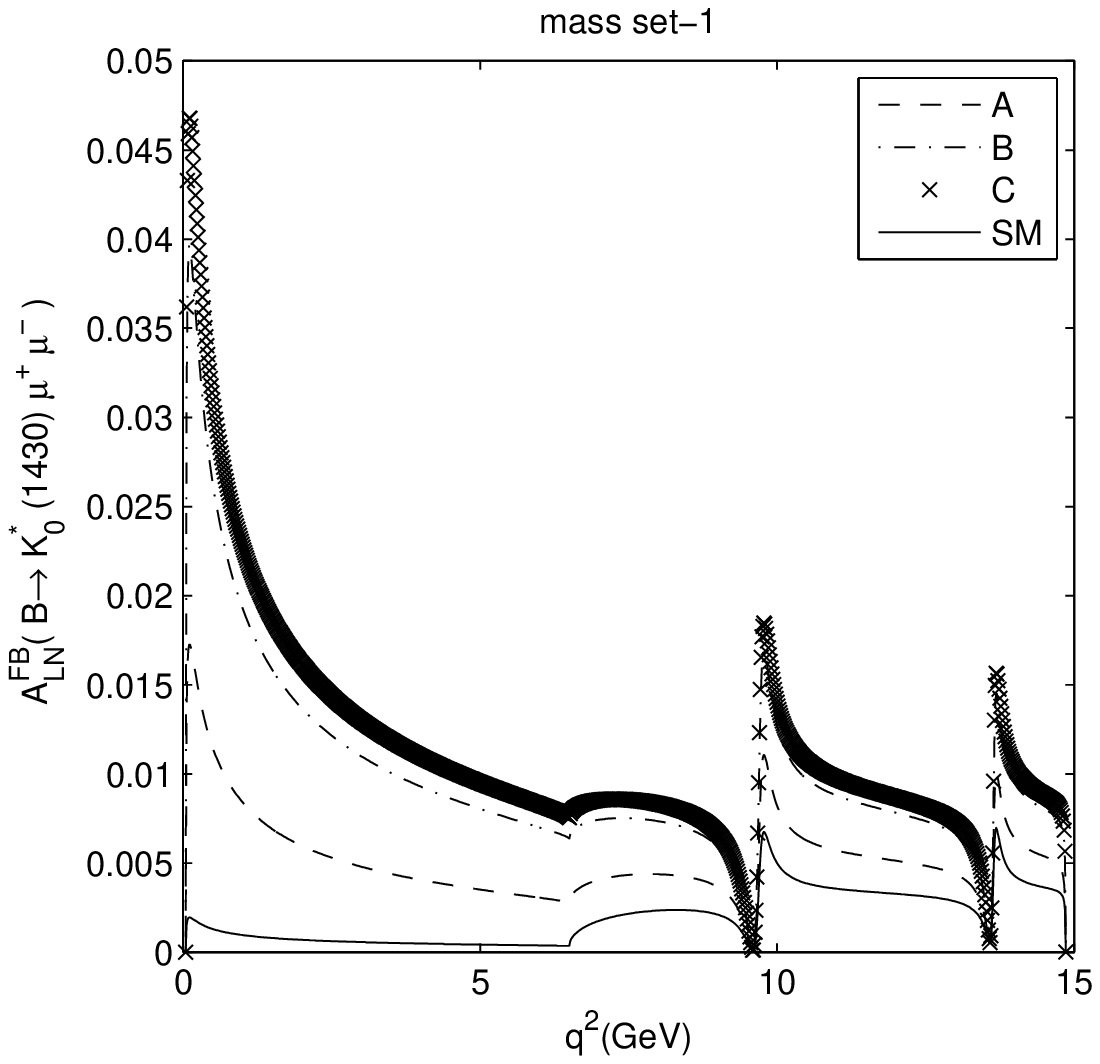}~
             \includegraphics[height=2in]{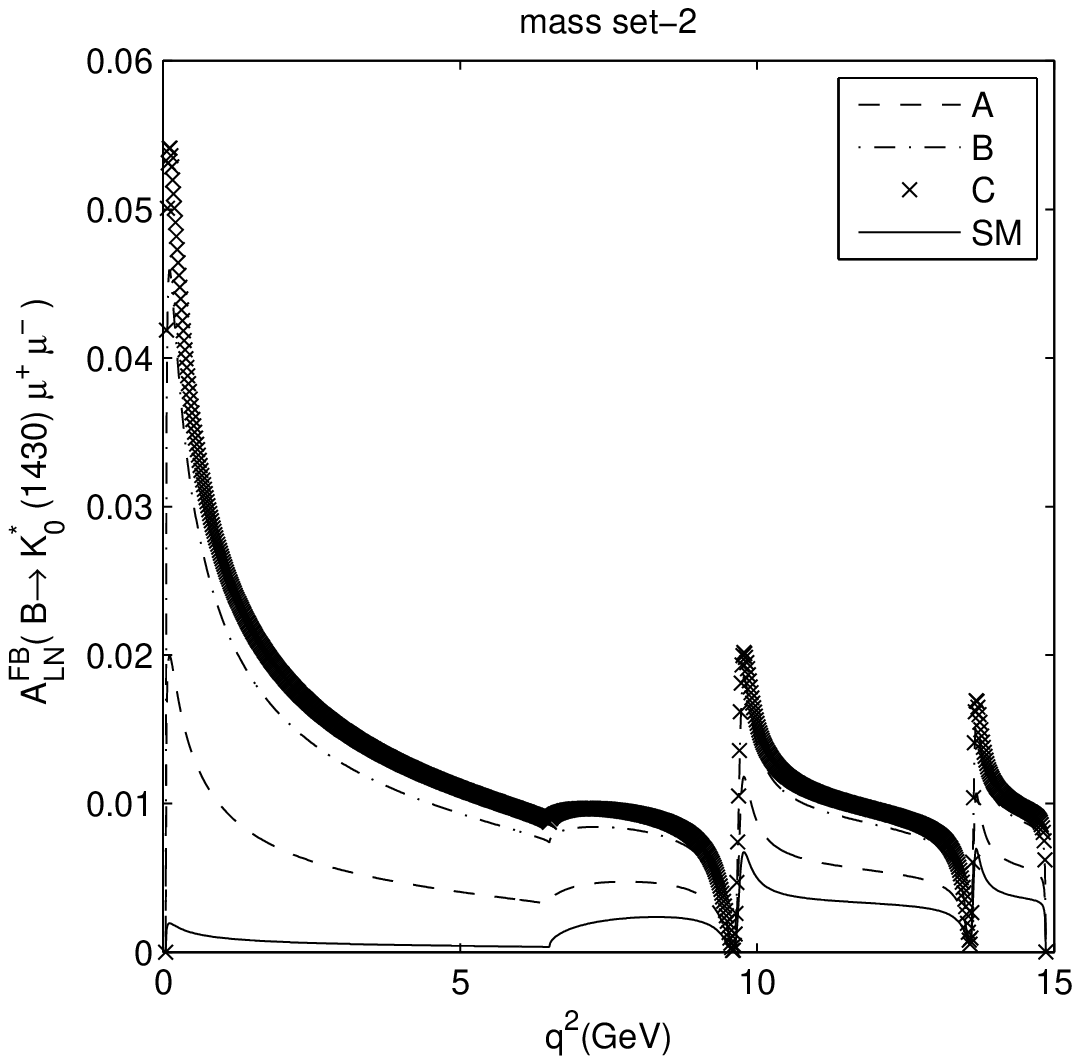}
             \includegraphics[height=2in]{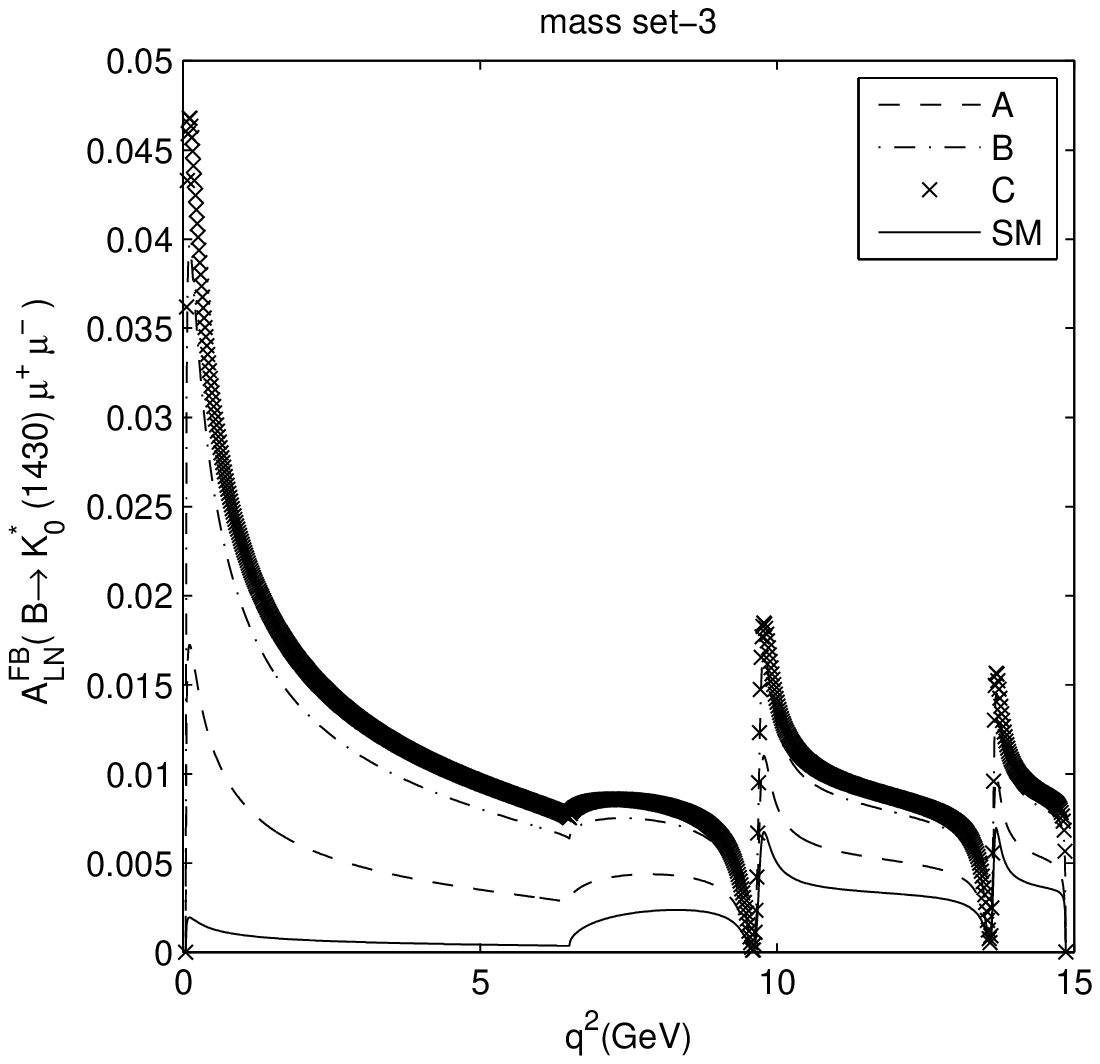}
             \includegraphics[height=2in]{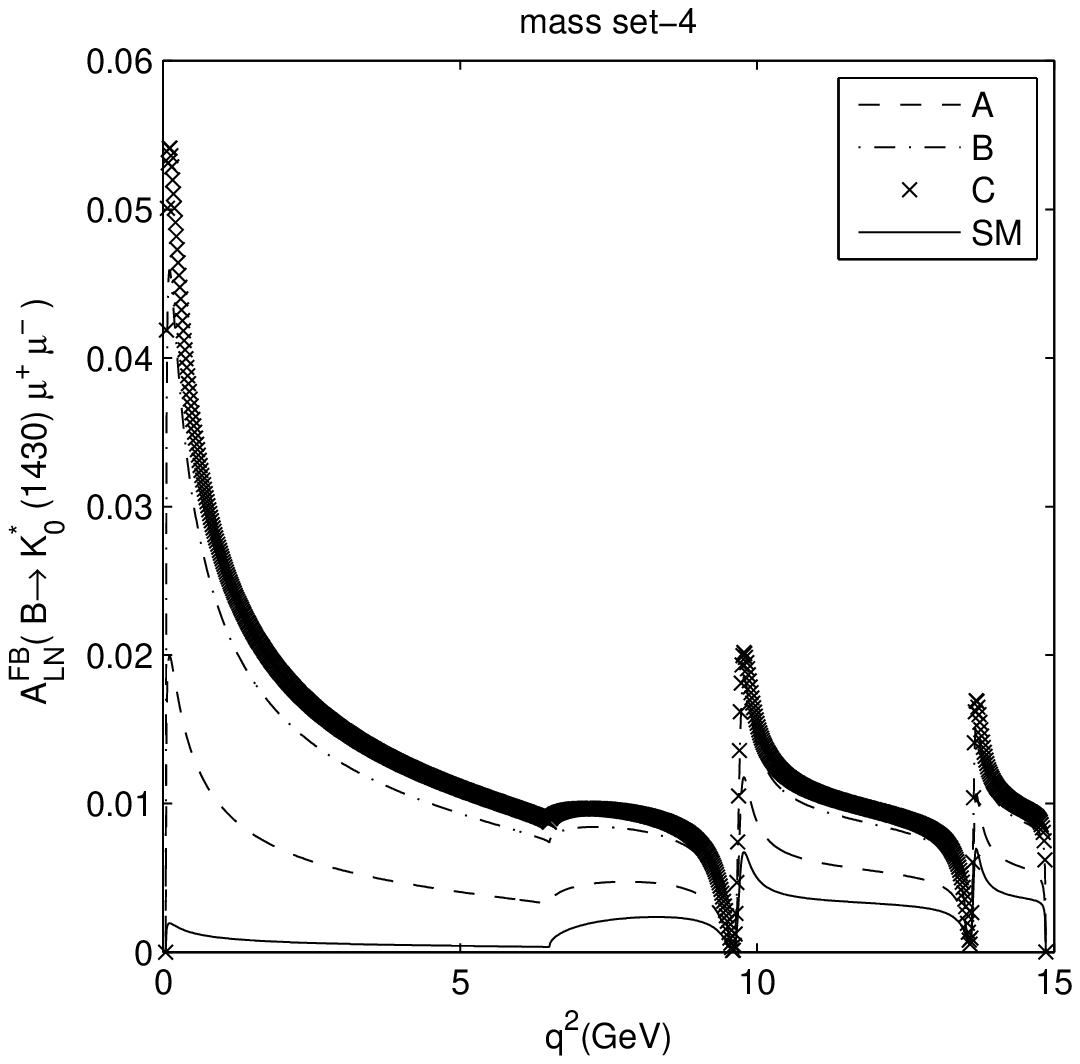}
               \caption{The dependence of the $ {\cal A}_{FB}^{LN}$ polarization  on $q^2$  and the three typical cases of 2HDM, i.e.
               cases A, B and C and SM  for  the $\mu$  channel of  $\overline{B}\to\overline{K}_0^{*}$  transition for the  mass sets 1, 2, 3  and 4. } \label{ALNmKstar}
    \end{figure}

\begin{figure}[ht]
  \centering
  \setlength{\fboxrule}{2pt}
        \centering
                       \includegraphics[height=2in]{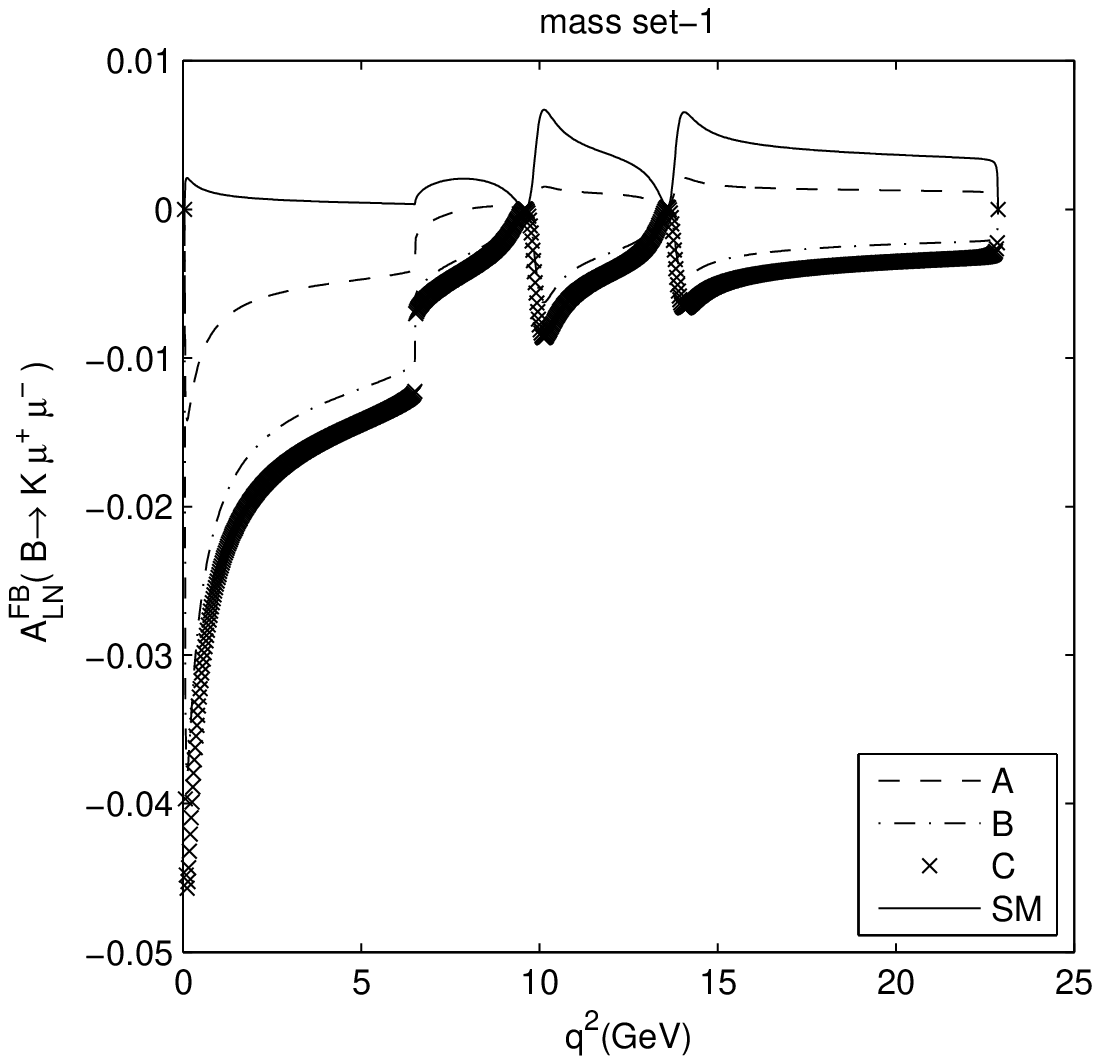}~
             \includegraphics[height=2in]{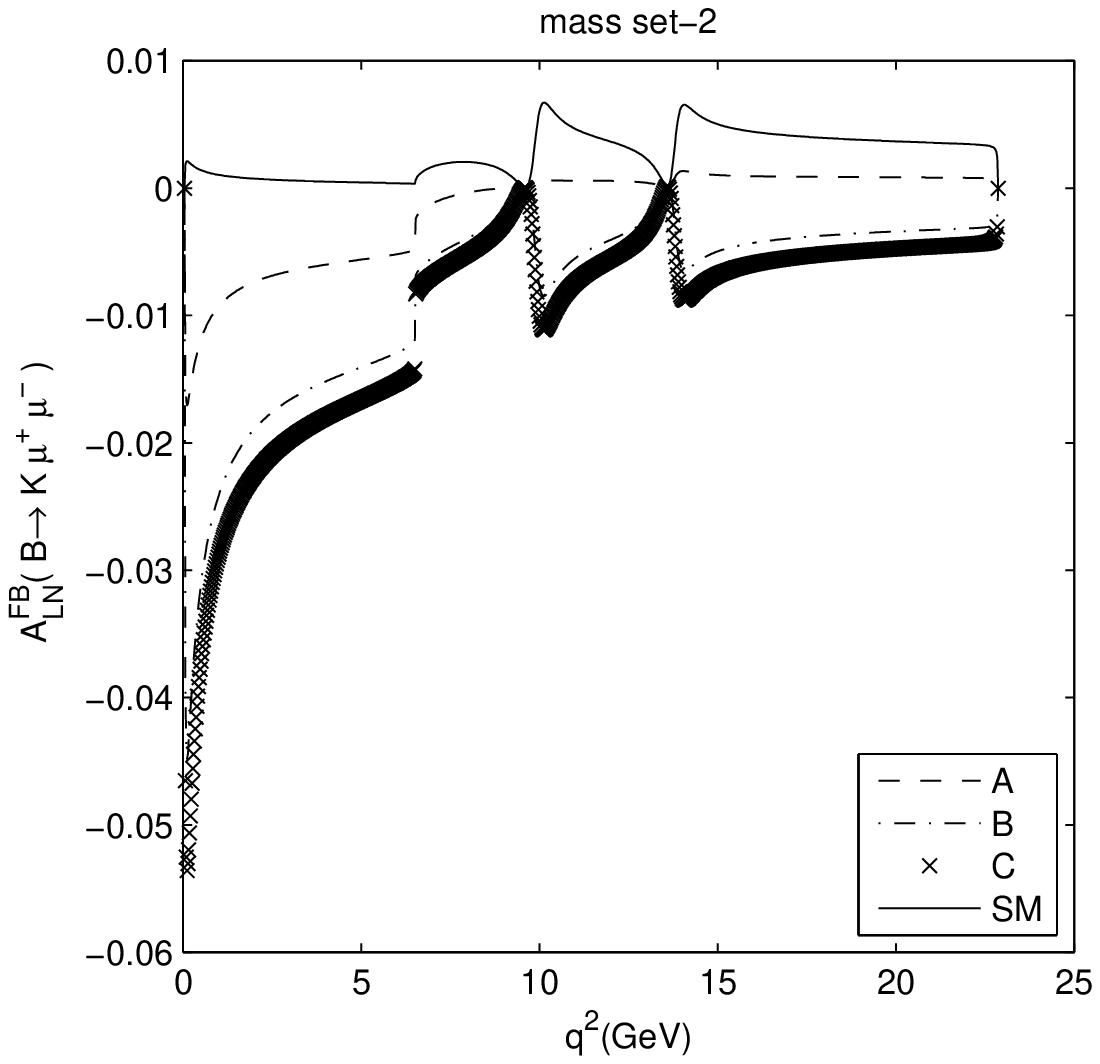}
             \includegraphics[height=2in]{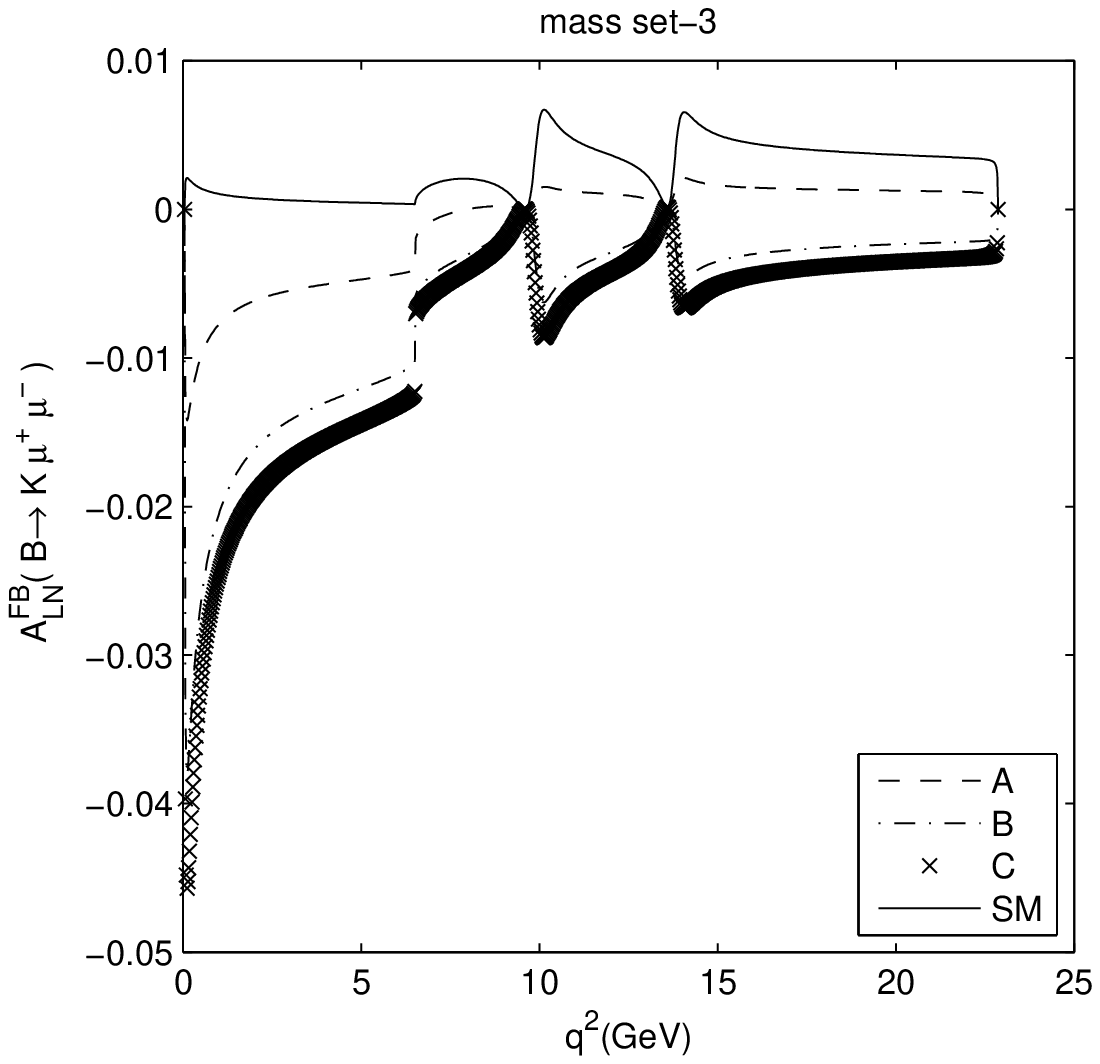}
             \includegraphics[height=2in]{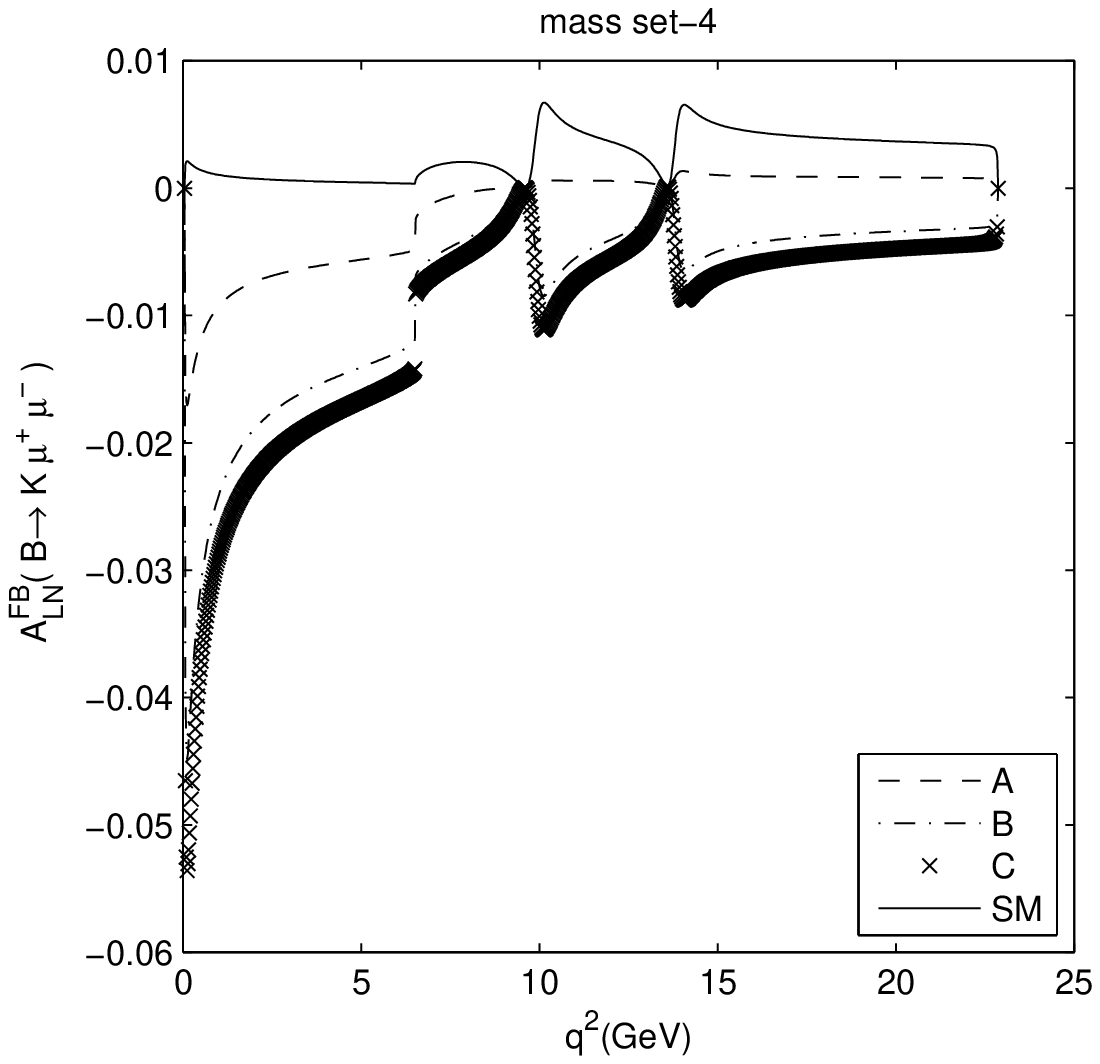}
               \caption{The dependence of the $ {\cal A}_{FB}^{LN}$ polarization  on $q^2$  and the three typical cases of 2HDM, i.e.
               cases A, B and C and SM  for  the $\mu$  channel of  $\overline{B}\to\overline{K}$ transition for the  mass sets 1, 2, 3  and 4. } \label{ALNmK}
    \end{figure}

    \begin{figure}[ht]
  \centering
  \setlength{\fboxrule}{2pt}
        \centering
                       \includegraphics[height=2in]{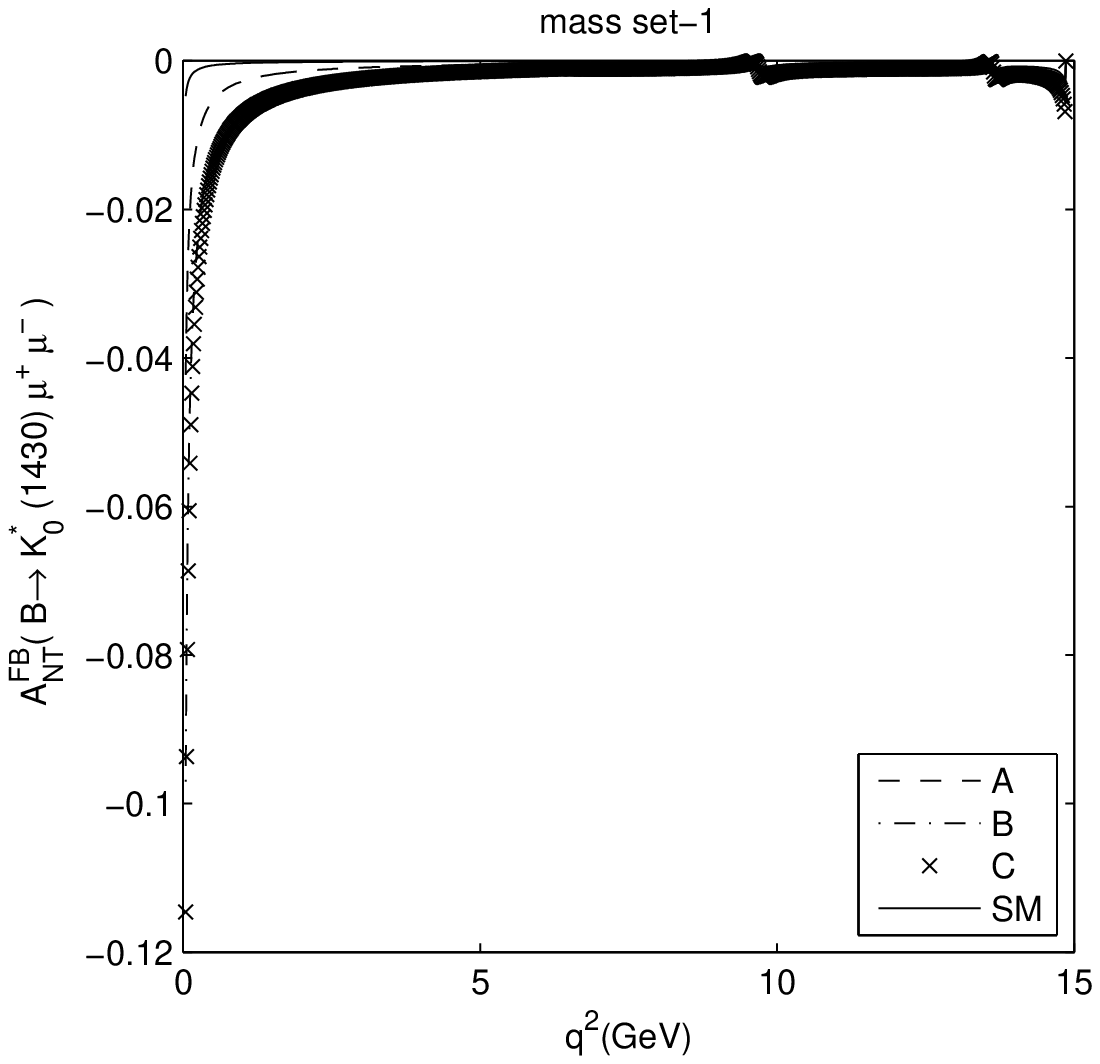}
             \includegraphics[height=2in]{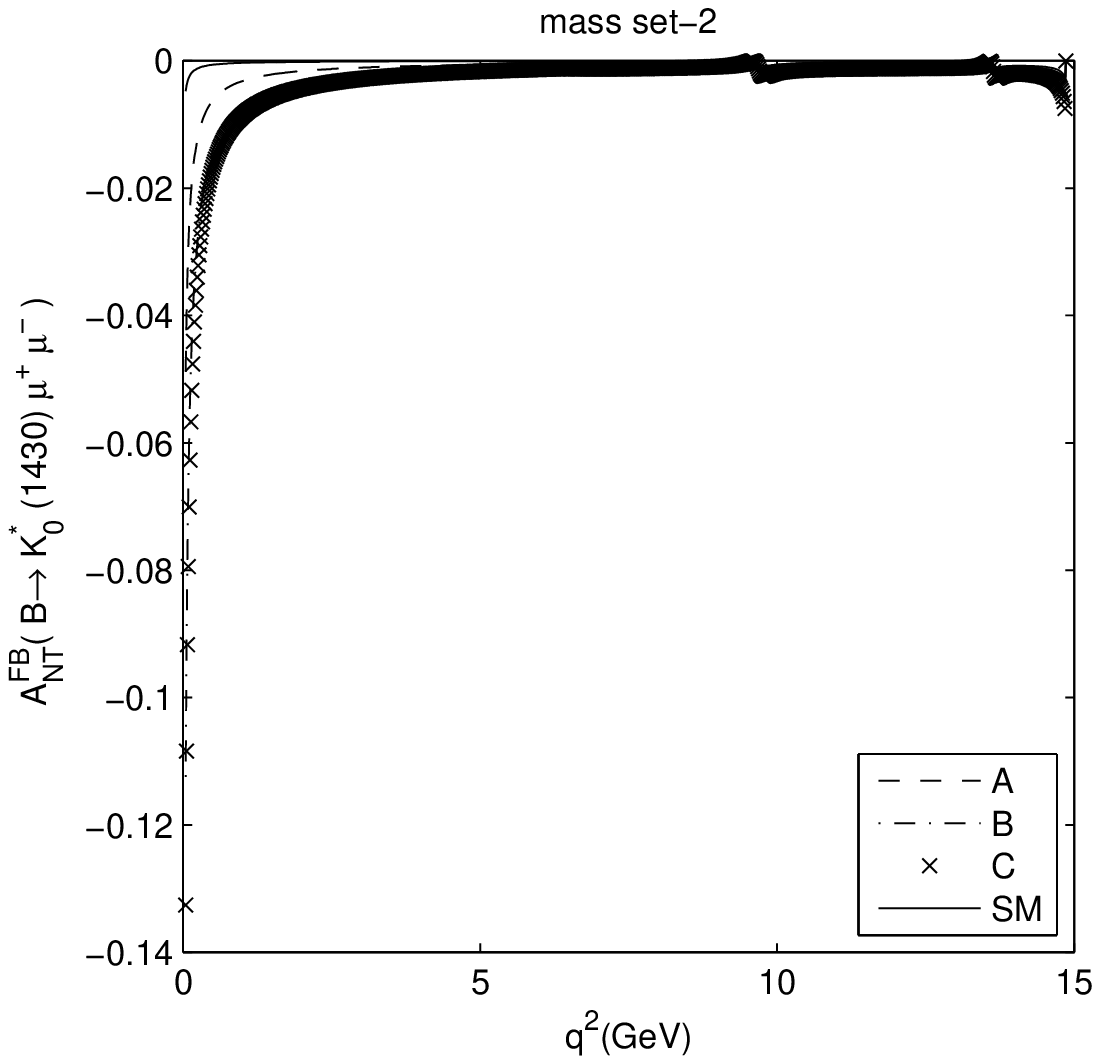}
             \includegraphics[height=2in]{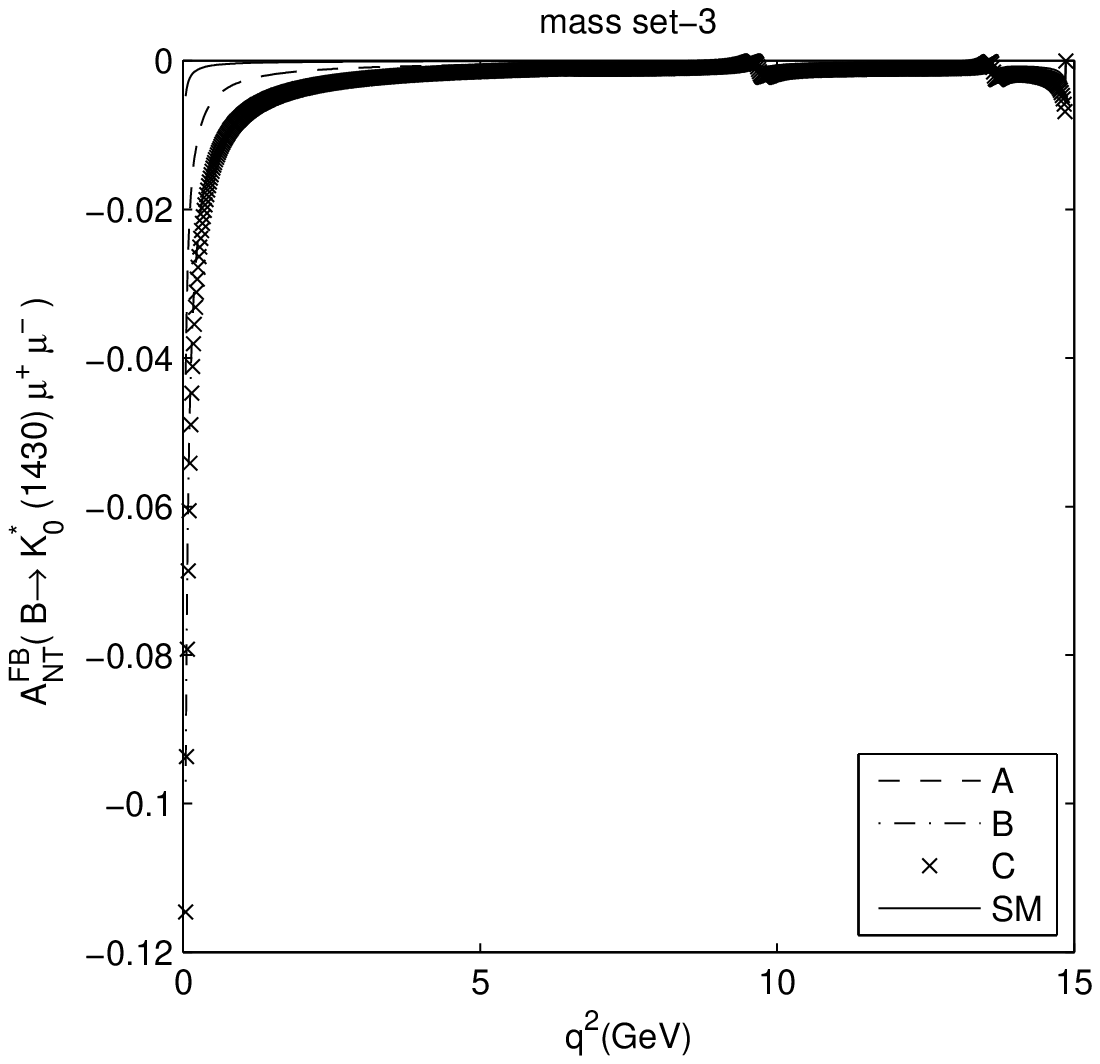}
             \includegraphics[height=2in]{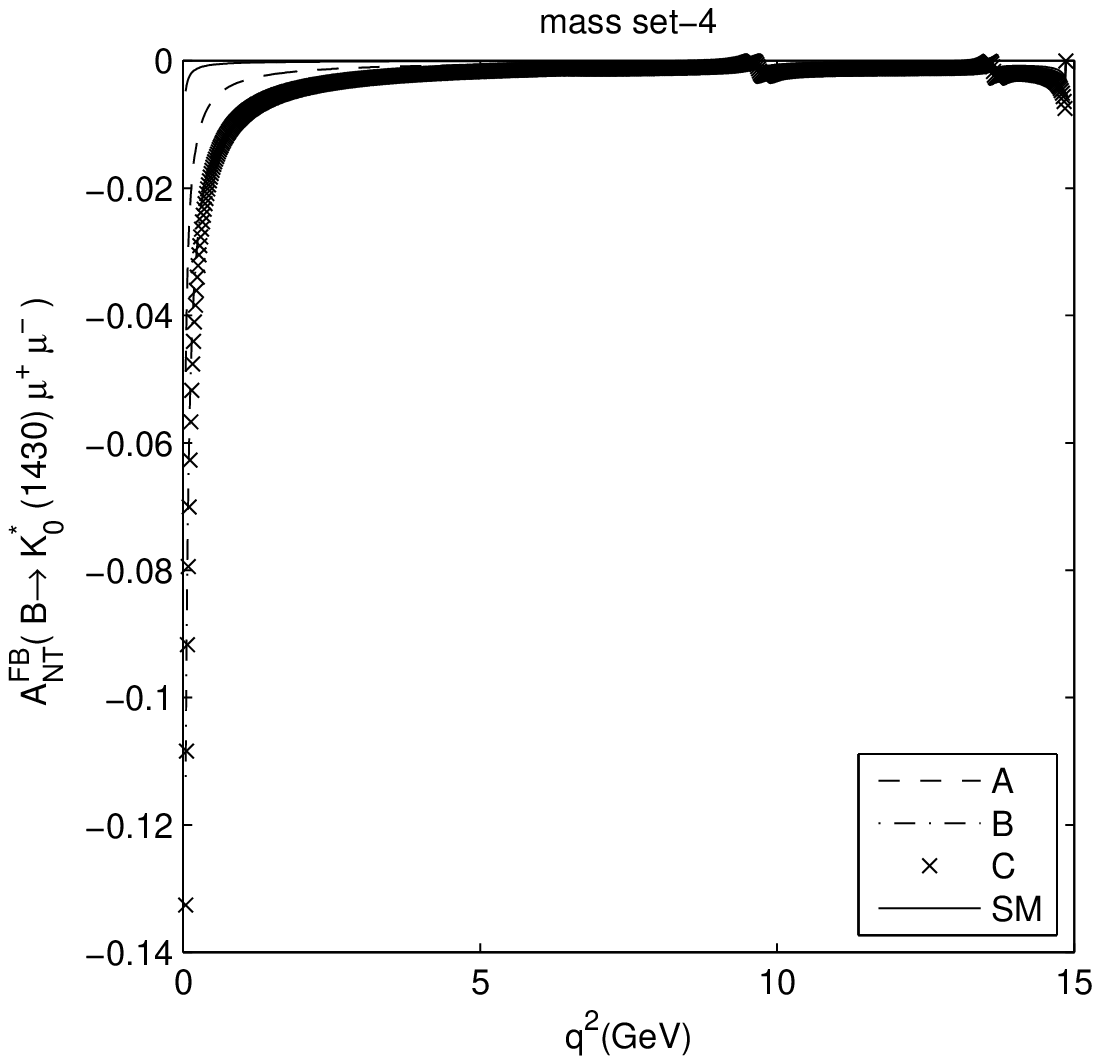}
               \caption{The dependence of the $ {\cal A}_{FB}^{NT}$ polarization  on $q^2$  and the three typical cases of 2HDM, i.e.
               cases A, B and C and SM  for  the $\mu$  channel of  $\overline{B}\to\overline{K}_0^{*}$  transition for the  mass sets 1, 2, 3 and 4. } \label{ANTmKstar}
    \end{figure}

\begin{figure}[ht]
  \centering
  \setlength{\fboxrule}{2pt}
        \centering
                       \includegraphics[height=2in]{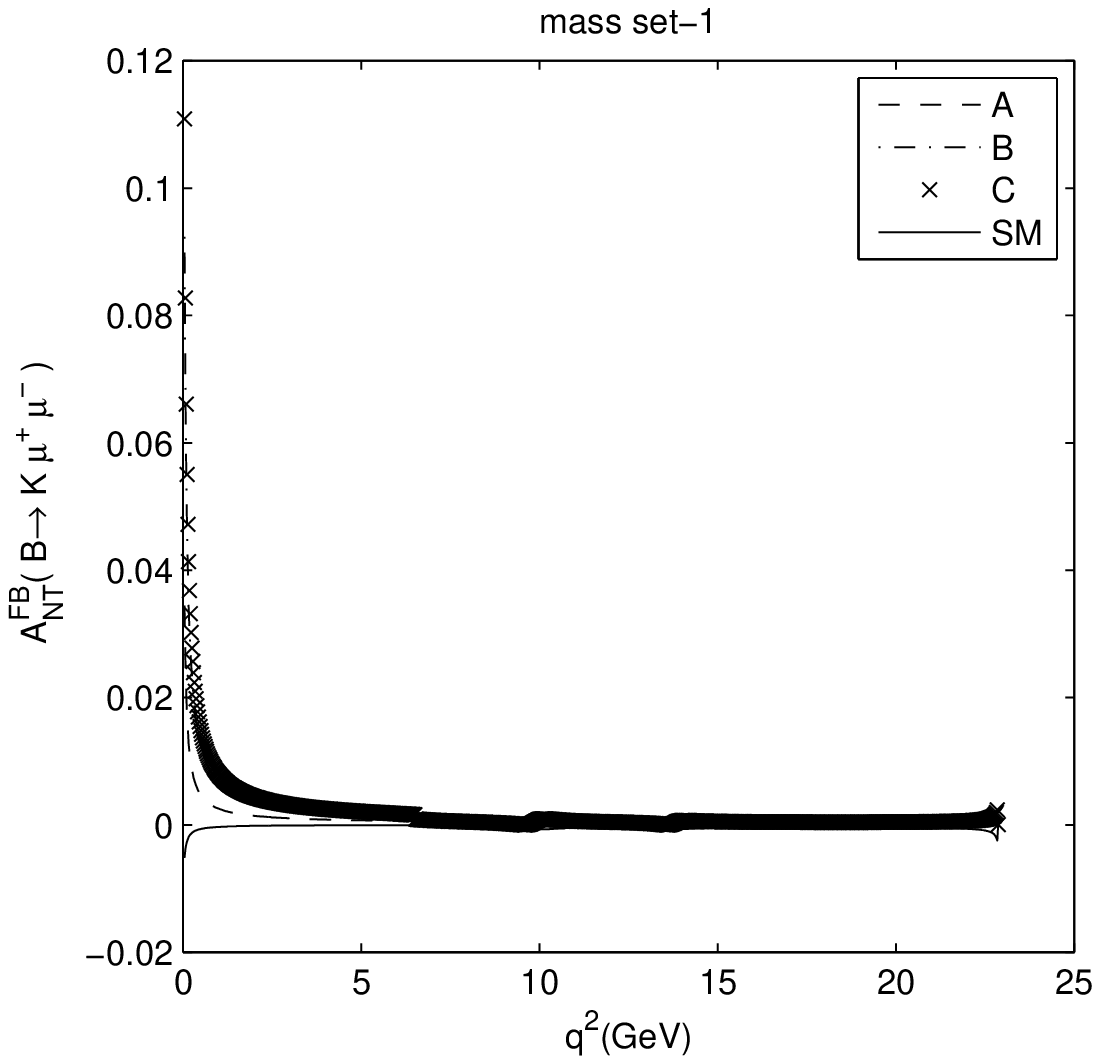}
             \includegraphics[height=2in]{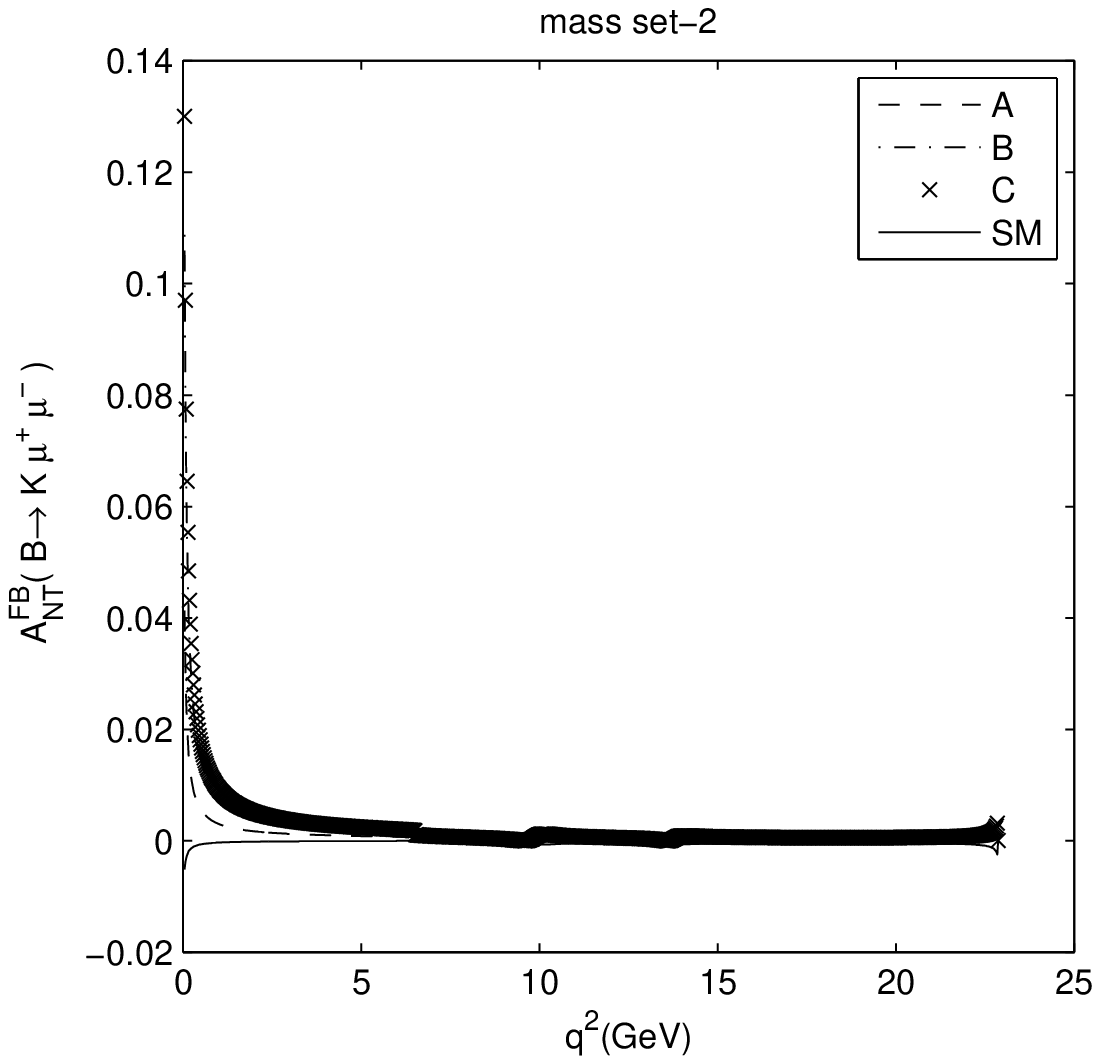}
             \includegraphics[height=2in]{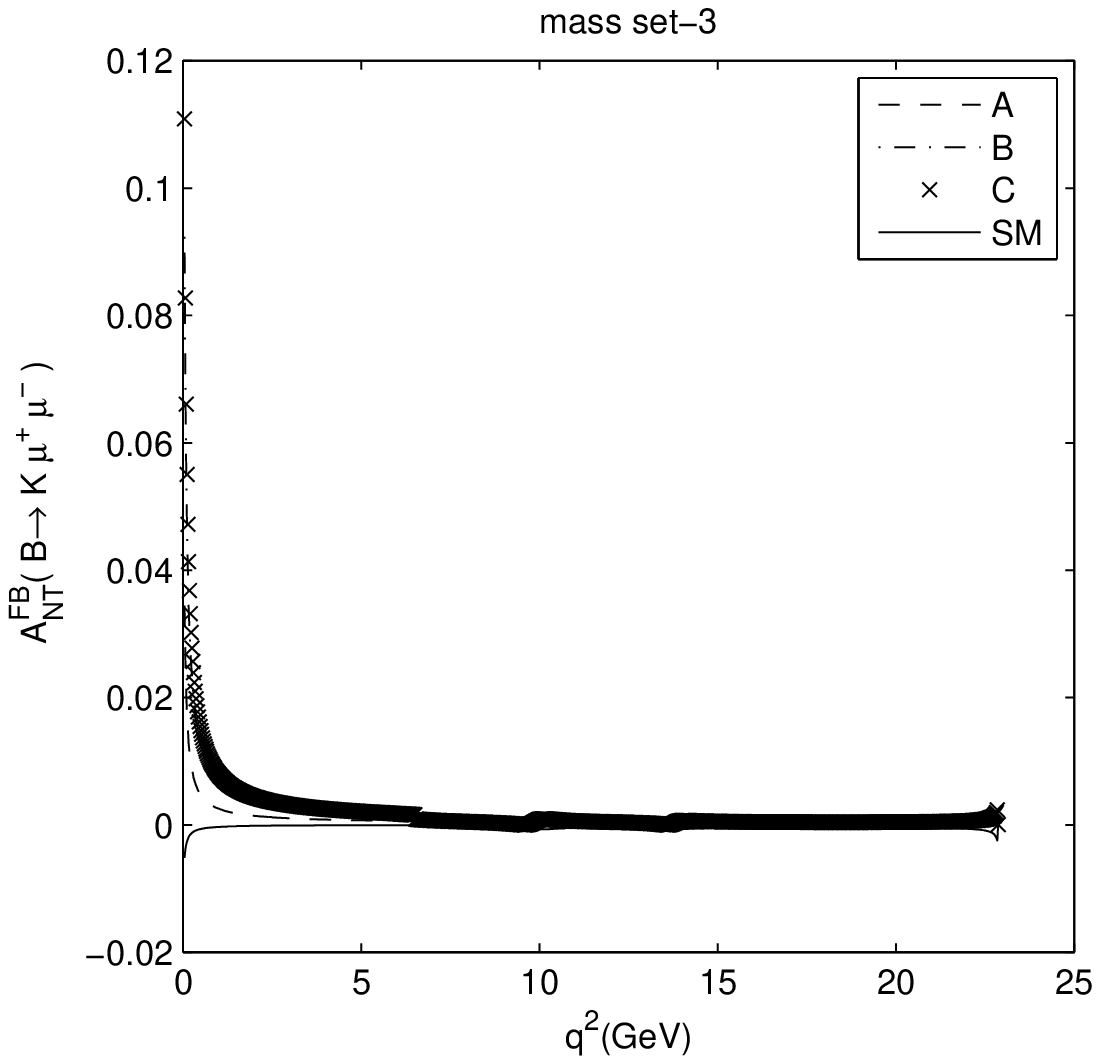}
             \includegraphics[height=2in]{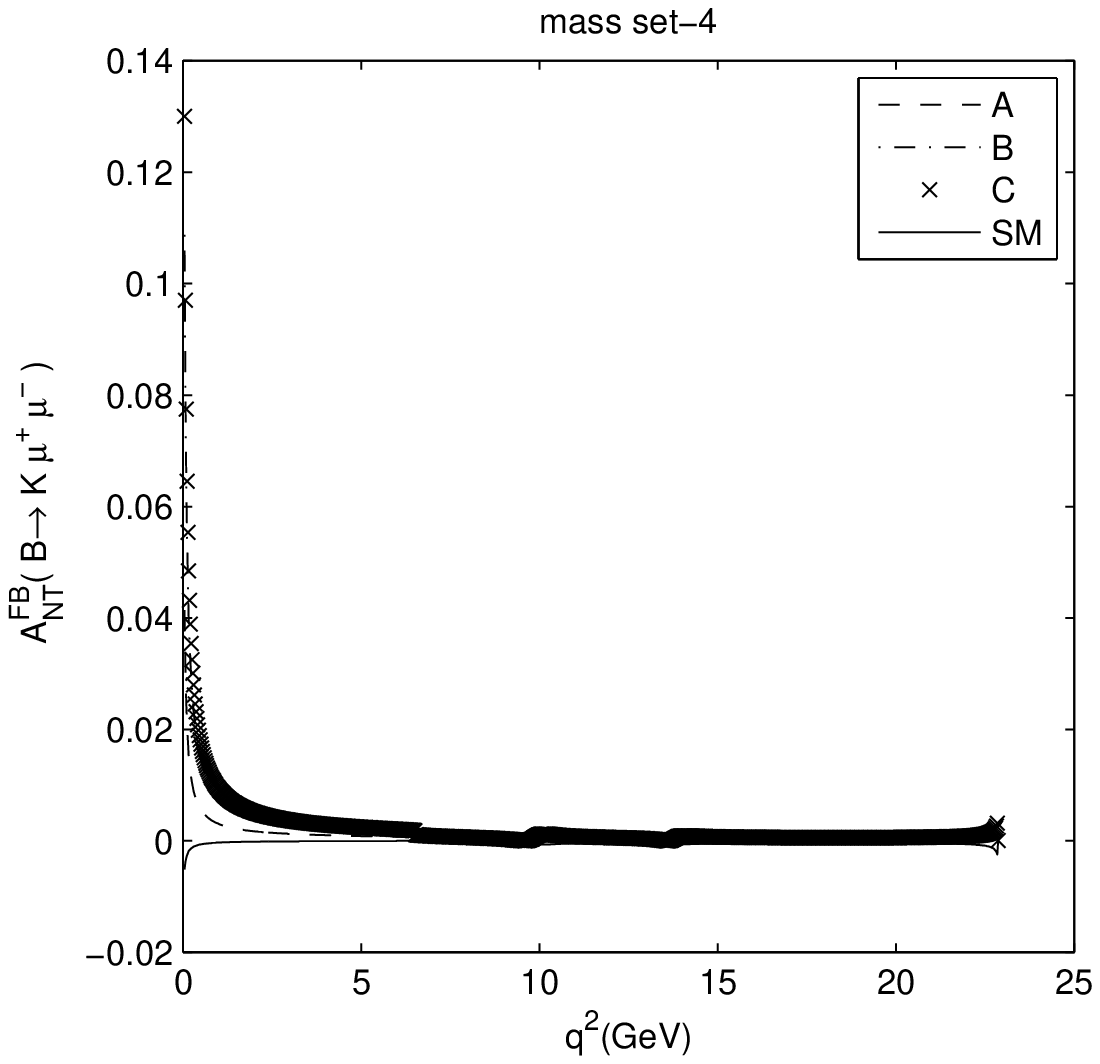}
               \caption{The dependence of the $ {\cal A}_{FB}^{NT}$ polarization  on $q^2$  and the three typical cases of 2HDM, i.e.
               cases A, B and C and SM  for  the $\mu$  channel of $\overline{B}\to\overline{K}$ transition for the  mass sets 1, 2, 3 and 4. } \label{ANTmK}
    \end{figure}
\begin{figure}[ht]
  \centering
  \setlength{\fboxrule}{2pt}
        \centering
                \includegraphics[height=2in]{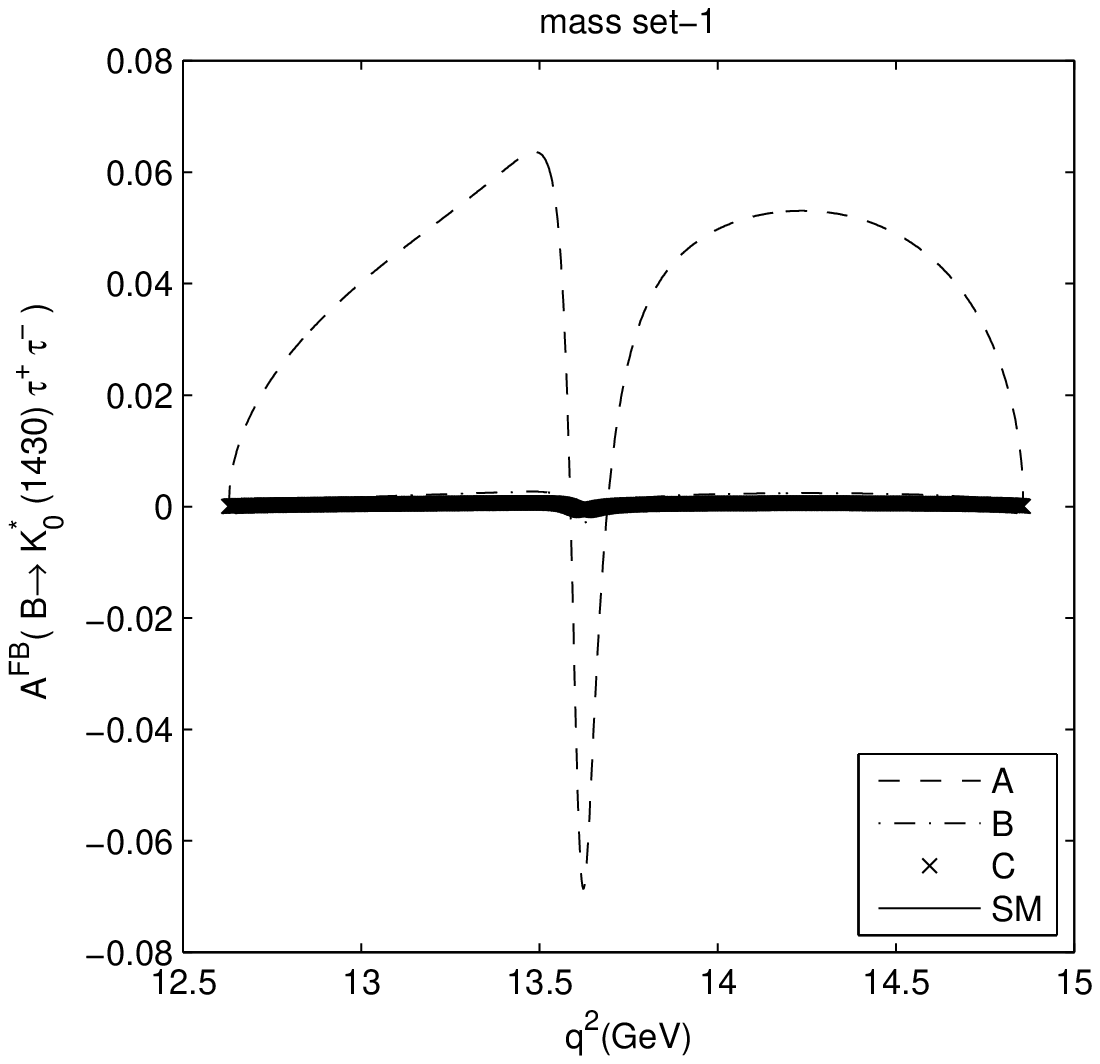}~
             \includegraphics[height=2in]{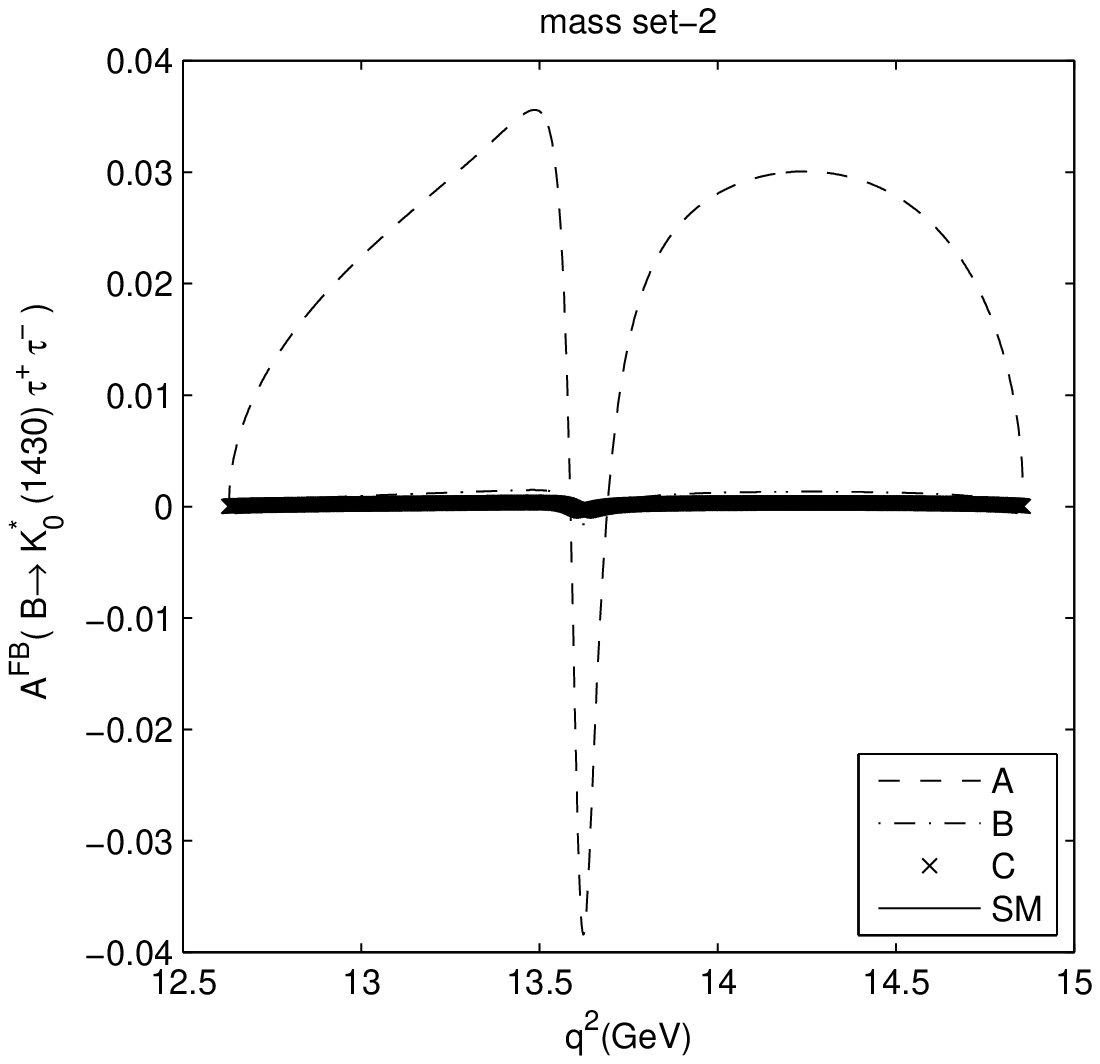}
              \includegraphics[height=2in]{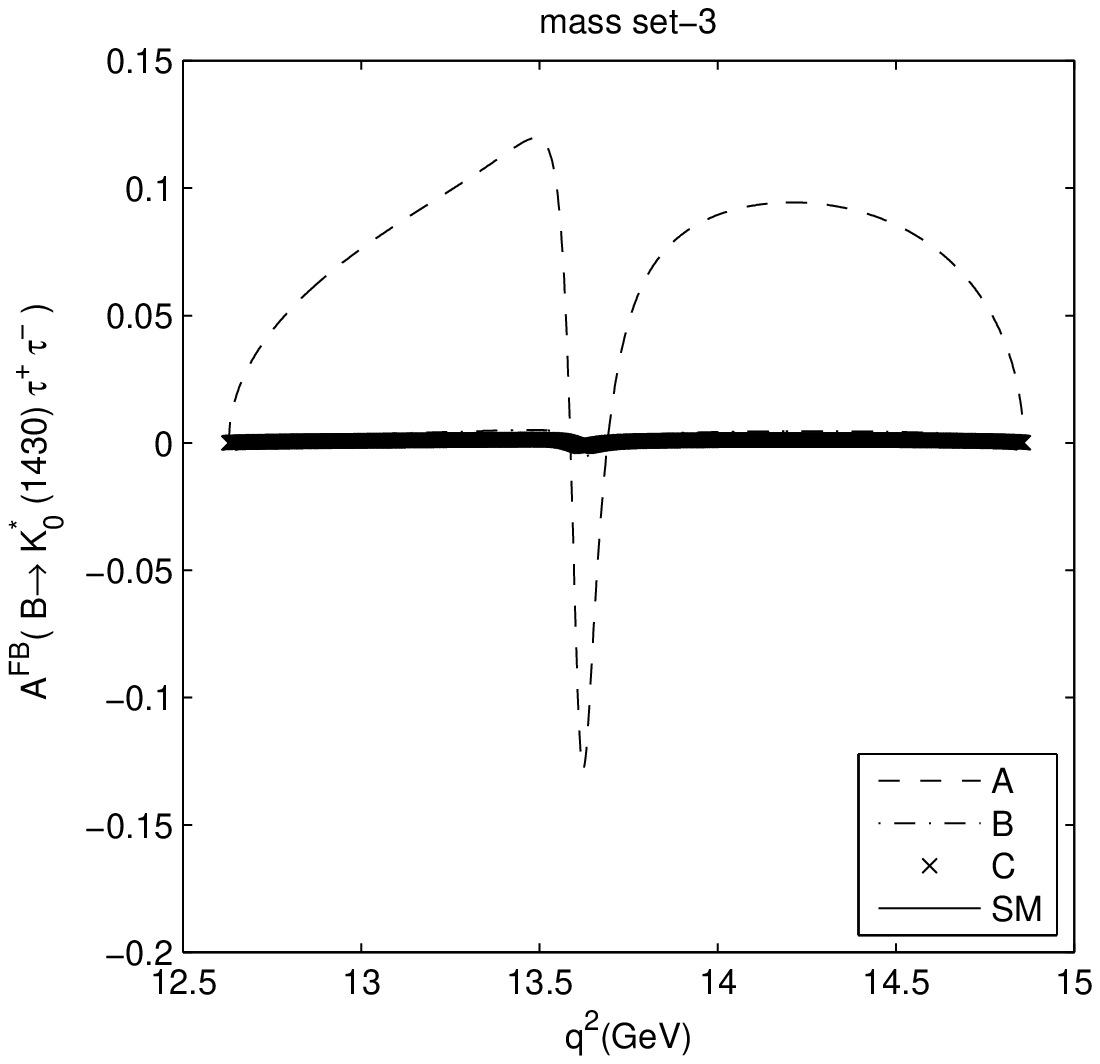}~
             \includegraphics[height=2in]{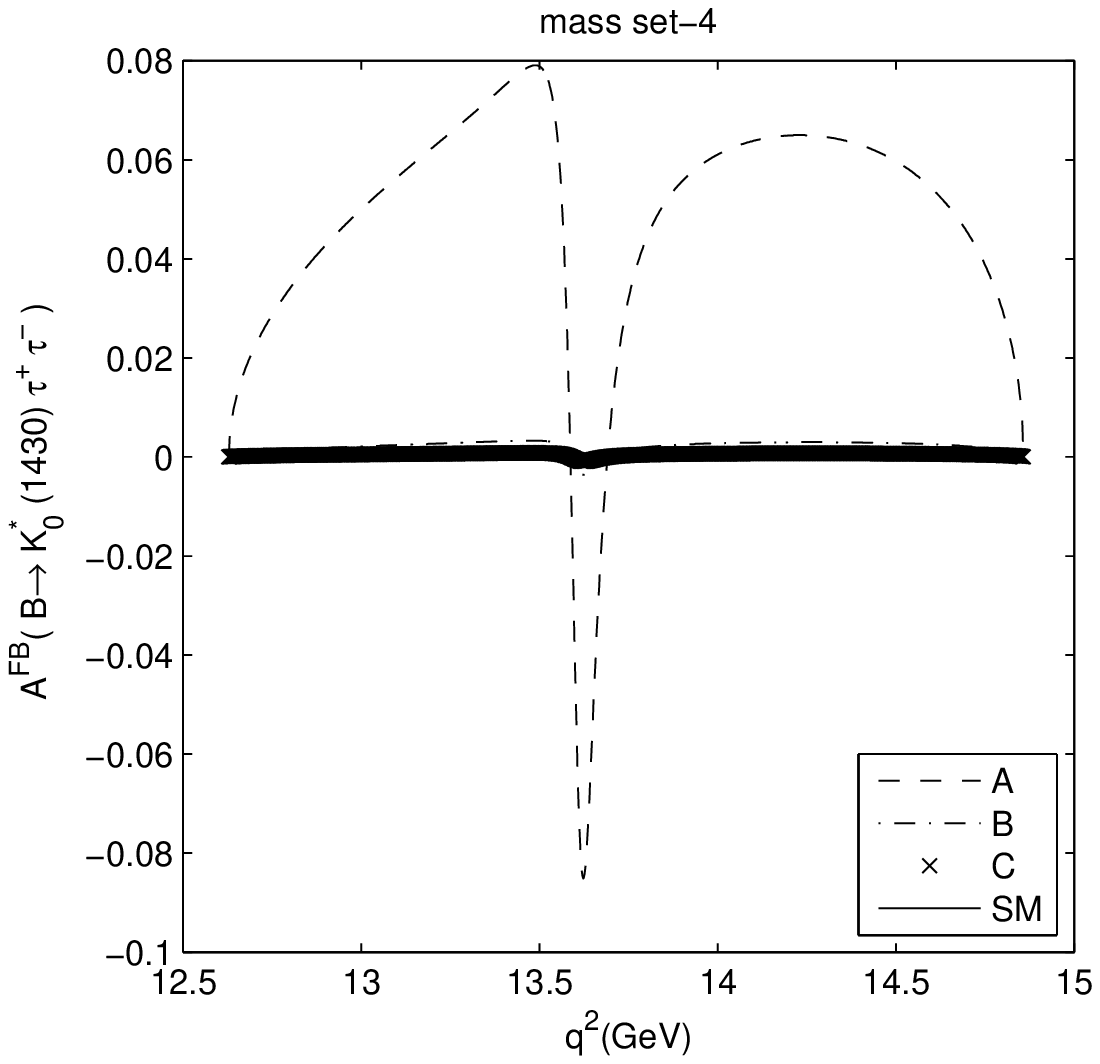}
               \caption{The dependence of the $ {\cal A}_{FB}$ polarization  on $q^2$  and the three typical cases of 2HDM, i.e.
               cases A, B and C and SM  for  the $\tau$  channel of $\overline{B}\to\overline{K}_0^{*}$  transition for the mass sets 1, 2, 3 and 4.} \label{AFBtKstar}
    \end{figure}
    \begin{figure}[ht]
  \centering
  \setlength{\fboxrule}{2pt}
        \centering
                \includegraphics[height=2in]{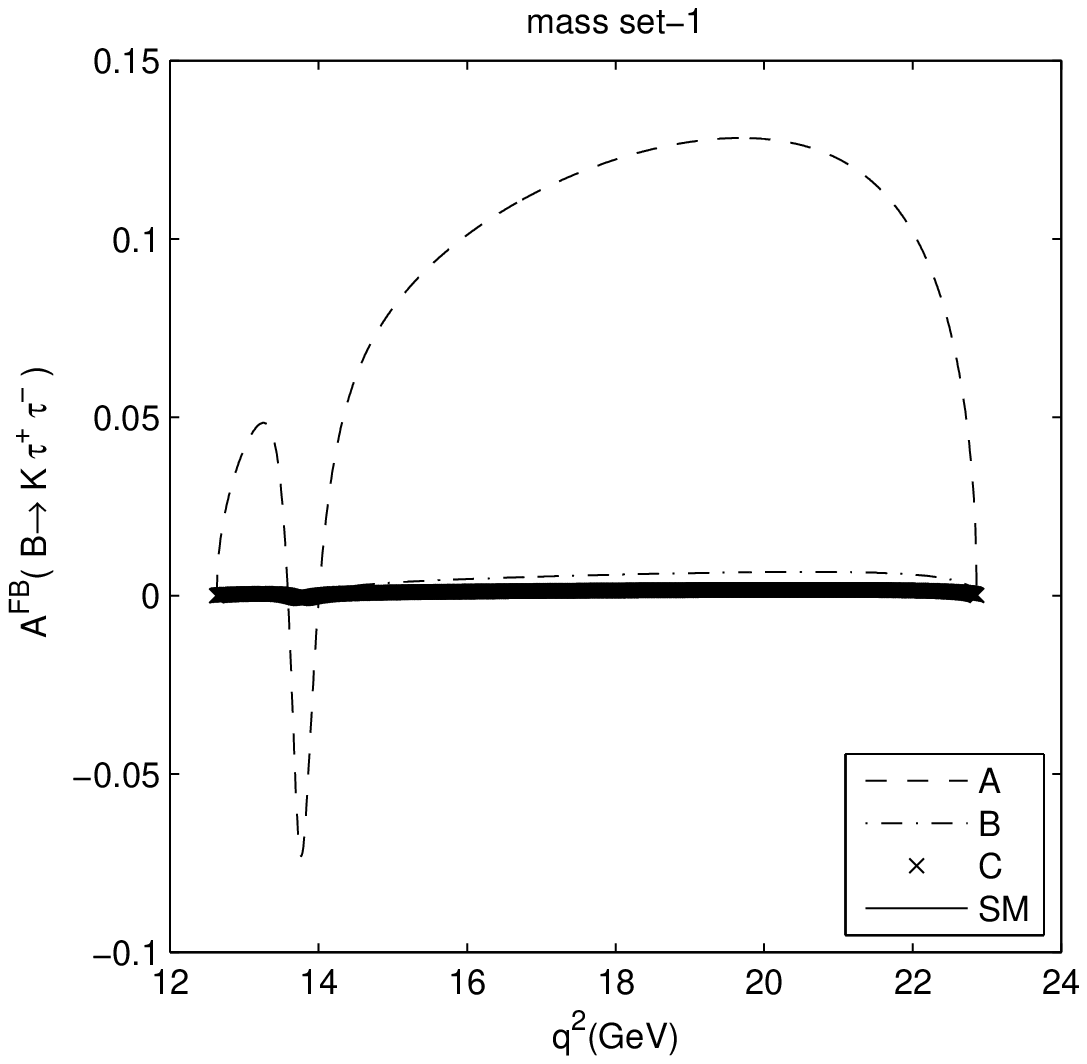}~
             \includegraphics[height=2in]{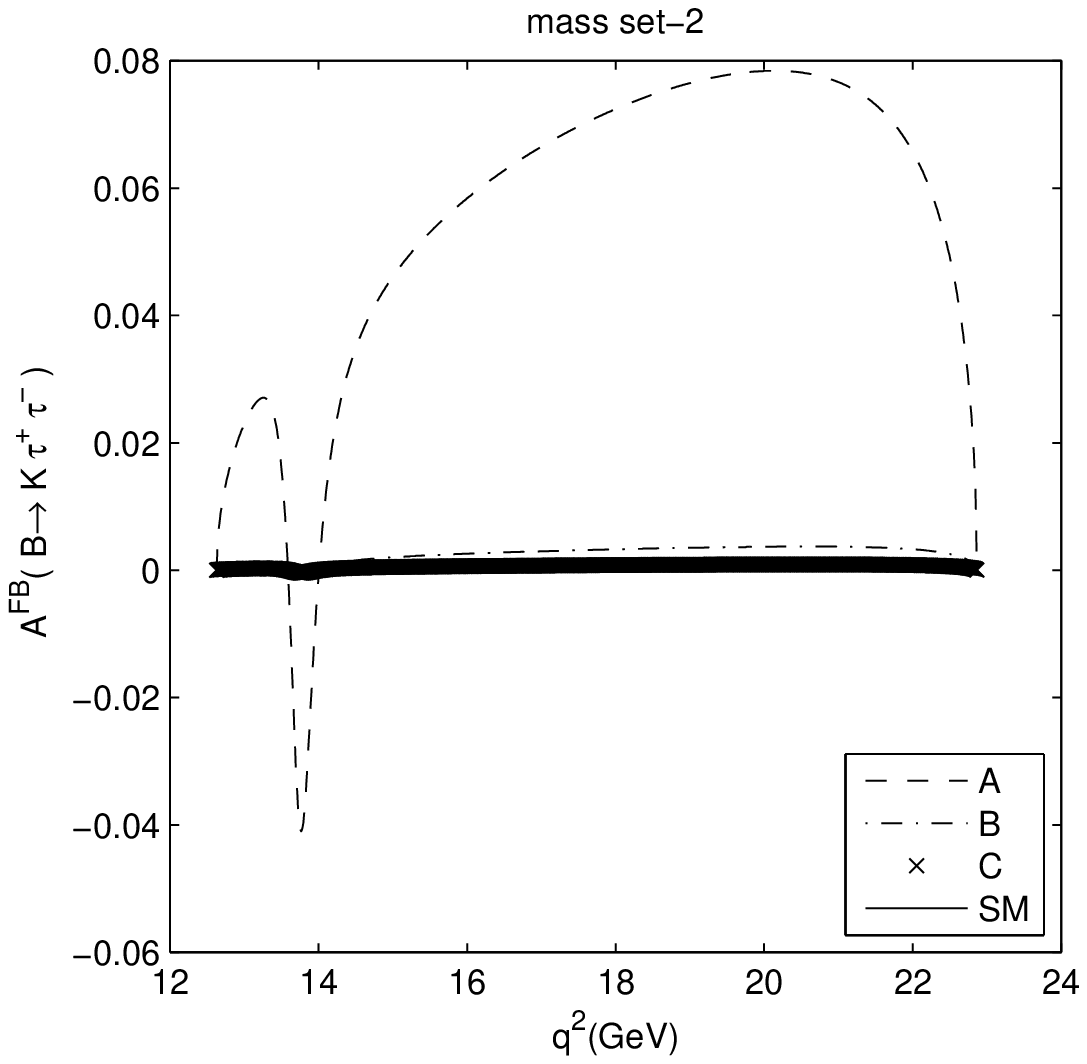}
              \includegraphics[height=2in]{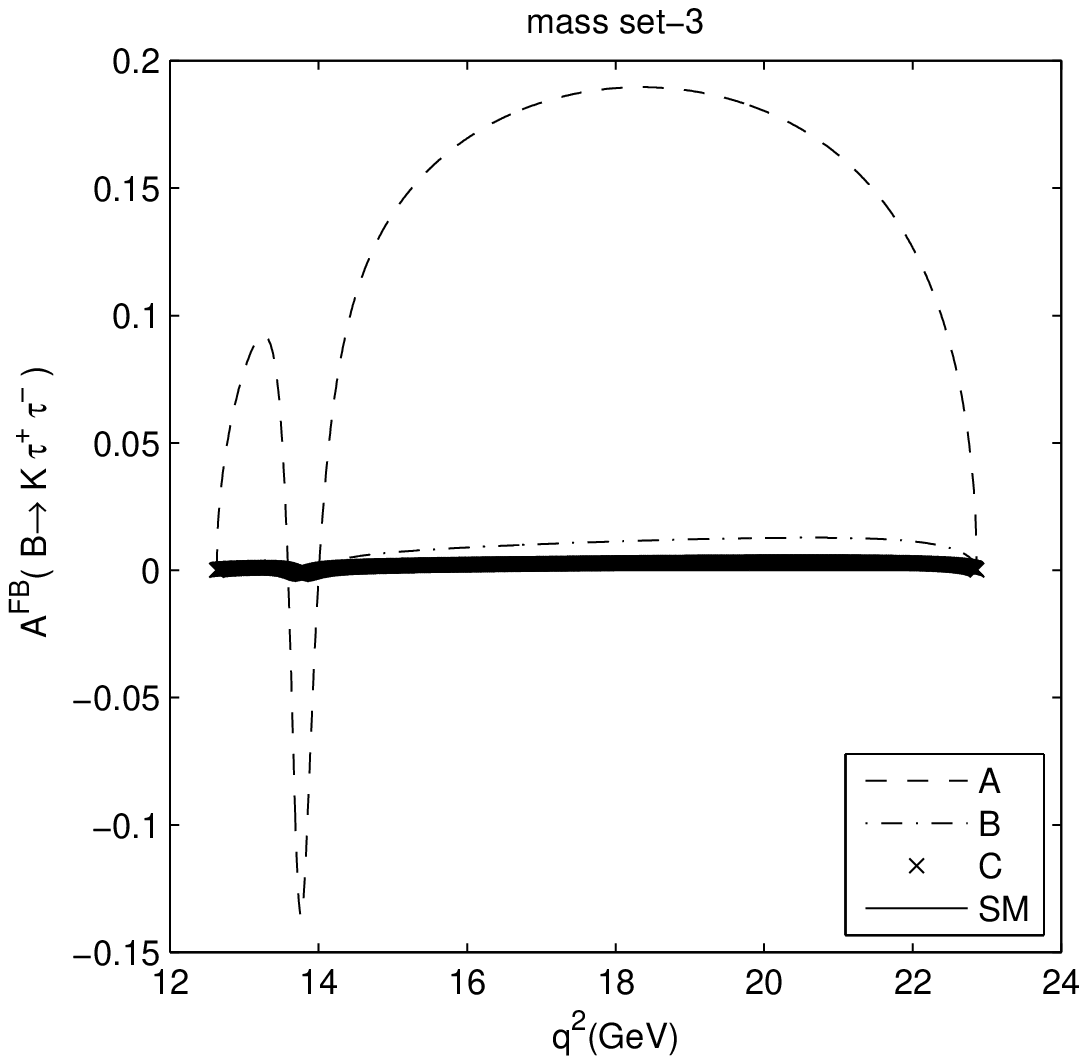}~
             \includegraphics[height=2in]{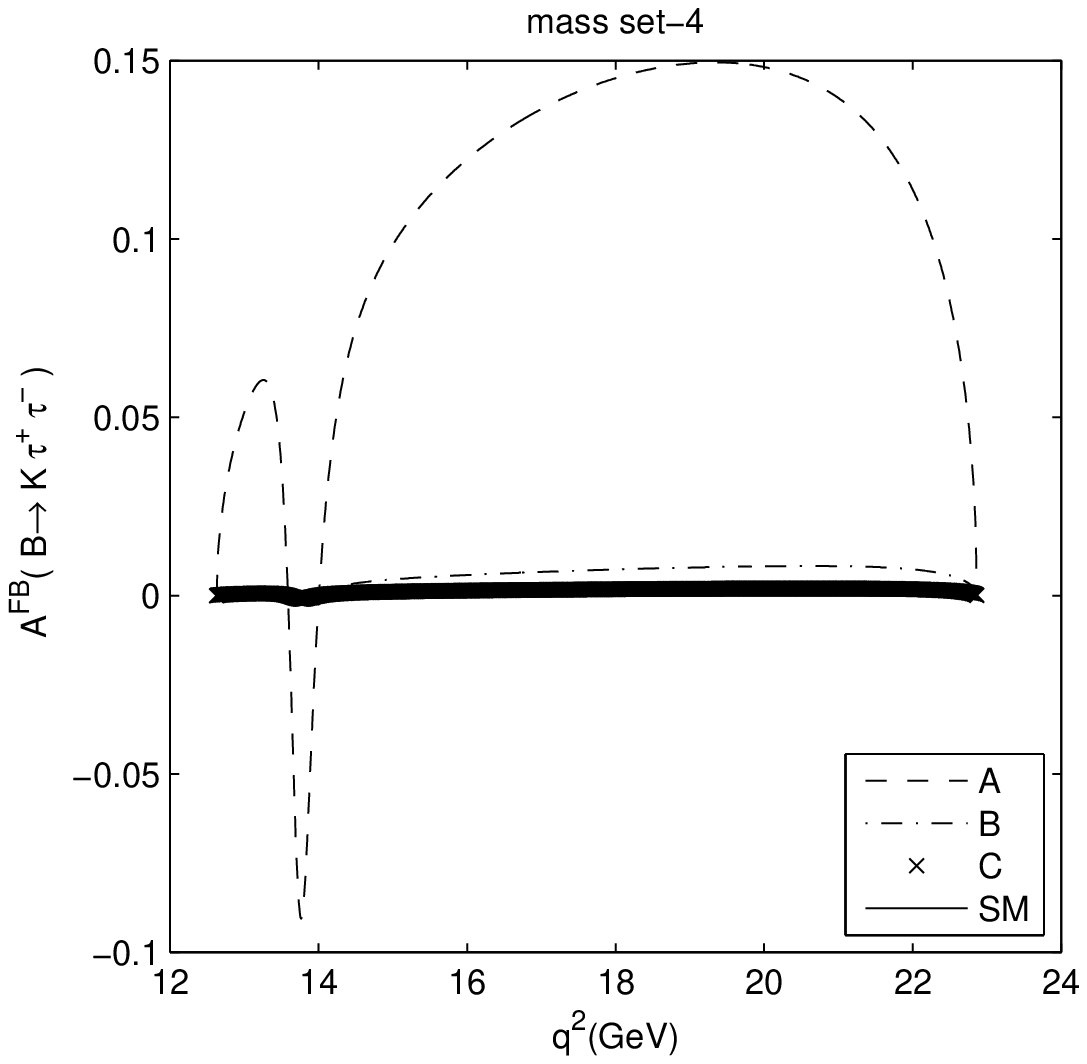}
               \caption{The dependence of the $ {\cal A}_{FB}$ polarization  on $q^2$  and the three typical cases of 2HDM, i.e.
               cases A, B and C and SM  for  the $\tau$  channel of  $\overline{B}\to\overline{K}$ transition for  mass sets 1, 2, 3 and 4. } \label{AFBtK}
    \end{figure}

 \begin{figure}[ht]
  \centering
  \setlength{\fboxrule}{2pt}
        \centering
                \includegraphics[height=2in]{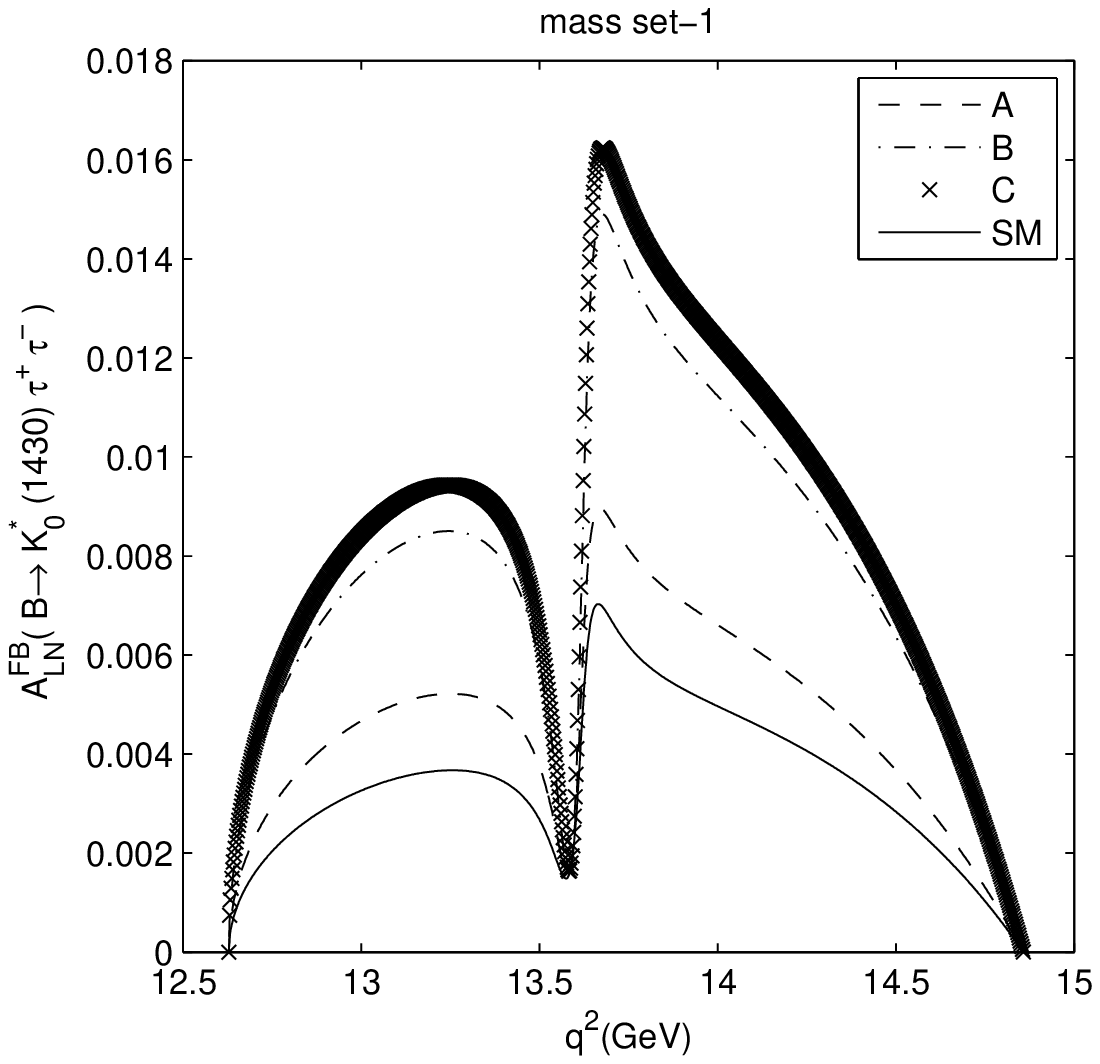}~
             \includegraphics[height=2in]{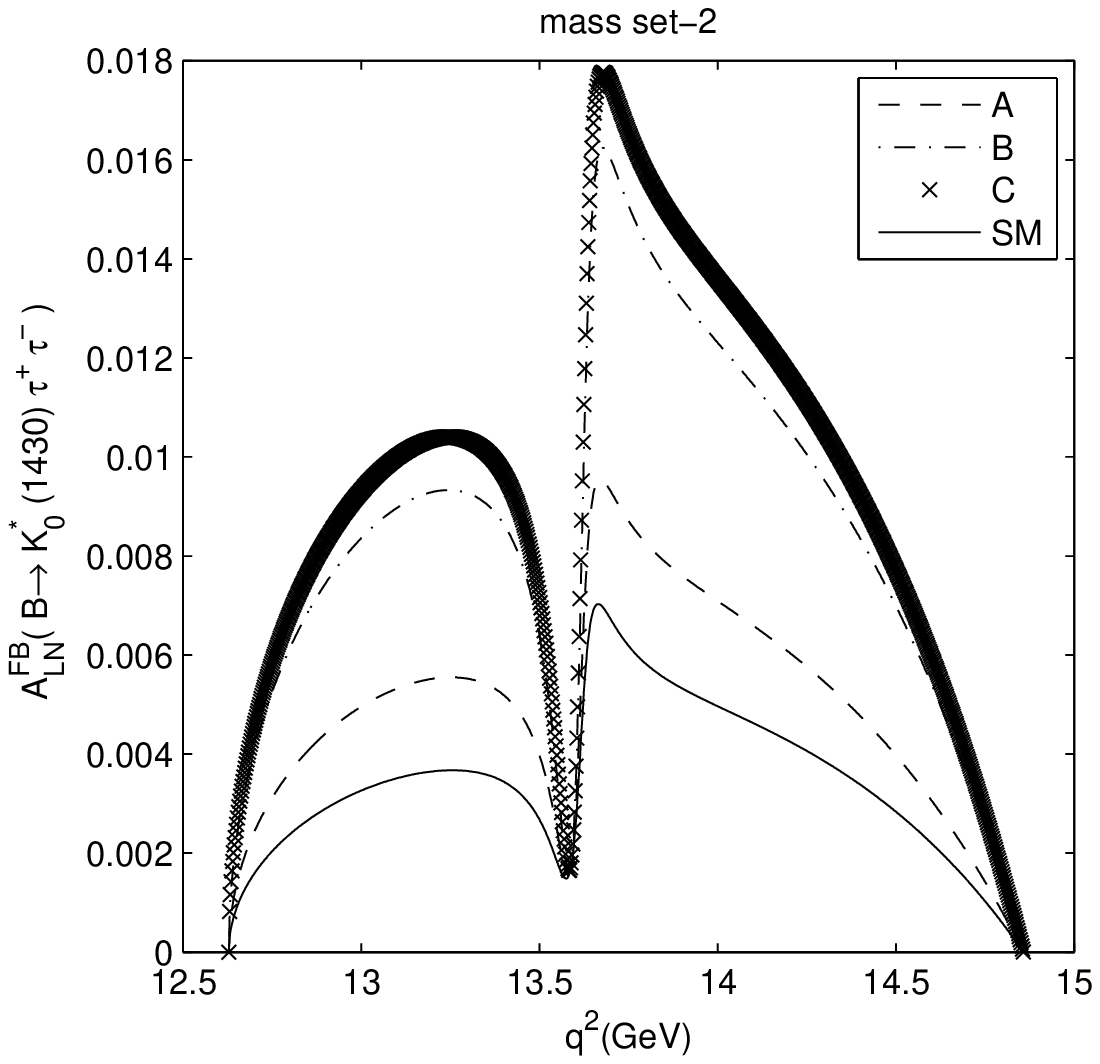}
              \includegraphics[height=2in]{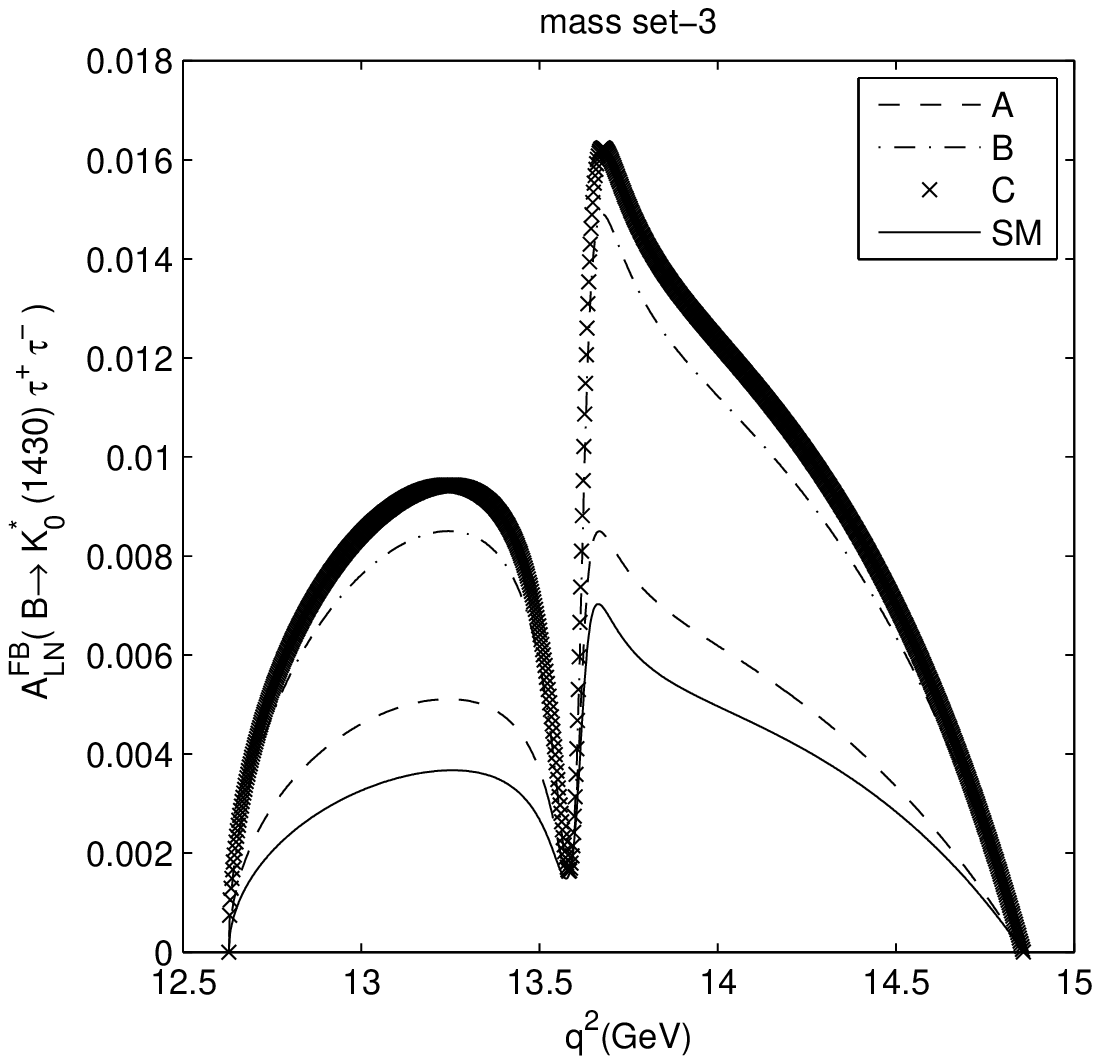}~
             \includegraphics[height=2in]{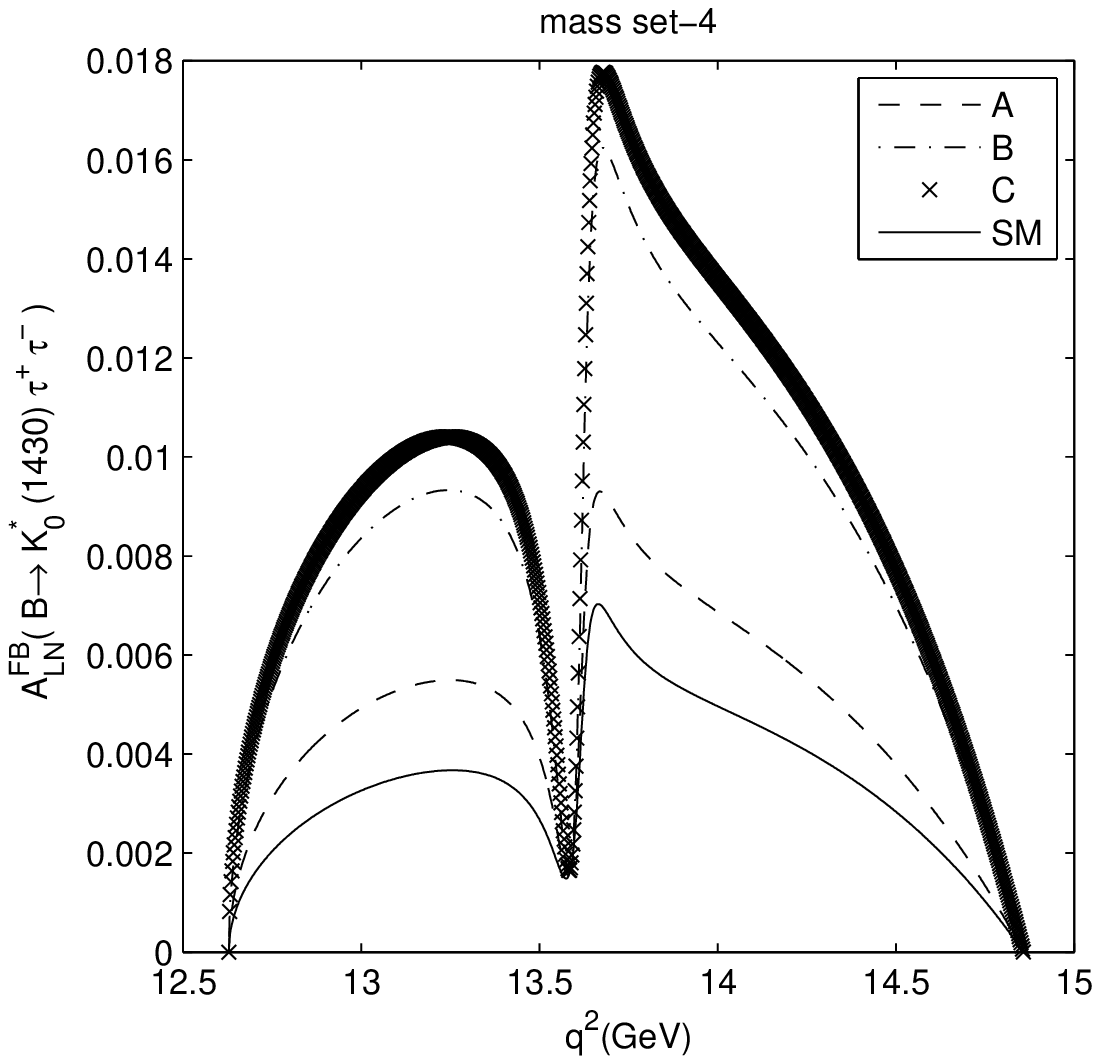}
               \caption{The dependence of the $ A_{LN}$ polarization  on $q^2$  and the three typical cases of 2HDM, i.e.
               cases A, B and C and SM  for  the $\tau$  channel of $\overline{B}\to\overline{K}_0^{*}$  transition for the mass sets 1, 2, 3 and 4.} \label{ALNtKstar}
    \end{figure}
    \begin{figure}[ht]
  \centering
  \setlength{\fboxrule}{2pt}
        \centering
                \includegraphics[height=2in]{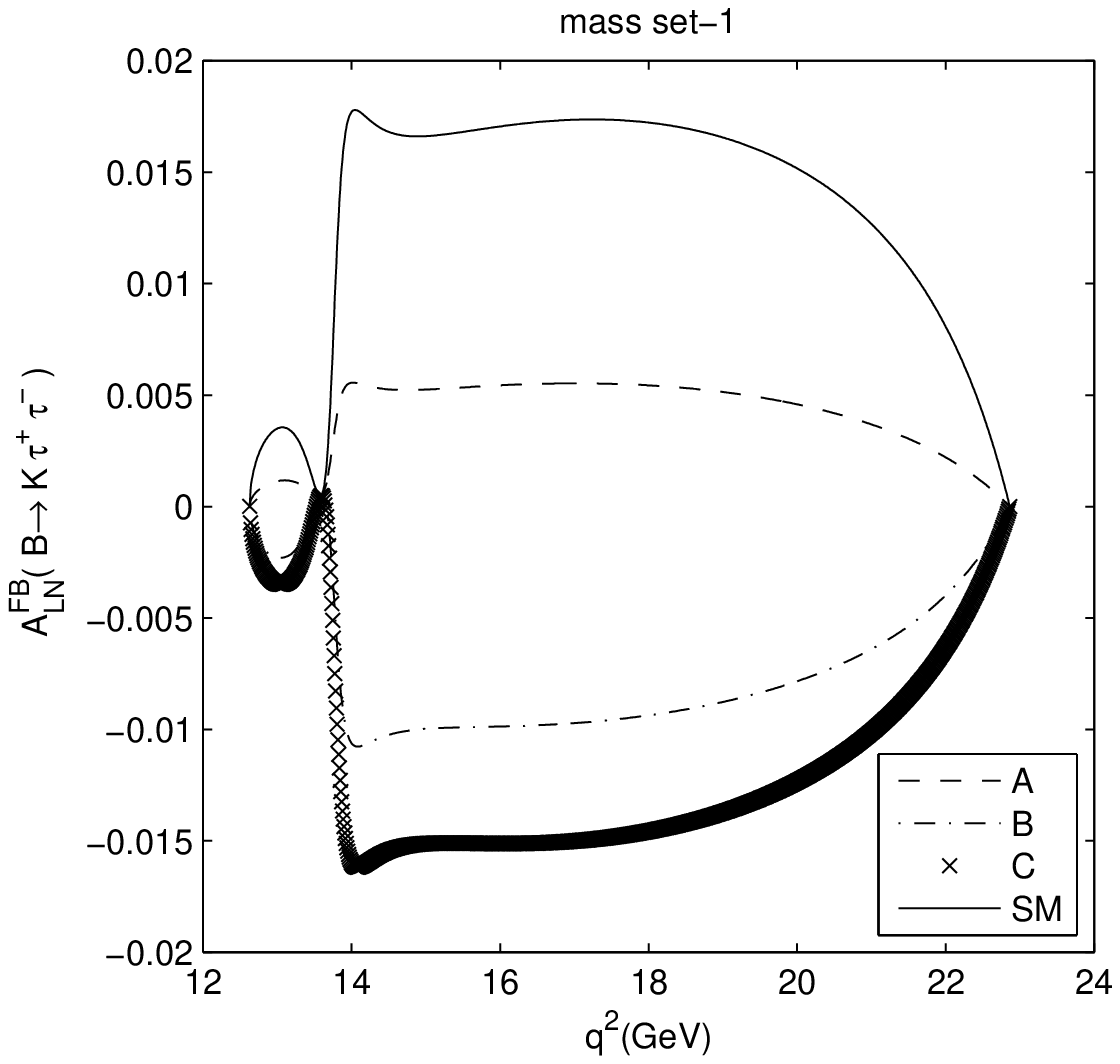}~
             \includegraphics[height=2in]{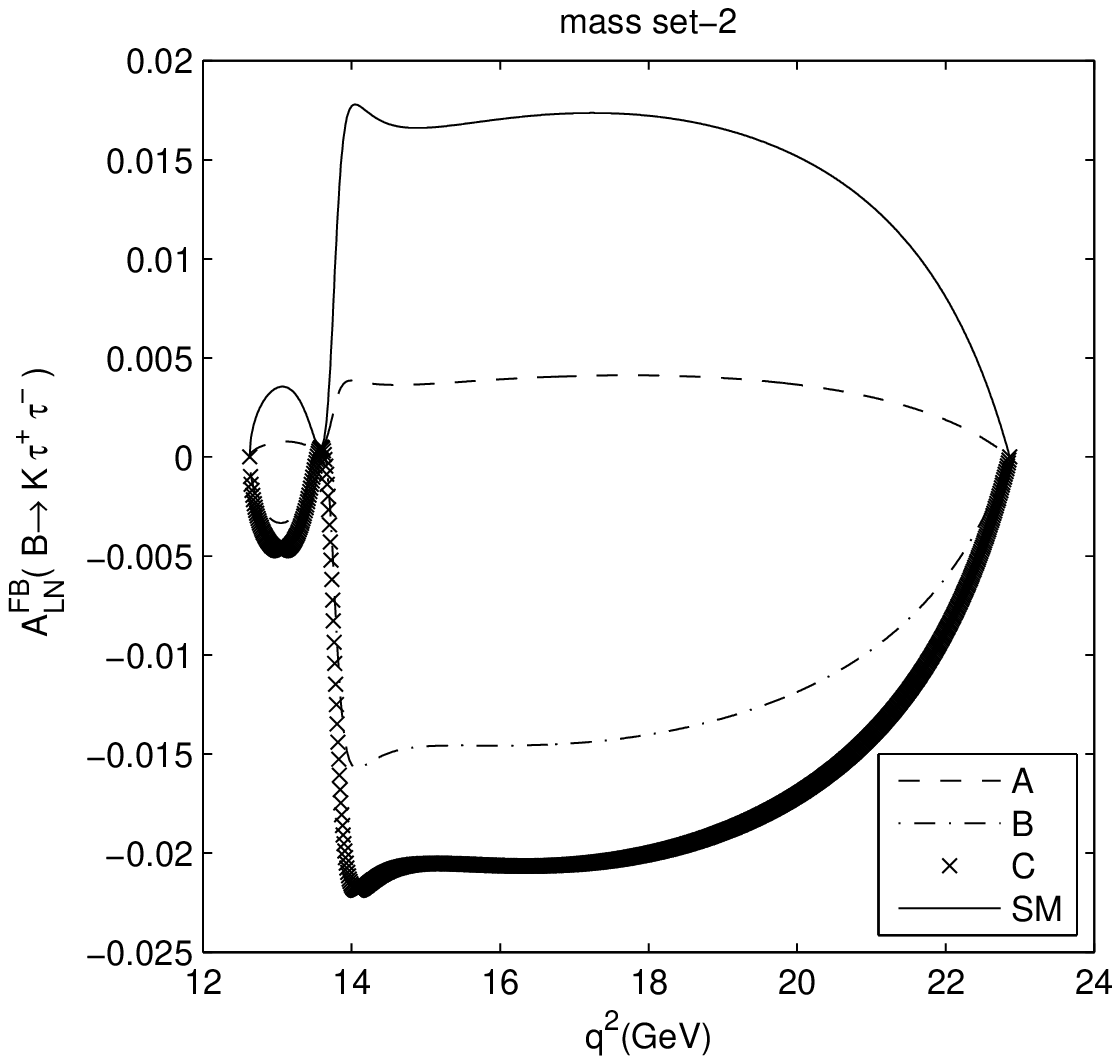}
              \includegraphics[height=2in]{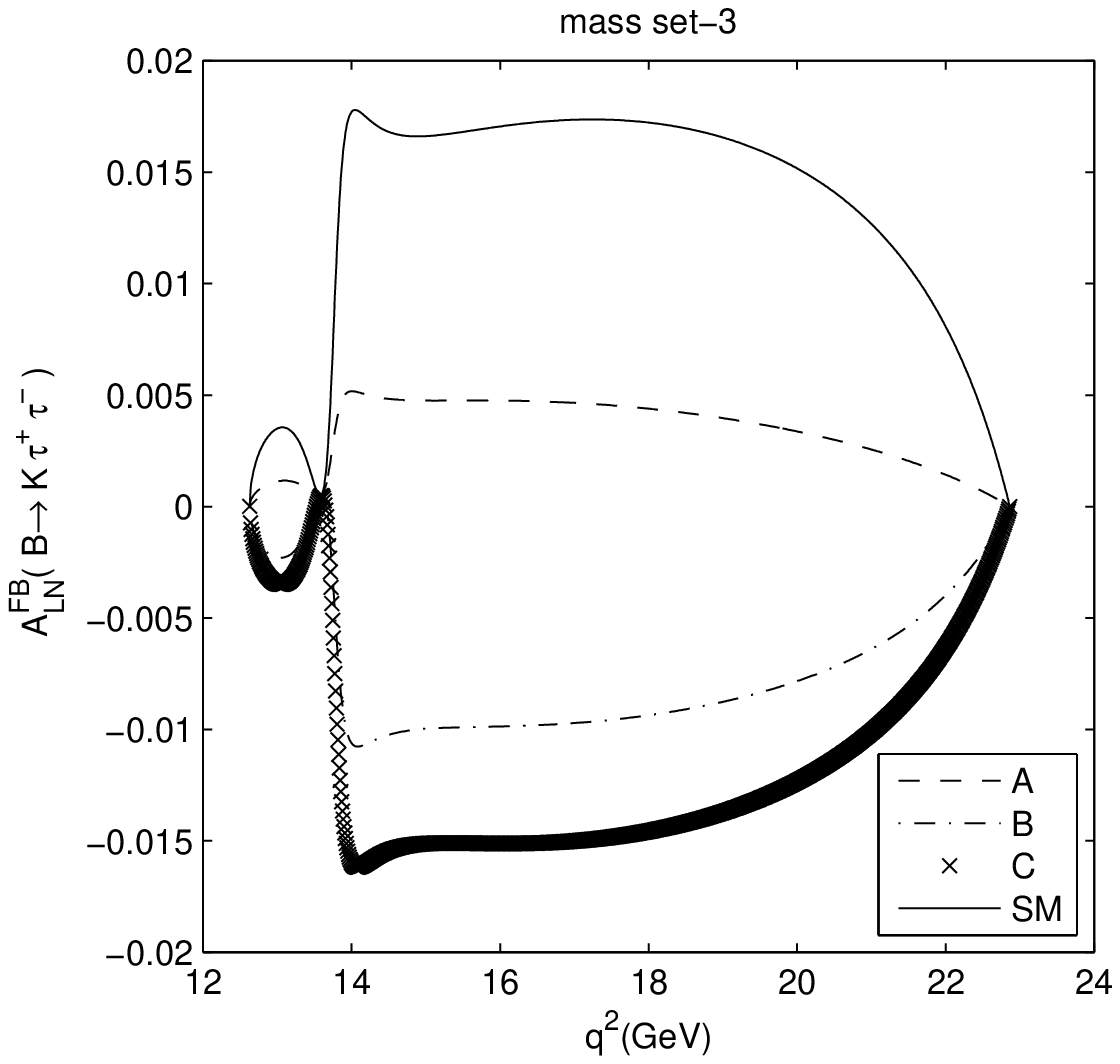}~
             \includegraphics[height=2in]{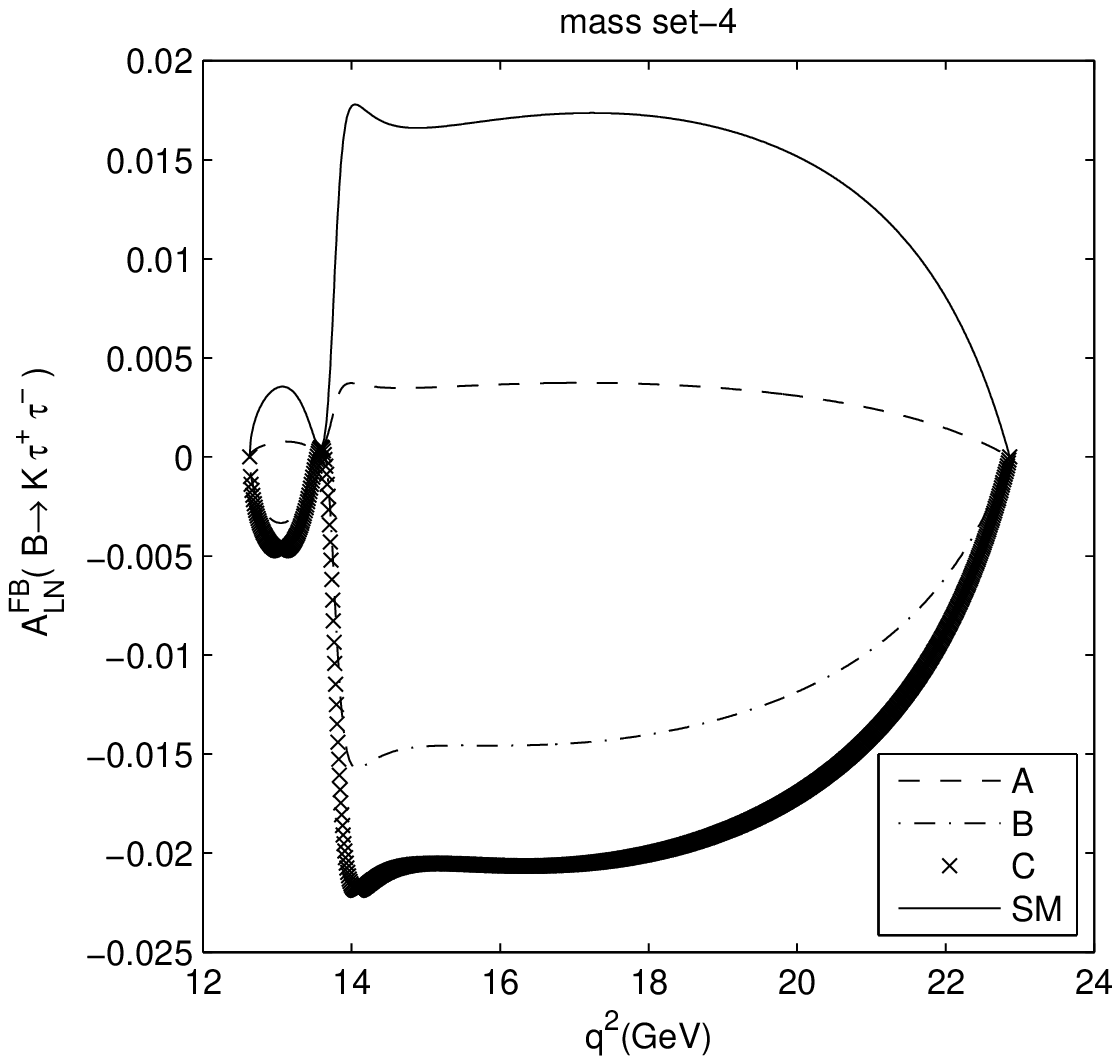}
               \caption{The dependence of the $ A_{LN}$ polarization  on $q^2$  and the three typical cases of 2HDM, i.e.
               cases A, B and C and SM  for  the $\tau$  channel of $\overline{B}\to\overline{K}$  transition for the mass sets 1, 2, 3 and 4.} \label{ALNtK}
    \end{figure}
     \begin{figure}[ht]
  \centering
  \setlength{\fboxrule}{2pt}
        \centering
                \includegraphics[height=2in]{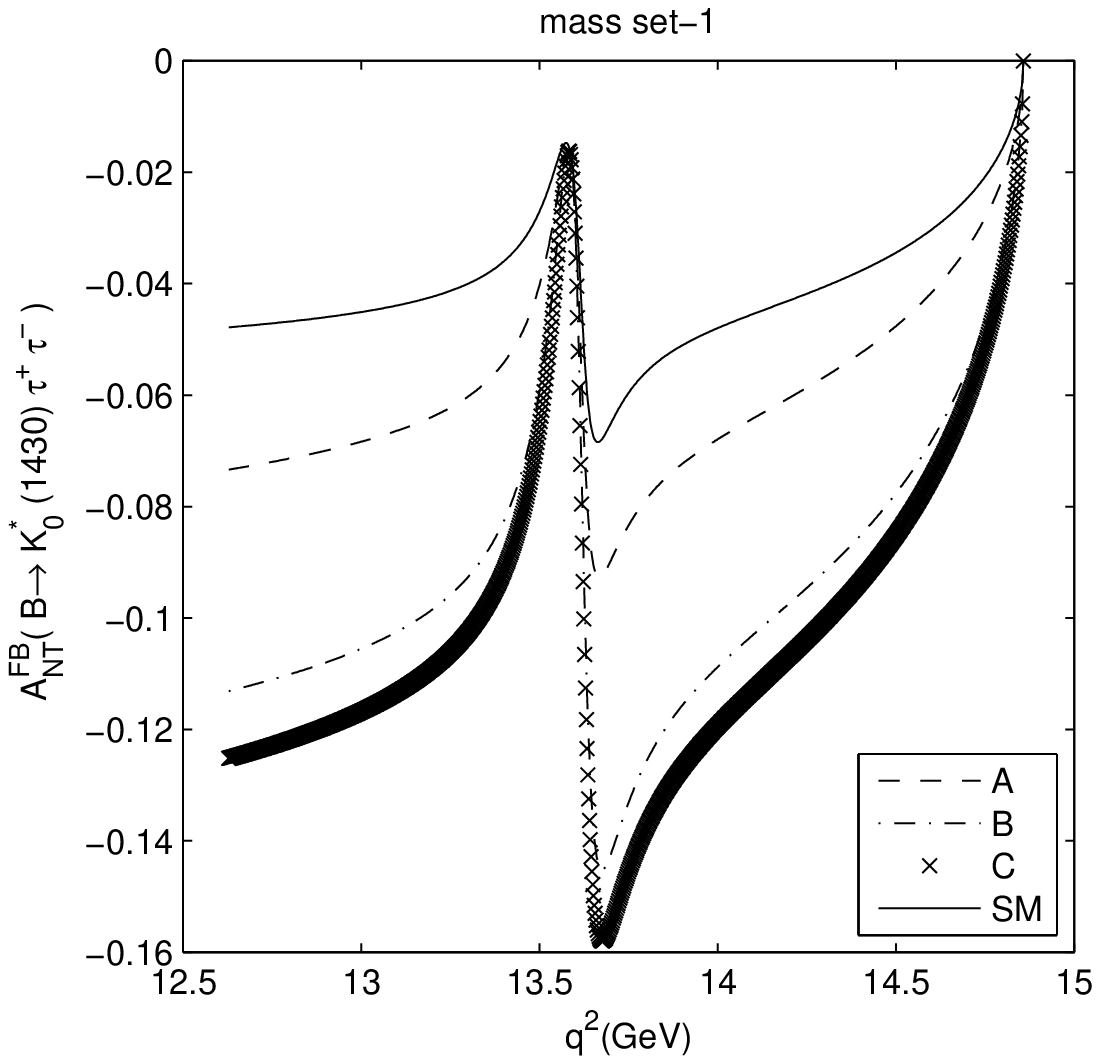}~
             \includegraphics[height=2in]{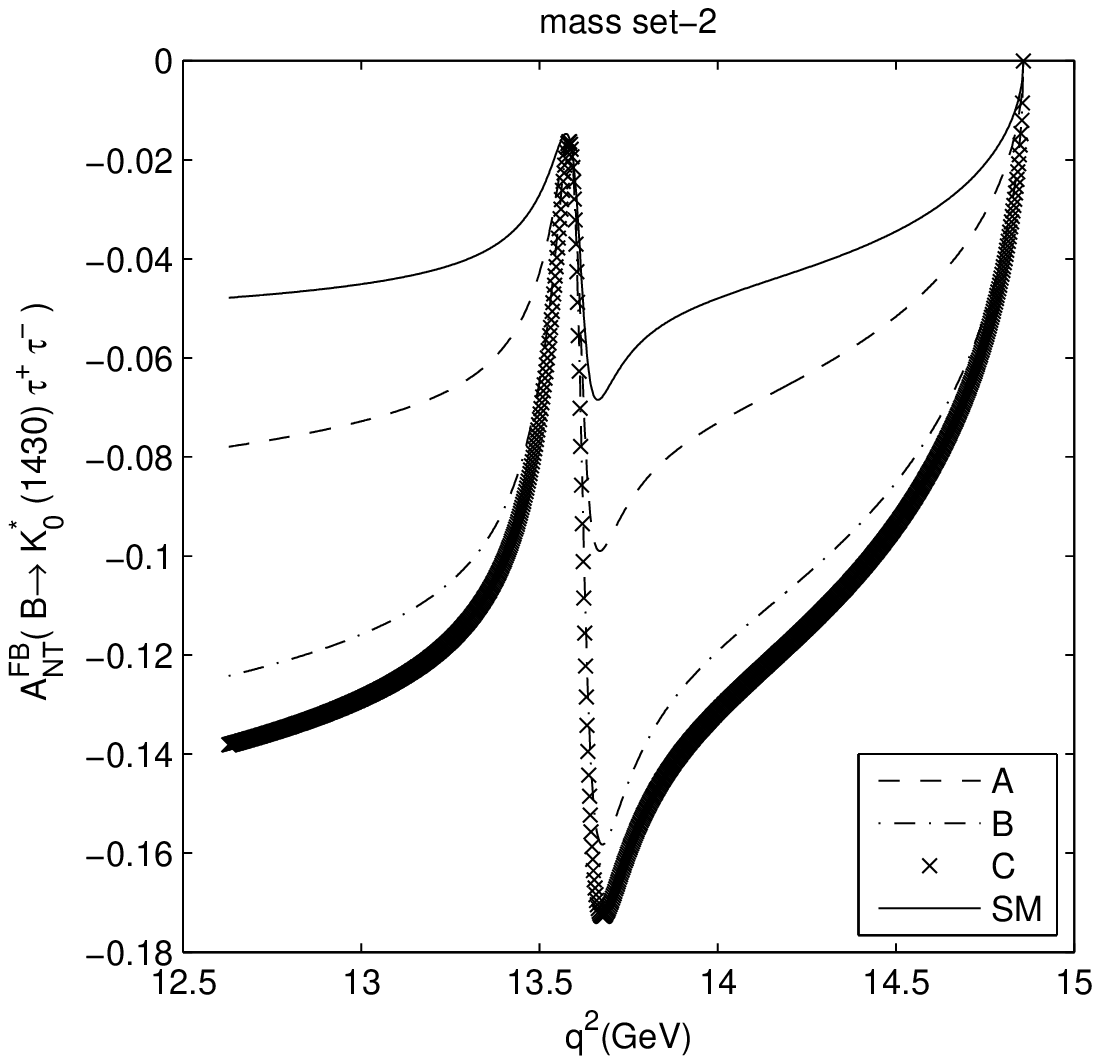}
              \includegraphics[height=2in]{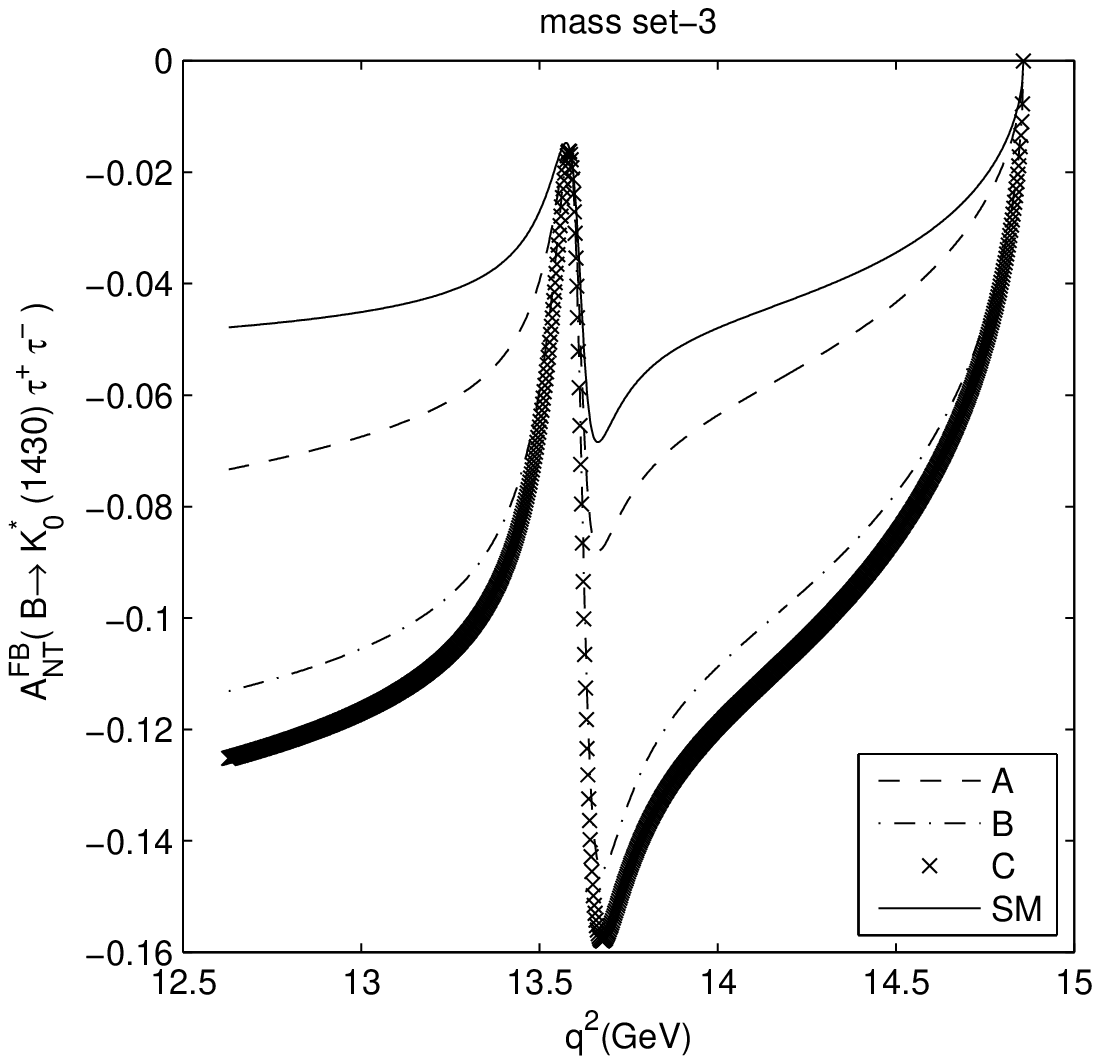}~
             \includegraphics[height=2in]{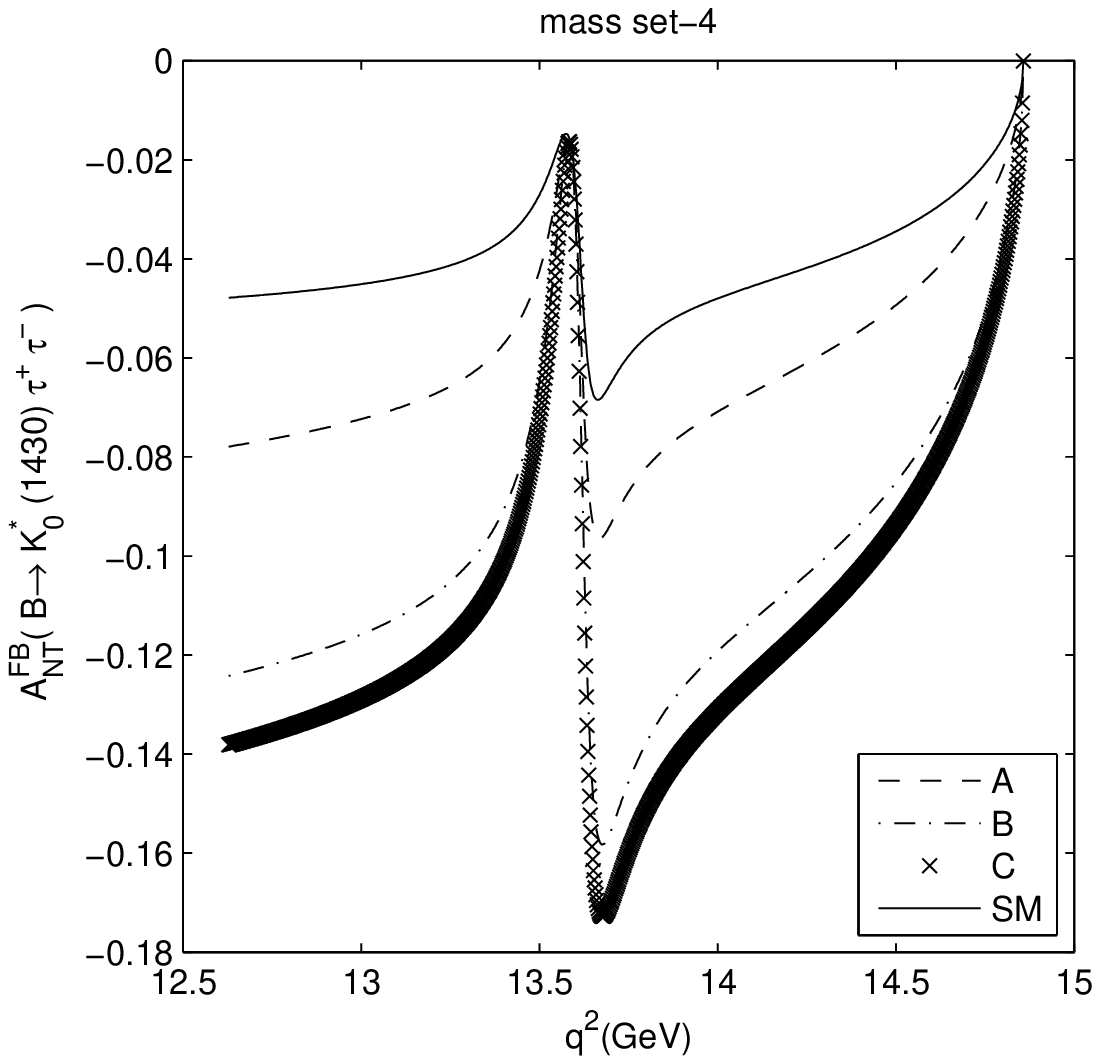}
               \caption{The dependence of the $ A_{NT}$ polarization  on $q^2$  and the three typical cases of 2HDM, i.e.
               cases A, B and C and SM  for  the $\tau$  channel of $\overline{B}\to\overline{K}_0^{*}$  transition for the mass sets 1, 2, 3 and 4.} \label{ANTtKstar}
    \end{figure}
     \begin{figure}[ht]
  \centering
  \setlength{\fboxrule}{2pt}
        \centering
                \includegraphics[height=2in]{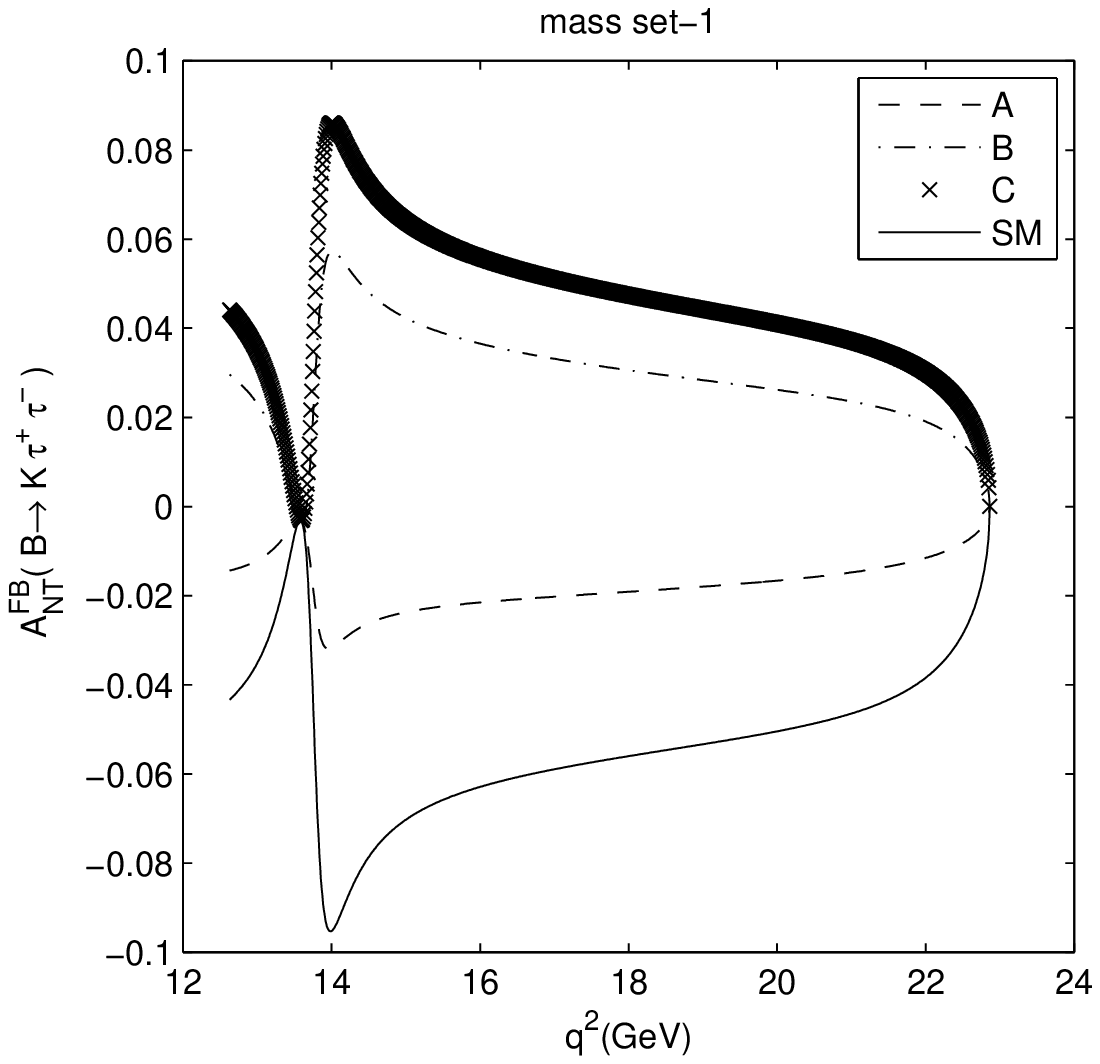}~
             \includegraphics[height=2in]{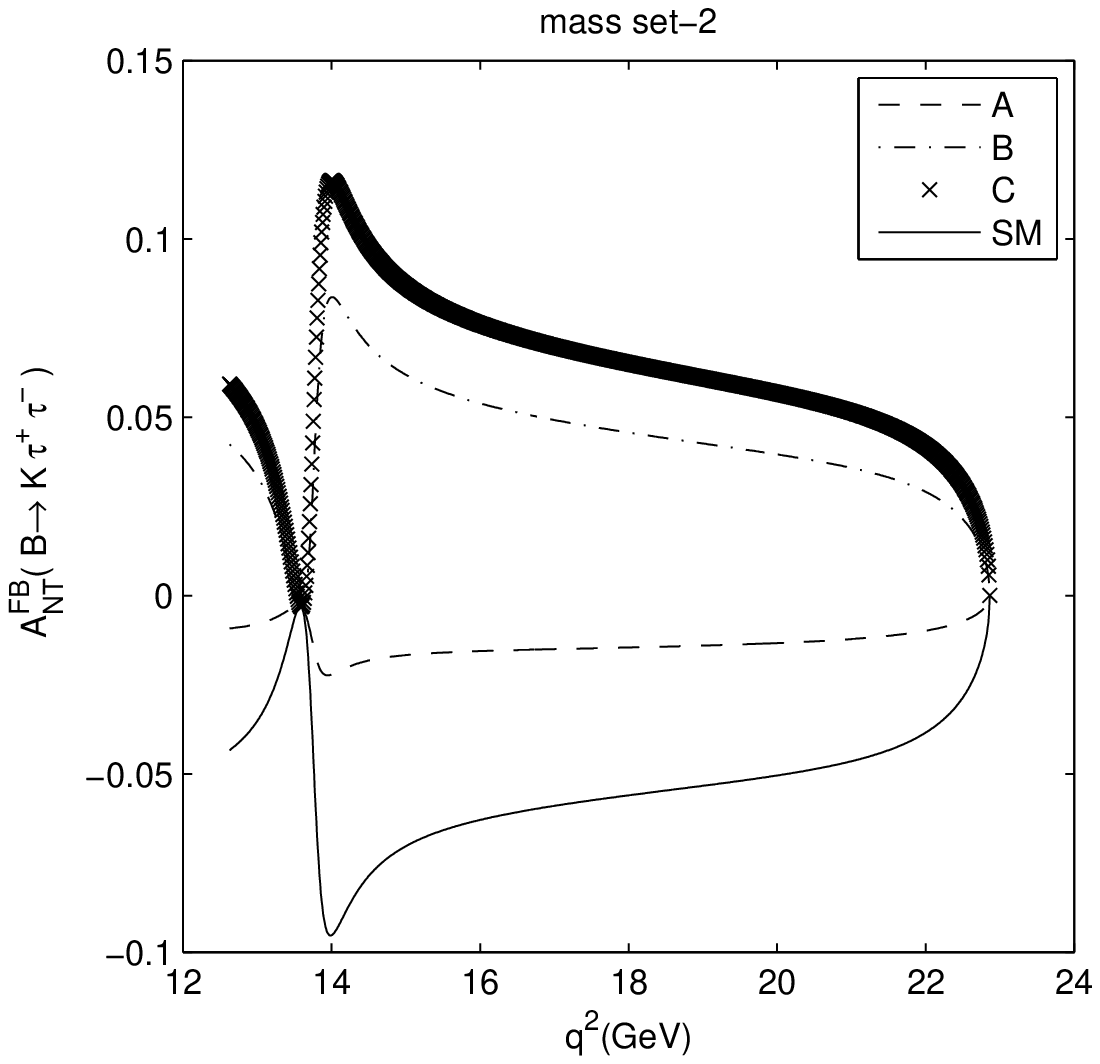}
              \includegraphics[height=2in]{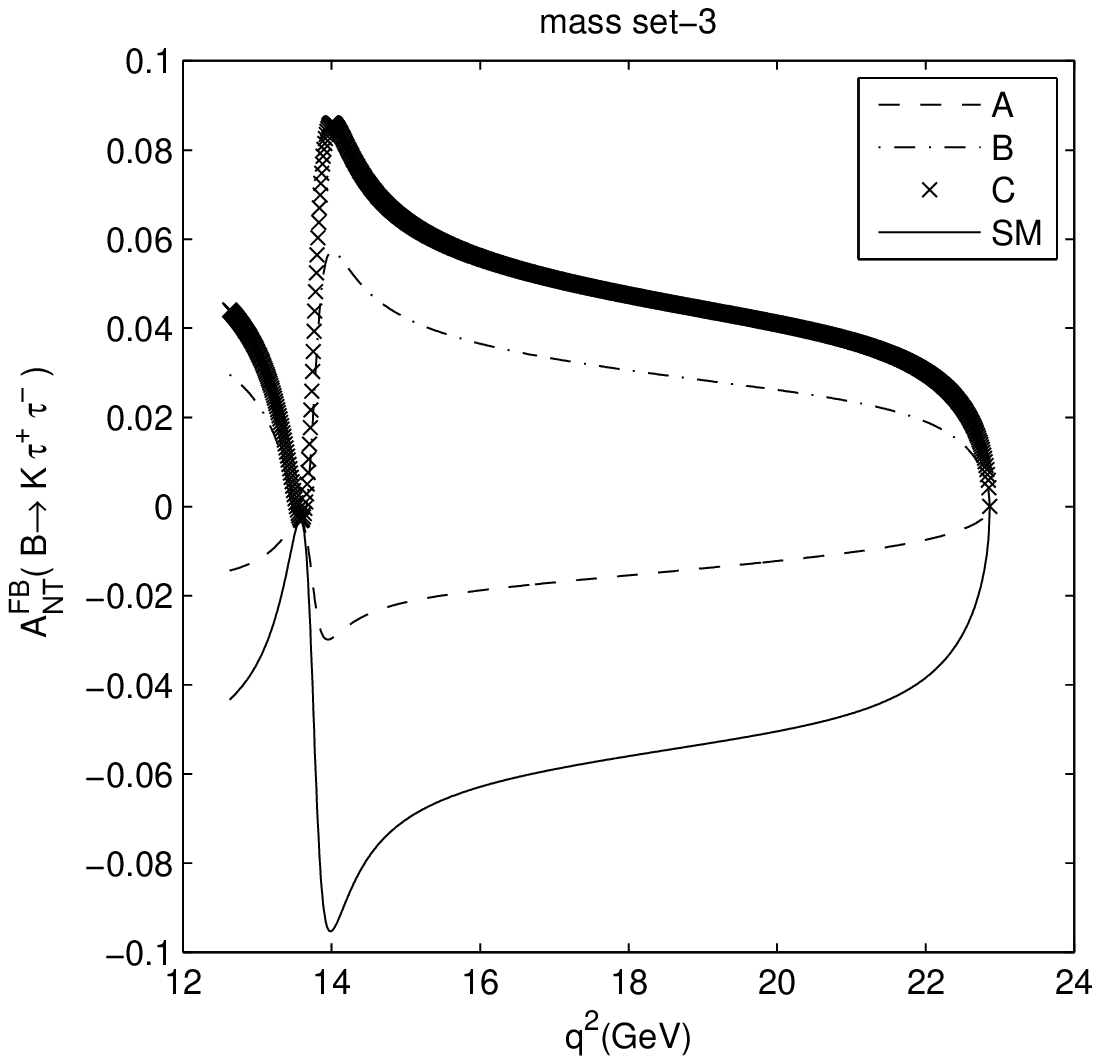}~
             \includegraphics[height=2in]{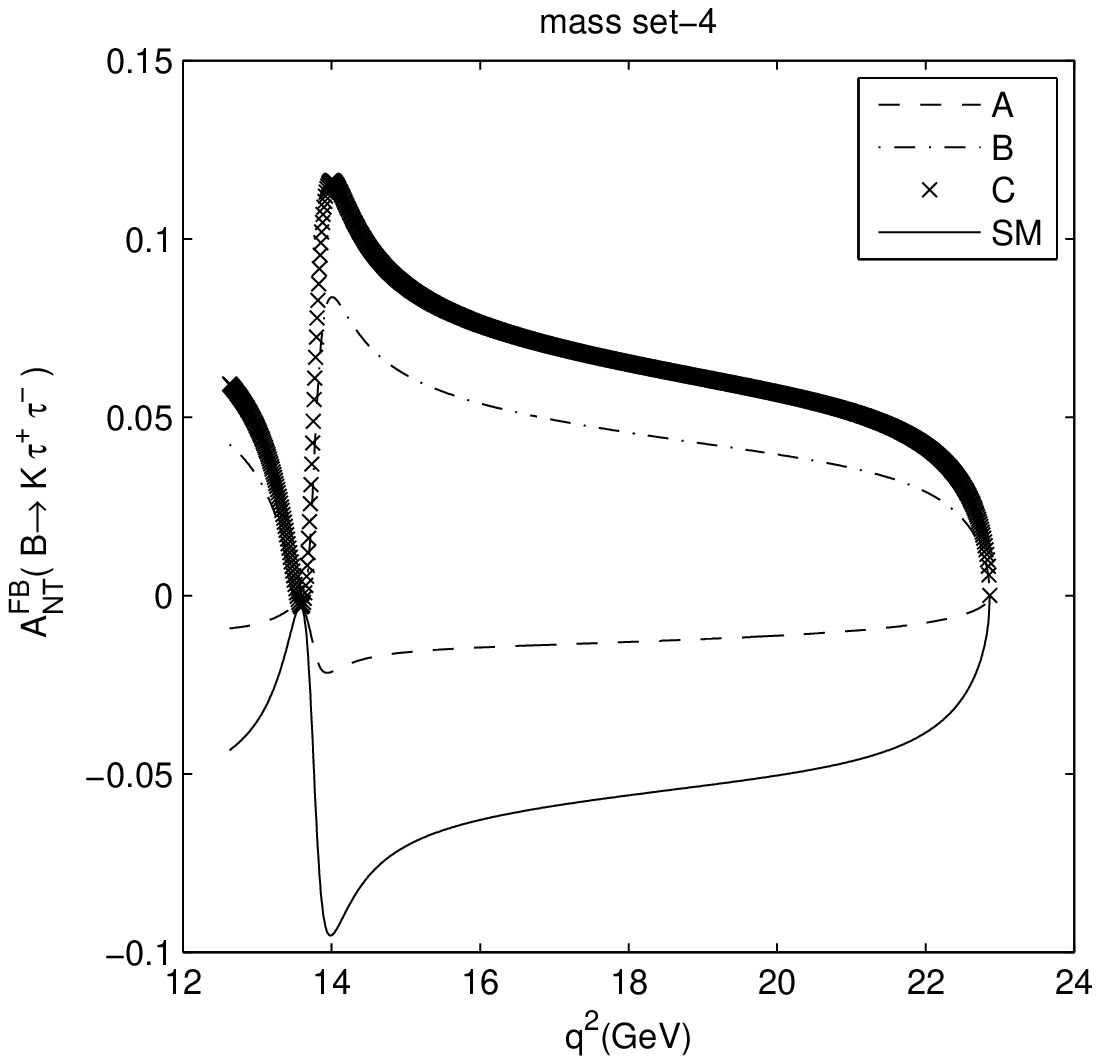}
               \caption{The dependence of the $ A_{NT}$ polarization  on $q^2$  and the three typical cases of 2HDM, i.e.
               cases A, B and C and SM  for  the $\tau$  channel of $\overline{B}\to\overline{K}$  transition for the mass sets 1, 2, 3 and 4.} \label{ANTtK}
    \end{figure}
    \end{document}